\documentclass[12pt, titlepage]{article}

\usepackage{color}
\usepackage{float}
\usepackage[bf,small]{caption2}
\usepackage{comment}
\usepackage{amsmath}
\usepackage{bigints}
\usepackage{rotating, booktabs}
\usepackage{amsopn}
\usepackage{amsfonts}
\usepackage{amsthm}
\usepackage{amssymb}
\usepackage{amsbsy}
\usepackage{gensymb}
\usepackage{multirow}
\usepackage{titling}
\usepackage{natbib}
\usepackage{tocbibind}
\usepackage[a4paper,colorlinks,breaklinks,bookmarksopen,bookmarksnumbered]{hyperref}
\usepackage{epsfig,psfrag}
\usepackage{graphicx}
\usepackage{amsmath}
\usepackage{epstopdf}
\usepackage{tabularx}
\usepackage{subfigure}
\usepackage[mathscr]{euscript}
\usepackage[margin=1in]{geometry}
\usepackage{xr-hyper}

\usepackage{amsfonts} 
\usepackage{calc}
\usepackage{titlesec}

\newcommand{\ueq}[1][]{%
  \if\relax\detokenize{#1}\relax
    \sbox0{$\underbrace{=}_{}$}%
    \mathrel{\mathmakebox[\wd0]{=}}
  \else
    \mathrel{\underbrace{=}_{\mathclap{#1}}}
  \fi}

\newcommand{\bzero}{\boldsymbol{0}}
\newcommand{\bone}{\boldsymbol{1}}

\newcommand {\ctn}{\cite}

\usepackage{url}
\renewcommand{\t}{\ensuremath{\theta}}

\newcommand{\ba}{\boldsymbol{a}}
\newcommand{\bb}{\boldsymbol{b}}

\newcommand{\btheta}{\boldsymbol{\theta}}
\newcommand{\bta}{\boldsymbol{\theta}}

\newcommand{\bbeta}{\boldsymbol{\beta}}

\newcommand{\bphi}{\boldsymbol{\phi}}

\newcommand{\bTheta}{\boldsymbol{\Theta}}

\newcommand{\bSigma}{\boldsymbol{\Sigma}}

\newcommand{\bmu}{\boldsymbol{\mu}}
\newcommand{\bzeta}{\boldsymbol{\zeta}}

\newcommand{\bvartheta}{\boldsymbol{\vartheta}}

\newcommand{\bB}{\boldsymbol{B}}
\newcommand{\bC}{\boldsymbol{C}}

\newcommand{\bV}{\boldsymbol{V}}

\newcommand{\bh}{\boldsymbol{h}}

\newcommand{\bI}{\boldsymbol{I}}
\newcommand{\bJ}{\boldsymbol{J}}

\newcommand{\bA}{\boldsymbol{A}}

\newcommand{\bu}{\boldsymbol{u}}

\newcommand{\bs}{\boldsymbol{s}}
\newcommand{\bU}{\boldsymbol{U}}
\newcommand{\bx}{\boldsymbol{x}}
\newcommand{\bX}{\boldsymbol{X}}

\newcommand{\bY}{\boldsymbol{Y}}

\newtheorem{theorem}{Theorem}

\newtheorem{definition}[theorem]{Definition}

\newcommand{\e}{\ensuremath{\epsilon}}

\newcommand{\bM}{\mathbf M}

\renewcommand{\t}{\ensuremath{\theta}}

\newcommand{\topline}{\hrule height 1pt width \textwidth \vspace*{2pt}}
\newcommand{\botline}{\vspace*{2pt}\hrule height 1pt width \textwidth \vspace*{4pt}}
\newtheorem{algo}{Algorithm} 

\newcommand{\bi}[1]{\mbox{\boldmath{$ #1 $}}}

\begin{document}

\title{\vspace{-0.8in}
\textbf{Bayesian L\'{e}vy-Dynamic Spatio-Temporal Process: Towards Big Data Analysis}}
\author{Sourabh Bhattacharya\thanks{
Sourabh Bhattacharya is a Professor in Interdisciplinary Statistical Research Unit, Indian Statistical
Institute, 203, B. T. Road, Kolkata 700108.
Corresponding e-mail: sourabh@isical.ac.in.}}
\date{\vspace{-0.5in}}
\maketitle%
	
\begin{abstract}

In this era of big data, all scientific disciplines are evolving fast to cope up with the enormity of the available information. So is statistics, the queen of science.
Big data are particularly relevant to spatio-temporal statistics, thanks to much-improved technology in satellite based remote sensing and Geographical Information Systems.
As can be anticipated, a plethora of methods to take on the challenges of big data, have been poured into the statistical literature. Our survey reveals that
the purpose of entire chunk of new methods is to simplify models and methods based on Gaussian processes, and even so, application to any significantly large data
does not seem to be reported in the literature.

Since Gaussian process models and methods are somewhat limited in the sense that real phenomena are usually non-Gaussian and exhibit nonstationarity and nonseparability
with respect to space and time, it is pertinent to come up with new ideas that emulate the reality and amenable to cheap computation for dealing with large data.
In this regard, with the L\'{e}vy random fields as the starting point, we construct a new nonparametric, nonstationary and nonseparable dynamic  
spatio-temporal process with the additional realistic property that the lagged spatio-temporal correlations converge to zero as the lag tends to infinity. 
The process is flexibly applicable even to phenomena with weak temporal dynamics and purely spatial setups. 
We refer to this new process as L\'{e}vy-dynamic spatio-temporal process.
We incorporate spatio-temporal random effects in the model to further enhance its applicability and effectiveness, and adopt the Bayesian paradigm for our purpose.

Although our Bayesian model seems to be intricately structured and is variable-dimensional with respect to each time index, we are able to devise a fast and efficient
parallel Markov Chain Monte Carlo (MCMC) algorithm for Bayesian inference. Our simulation experiment brings out quite encouraging performance
from our Bayesian L\'{e}vy-dynamic approach.

We finally apply our Bayesian L\'{e}vy-dynamic model and methods to a sea surface temperature dataset consisting of $139,300$ data points in space and time.
Although not big data in the true sense, this is a large and highly structured data by any standard. Even for this large and complex data, 
our parallel MCMC algorithm, implemented on $80$ processors, generated $11\times 10^4$ MCMC realizations from the L\'{e}vy-dynamic posterior within a single day,
and the resultant Bayesian posterior predictive analysis turned out to be encouraging. Thus, it is not unreasonable to expect that with significantly more computing
resources, it is feasible to analyse terabytes of spatio-temporal data with our new model and methods.
\\[2mm]

\end{abstract}
	
\tableofcontents
\pagebreak

\section{Introduction}

This is the era of ``big data" and the scientific community, including the statistical community, is mesmerized by the sheer charm of the phrase! 
So much so that editors and reviewers of so-called reputed journals keep rejecting papers submitted to those journals 
on account of their perception of inapplicability of the papers' contribution to big data (in our experiences)! 
Such papers are often related to genetics and spatio-temporal statistics where complex dependence structures play the key roles.
However, big data refers to at least one terabyte of data,
and from that perspective, no sensible modeling approach to account for complex dependence in the underlying data-generating process can be feasible without
supercomputing resources. Even with much less amount of such structured data, very powerful and well-maintained computing facilities are necessary, which are usually
unavailable to individual researchers. The response of the statistical community to the big data challenge (and indeed the advices of the editors and reviewers!) is to simplify
the model by ignoring most of the dependence structures, assuming linearity, and so on. One reviewer also very kindly advised not to develop sophisticated
theories and methods, since in his opinion, the linearity assumption is almost always sufficient! Breaking up the data into as many parts as possible despite
its highly dependent structure, implementing the model
on broken-up sub-datasets in the embarrassingly parallel way treating the sub-datasets as independent, 
and then combining the results in some manner, is another response being adopted by the statistical community.
Unfortunately, despite the existence of such magic tricks of the trade, much to the likings of the editors and reviewers, there does not seem 
to exist any model-based statistical work that analyzes any structured big data of the order of terabytes.

In this article, we confine ourselves to the area spatio-temporal statistics, and propose and develop a new, highly structured Bayesian nonparametric and dynamic 
spatio-temporal model based on L\'{e}vy random fields for analyzing reasonably large spatio-temporal data. 
We refer to the new underlying process as L\'{e}vy-dynamic spatio-temporal process.
With the very limited computing facilities of Indian Statistical Institute, which provides us access to only $80$ VMWare cores,
analyzing terabytes of spatio-temporal data is still infeasible. But we are able to analyze a highly structured spatio-temporal sea surface temperature data
consisting of $139,300$ observations (we chose $300$ spatial locations, each with $398$ time points as training points and set aside another 
$50$ spatial locations, each with $398$ time points for prediction) in less than $24$ hours. 
It is important to remark in this context that although our Markov Chain Monte Carlo (MCMC) algorithm
is highly intricate, it consists of parallelizable structures exploiting which we have parallelized the algorithm over
the available cores with the C language and the Message Passing Interface (MPI) protocol, leading to significant computational savings.  
We point out that our approach is flexible enough to model space-time data where the dynamic structure is less pronounced, or even purely spatial data.

In general, and definitely in the big data context, researchers do not concern themselves with the theoretical properties of their spatio-temporal models.
For instance, most real datasets are expected to arise from nonstationary, non-Gaussian stochastic processes, but it is common practice for the sake
of convenience to assume stationary Gaussian processes, usually with isotropic covariance structures. Assumption of
separability of the covariance structure with respect to spatial and temporal structures is also very much common in the literature. 
The drawbacks of such simplistic approaches did motivate researchers to develop nonstationary, nonparametric and nonseparable approaches to modeling
space-time data. However, a common limitation in all such approaches is the failure to account for the realistic property that the lagged
spatio-temporal correlations converge to zero as the lags tend to infinity, despite nonstationarity, non-Gaussianity and nonseparability.
\ctn{Das20} provide an example of such lagged correlation property in the case of a PM10 pollution dataset, which they also established to be nonstationary
and non-Gaussian. In this article, we demonstrate the same properties in the case of the sea surface temperature data. Thus, all the published approaches  
seem to be inadequate for modeling realistic spatial/spatio-temporal data.
A comprehensive account of the strengths and limitations of the existing spatio-temporal approaches is detailed in \ctn{Das20}.
In this endeavor, we show that our L\'{e}vy-dynamic space-time process possesses all the aforementioned 
realistic properties. Thus, our new model harnesses powerful parallel computing ability with desirable realistic properties for analyzing large datasets.

\subsection{Other approaches with the desirable spatio-temporal properties}
\label{subsec:other_approaches}

It is important to point out that our L\'{e}vy-dynamic approach is not the first one to consist of the desirable spatio-temporal properties. 
Indeed, the spatio-temporal process of \ctn{Das20} is a very flexible process in this regard. The process, which results from
an appropriate kernel convolution of order-based dependent Dirichlet processes (\ctn{Griffin06}), is nonstationary, nonparametric, nonseparable, and possesses
the property that the lagged spatio-temporal correlations converge to zero as the lags tend to infinity. The continuity and smoothness properties are also accounted for.  
However, the temporal part of the process does not have the dynamic structure, and considers time as an argument of the functional form of the stochastic process.
Note that such a strategy is very appropriate for various datasets where the numbers of time points vary significantly with the spatial locations, with many locations
having only a few time points. Pollution datasets on PM10 and PM2.5, for instance, are of this nature, and have been analyzed by \ctn{Das20}. In such cases,
temporal dynamics are inappropriate. However, for other cases, incorporation of temporal dynamics is important. Computationally, the method is not too demanding,
but analysis of very large data in reasonable time still seems to be infeasible. Importantly, several aspects of the MCMC algorithm can be parallelized, which might
make analyses of many large spatio-temporal datasets feasible. 

\ctn{Guha17} proposed a nonstationary, nonparametric, nonseparable dynamic state space spatio-temporal model, based on compositions of Gaussian processes 
in both the observational and the latent evolutionary levels. Under suitable conditions, the lagged spatio-temporal correlations also converge to zero. 
Continuity and smoothness properties of the process are investigated as well. But as it stands, the computational aspects seem to be too demanding 
to allow analysis of very large space-time datasets within reasonable time. However, we do have ideas to significantly improve the computational method,
along with suitable parallelization.

To our knowledge, other than our L\'{e}vy-dynamic process, 
the approaches of \ctn{Das20} and \ctn{Guha17} are the only available ones that realistically account for nonstationarity, non-Gaussianity,
nonseparability and convergence of the lagged correlations to zero. That all these properties are to be expected of real data, has been aptly demonstrated
in \ctn{Das20} with the pollution data, as already referred to. 
In this article, we shall demonstrate all these properties in detail, with respect to the sea surface temperature data that we analyze.

\subsection{Existing approaches for large spatio-temporal data analyses}
\label{subsec:large_overview}
Although our intention is to provide an overview of the existing models and methods for large space-time data, most of the relevant existing literature seems to be
exclusively concentrated on spatial data. Hence, we shall include mostly spatial methods in this brief review. 

\ctn{Banerjee17} reviews methods based on Gaussian processes for large spatio-temporal data, with focus on low-rank models and methods based on sparse covariance matrices
associated with Gaussian processes. The essence of low-rank models is to represent the underlying (Gaussian) process in terms of realizations 
of some latent process with a relatively small number
of points, so that dimension is effectively reduced. There are various approaches in this regard based on kernel convolutions and posterior expectations of the
original process given the process values at a small set of points. Sparsity in covariance matrices is induced by specifying that the spatio-temporal distance
between two points in space and time is zero beyond some specified threshold. Various issues related to the basic methods are discussed, with references to computational
gains in large datasets of size of the order $10^5$.

\ctn{Raj11} analyse a forest biomass spatial dataset consisting of about $6000$ observations, using Gaussian predictive process model, based on $25,000$ MCMC iterations.
In another work, \ctn{Raj18} divide up the available spatial data into several sub datasets, fit Bayesian Gaussian process model to each sub dataset in parallel, 
and combine the results using geometric mean of the posteriors given the sub datasets. The procedure allowed them to analyse a spatial sea surface temperature dataset
consisting of $120,000$ spatial observations (they used $117,600$ observations as training data points and set aside the rest for prediction).
The approach does not have a temporal component and that is the reason that they were forced to consider only spatial analysis of the data, confining attention the same month
(October) across the domain. 
It would have been useful if the MCMC details and computing time for this dataset were also reported in the paper.
More recently, \ctn{Guhaniyogi19} and \ctn{Guhaniyogi20} consider general Gaussian process and varying coefficient Gaussian process models
and apply the aforementioned divide-and-conquer principle, combining the sub data-specific results obtained by embarrassing parallelization, by somewhat more sophisticated
methods. In the purely spatial framework, with a large number of sub datasets the method of \ctn{Guhaniyogi19} allowed the authors to deal with datasets of size about $10^6$; 
however, their sub data-specific MCMC runs of size only $1000$, which were based on $15,000$ iterations 
(discarding the first $10,000$ and storing every fifth in the next $5,000$), are perhaps much smaller than adequate. 
Again, it would have been useful if the computing times were also reported.
The method of \ctn{Guhaniyogi20} aims for spatio-temporal models, including purely spatial and purely temporal, but dynamic structures for the spatio-temporal setup
is yet to be considered. Here the authors consider an application to the sea surface temperature data with $72,000$ space-time observations. The details of their algorithm
and the computing times would have been helpful here as well.

\ctn{Heaton18} presents a new flavour by not only reviewing Gaussian process based methods for large spatial data, but also reporting the details of a competition among various
research groups on the basis of their preferred methods for analysing given simulated and real datasets. Both the simulated and real datasets consisted of $105,569$ spatial
training observations, while the test data sets consisted of $44,431$ and $42,740$ observations, respectively.

The major concern in Gaussian process models is the
large matrix-based computations which are necessarily inefficient, and much effort of the existing works has been directed towards simplification of such matrix computations,
by various means. Clearly, far efficient computational algorithms can be achieved for approaches that are matrix-free. 
The Whittle likelihood approach (\ctn{Whittle54}, \ctn{Guyon95}) associated with the spectral domain is also matrix-free and hence amenable to fast computation, 
but in reality the approach has limited application (see \ctn{Banerjee17}). In the realm of classical spatio-temporal 
linear dynamic Gaussian state-space models where the spatial points are on a lattice grid, efficient computational strategies, that are essentially matrix-free, can be designed;
see \ctn{Dutta15}, for instance. However, for other setups available in the literature, and particularly in the Bayesian paradigm, 
appropriate matrix-free methods are difficult to devise.
%
Furthermore, issues such as nonstationarity, non-Gaussianity, nonseparability and properties of lagged correlation structures do not seem to find 
importance in the existing works related to large data. 
The relatively small sizes of the datasets and alarmingly small MCMC sample sizes usually employed in the Bayesian spatial/spatio-temporal literature also
leaves much to be desired. 

The above issues provide the motivation for introduction of our L\'{e}vy-dynamic process. 
Indeed, our L\'{e}vy-dynamic approach is completely matrix-free, and hence, needless to mention, is a right candidate for analyzing large datasets. 
In addition, our model encapsulates all the realistic
properties of spatial/spatio-temporal processes that are overlooked by the existing methods.

The rest of our article is structured as follows.
In Section \ref{sec:levy_random_field} we provide a brief overview of L\'{e}vy random fields, and in Section \ref{sec:levy_dynamic} introduce our
L\'{e}vy-dynamic spatio-temporal process. 
In Section \ref{sec:comparison}, we provide a comparison of our approach with those based on mixtures of Dirichlet processes.
The properties of the covariance structure of our L\'{e}vy-dynamic process, as well as continuity and smoothness properties, are investigated in 
Section \ref{sec:theoretical_properties}.
Specifications of relevant stochastic processes driving 
our spatio-temporal process, are provided in Section \ref{sec:choices}.
In Section \ref{sec:re}, we introduce spatio-temporal random effects in our Bayesian model to account for finer details of the underlying real phenomenon,
and in Section \ref{sec:hier}, we provide the hierarchical form of our complete Bayesian L\'{e}vy-dynamic model, along with the prior specifications.
An overview of our parallel MCMC algorithm for implementing the Bayesian model is provided in Section \ref{sec:mcmc_overview}.
In Section \ref{sec:simstudy}, we provide details of our simulation experiment for assessing the performance of our proposed model and methodologies.
Details of our analysis of the large sea surface temperature dataset are provided in Section \ref{sec:realdata}. Finally, we summarize our contributions
and make concluding remarks in Section \ref{sec:conclusion}.

Proofs of our results, the forms of the joint posterior and the full conditionals, the complete parallel MCMC algorithm for Bayesian L\'{e}vy-dynamic inference 
and technical details regarding nonstationarity, convergence to zero of the lagged correlations and non-Gaussianity of the real sea surface temperature data
are provided in the supplement, whose sections, equations, algorithms and figures have the prefix ``S-" when referred to in this paper.

\section{An overview of L\'{e}vy random fields}
\label{sec:levy_random_field}

We proceed towards L\'{e}vy random fields by first providing a briefing on L\'{e}vy random measures.

\subsection{L\'{e}vy random measure}
\label{subsec:levy_random_measure}
For any set $\bA\in\mathcal B(\mathbb R^p)$, the Borel $\sigma$-field on $\mathbb R^p$, where $\mathbb R$ is the real line and $p\geq 1$, let us define the following:
\begin{equation}
	\mathcal L(\bA)=\sum_{0\leq j<J}I_{\bA}(\bmu_j)\beta_j,
	\label{eq:levy1}
\end{equation}
where $J\sim \mathcal P(\lambda)$, the Poisson distribution with mean $\lambda$ ($>0)$, and given $J$, for $j=1,\ldots,J$, 
$(\bmu_j,\beta_j)\stackrel{iid}{\sim}\pi(d\bmu,d\beta)$, where $\pi(\cdot,\cdot)$ is some measure, not necessarily a probability measure.
Here $\bmu=\left(\mu^{(1)},\ldots,\mu^{(p)}\right)^T$ and $\bmu_j=\left(\mu^{(1)}_j,\ldots,\mu^{(p)}_j\right)^T$, and $I_{\bA}$ is the indicator function of the set $\bA$.

Then, $\mathcal L(\cdot)$ is a random signed measure such that for disjoint Borel sets $\bA_i$, $\mathcal L(\bA_i)$ are 
independent, infinitely-divisible random variables. 
This random measure is referred to as the L\'{e}vy random measure, which is endowed 
with the following form of characteristic function (see, for example, \ctn{Tu11})
\begin{equation}
	E\left[\exp\left(i\zeta\mathcal L(\bA)\right)\right]=\exp\left[\int_{\bA}\int_{\mathbb R}\left\{\exp(i\zeta\beta)-1\right\}\nu(d\bmu,d\beta)\right],
	\label{eq:levy2}
\end{equation}
where $\nu(d\bmu,d\beta)=\lambda\pi(d\bmu,d\beta)$, is referred to as the L\'{e}vy measure. 
The L\'{e}vy measure is not required to be finite, provided that (\ref{eq:levy2}) is well-defined for all $\zeta\in\mathbb R$. For details regarding
integrability in the case of infinite L\'{e}vy measure, see \ctn{Tu11} and \ctn{Apple04}. For our purpose, we shall consider only finite L\'{e}vy measure. 

When (\ref{eq:levy2}) is well-defined, it is possible to construct (\ref{eq:levy1}) using integrals with respect to Poisson random measures. That is, let 
$\mathcal N(d\bmu,d\beta)\sim \mathcal P(\nu(d\bmu,d\beta))$ be the Poisson random measure, 
so that for disjoint Borel sets $\bC_i\subseteq\mathbb R^{p+1}$, $\mathcal N(\bC_i)\sim \mathcal P\left(\nu(\bC_i)\right)$
independently. Then for any Borel set $\bA$ with compact closure, given $J=\mathcal N(\mathbb R^{p+1})$,
\begin{equation}
	\mathcal L(\bA)=\int_{\bA}\int_{\mathbb R}\beta\mathcal N(d\bmu,d\beta)=\sum_{0\leq j<J}I_{\bA}(\bmu_j)\beta_j,
	\label{eq:levy3}
\end{equation}
where given $J$, $\left\{(\bmu_j,\beta_j):j=1,\ldots,J\right\}$ is the random set of support points of the Poisson random measure.

For examples of the L\'{e}vy measure and the corresponding L\'{e}vy random measures, see \ctn{Tu11} and \ctn{Apple04}.  

\subsection{L\'{e}vy random field}
\label{subsec:levy_random_field}
Consider any real-valued measurable function $g$ on $\mathbb R^p$. Then, again using integration with respect to Poisson random measure as in (\ref{eq:levy3}), 
consider the following representation (see \ctn{Tu11}):
\begin{equation}
	\mathcal L[g]=\int_{\mathbb R^p}\int_{\mathbb R}\beta g(\bmu)\mathcal N(d\bmu,d\beta)=\sum_{0\leq j<J}g(\bmu_j)\beta_j.
	\label{eq:levy4}
\end{equation}
The representation (\ref{eq:levy4}) constitutes the L\'{e}vy random field. This is well-defined for bounded measurable functions $g$ when the L\'{e}vy measure is finite.

Now, extending $g(\bmu)$ to $g(\bx,\bmu)$, where $\bx\in\mathbb R^p$, we obtain from (\ref{eq:levy4}):
\begin{equation}
	\mathcal L[g(\bx)]=\int_{\mathbb R^p}\int_{\mathbb R}\beta g(\bx,\bmu)\mathcal N(d\bmu,d\beta)=\sum_{0\leq j<J}g(\bx,\bmu_j)\beta_j.
	\label{eq:levy5}
\end{equation}
The extension (\ref{eq:levy5}), the similar form of which is provided in \ctn{Tu11}, will play an important role in our spatio-temporal modeling strategy.

Next, we introduce our proposed idea of nonstationary, nonseparable, dynamic spatio-temporal process that uses aspects of L\'{e}vy 
random fields of the form (\ref{eq:levy5}) as building blocks.

\section{L\'{e}vy-dynamic spatio-temporal process}
\label{sec:levy_dynamic}

For $i=1,\ldots,n$ and $k=1,\ldots,m$, let $y(\bs_i,t_k)$ denote the response at location $\bs_i=\left(s^{(1)}_{i},\ldots,s^{(p)}_{i}\right)^T\in\mathbb R^p$ ($p\geq 2$) 
and time point $t_k$. The time points $t_k$; $k=1,\ldots,m$ need not be equispaced.
Let us begin with the following model for $y(\bs_i,t_k)$:
\begin{equation}
	y(\bs_i,t_k)=f(\bs_i,t_k)+\epsilon_{ik},
\label{eq:model}
\end{equation}
where, for $i=1,\ldots,n$ and $k=1,\ldots,m$, $\epsilon_{ik}\stackrel{iid}{\sim}N(0,\sigma^2)$, for unknown $\sigma^2$.
We represent the spatio-temporal process $f(\bs,t)$ using the same principle as (\ref{eq:levy5}):
\begin{equation}
	f(\bs,t)=\sum_{0\leq j< J_t}K(\bM(\bs)-\bmu_{jt},t-\tau|\bSigma,\xi)\beta_{jt}.
	\label{eq:levy6}
\end{equation}
In the above, $K(\bs,t|\bSigma,\xi)$ is some appropriately chosen bounded kernel, for example, $K(\bs,t|\bSigma,\xi)=\exp\left\{-\frac{1}{2}\bs^T\bSigma\bs-\xi |t|\right\}$,
where $\xi>0$ and $\bSigma$ is positive definite (see also \ctn{Higdon98}, \ctn{Higdon99}, \ctn{Higdon02}, \ctn{Tu11}). 
In this work, we shall concentrate on kernels of the above form.

In (\ref{eq:levy6}), 
$$\bM(\bs)=\left(M_1\left(s^{(1)}\right),M_2\left(s^{(2)}\right),\ldots,M_p\left(s^{(p)}\right)\right)^T;$$ 
each $M_\ell(\cdot)$; $\ell=1,\ldots,p$, being an almost surely monotonically increasing stochastic process. 
Also, for $j=1,2,\ldots$, $\left\{(\bmu_{jt},\beta_{jt}):t=1,2,\ldots\right\}\stackrel{iid}{\sim}\pi$, where $\pi$ denotes some appropriate stationary stochastic process
and $J_t\stackrel{iid}{\sim}P(\lambda)$, for $t=1,2,\ldots$, where $P(\lambda)$ denotes the Poisson distribution with parameter $\lambda$. 

Since $\pi$, the stochastic process for $\left\{(\bmu_{jt},\beta_{jt}):t=1,2,\ldots\right\}$, is stationary, the marginal distributions of 
$(\bmu_{jt},\beta_{jt})$ are the same for all $t$ and $j$. It follows that when $\bM(\bs)=\bs$, the 
marginal distribution $f(\bs,t)$ reduces to the same form as that of \ctn{Tu11}, when $\bSigma_j=\bSigma$ and $\tau_j=\tau$ in their case.

Now, there might arise the question regarding independence of $\bSigma$, $\tau$ and $\xi$ of $j$ and $t$.
Note that, since $J_t$ depends upon $t$, it is not possible to assume that $\bSigma=\bSigma_j$, $\tau=\tau_j$ and $\xi=\xi_j$, as this would imply that for different $t$,
the dimensions of $\left\{(\bSigma_j,\tau_j,\xi_j):j=1,\ldots,J_t\right\}$ are different, which would not make sense. 
The assumptions $\bSigma=\bSigma_{jt}$, $\tau=\tau_{jt}$ and $\xi=\xi_{jt}$ 
are sensible in this regard, but since $\bmu_{jt}$ and $\beta_{jt}$ are already time-dependent, time-dependence of $\bSigma$, $\tau$ and $\xi$ might lead to temporal bias
emerging from too many temporal dependence structures. Moreover, $\tau$ and $\xi$ are anyway associated with the temporal part of the kernel, which is a function
of the time index $t$.

There might also arise the question on not allowing $J_t$ to depend upon $\bs$. Again, by the same argument as above, this does not make sense unless
$\bmu_{jt}$ and $\beta_{jt}$ are also made $\bs$-dependent, which might lead to spatial bias in this case since the kernel is already spatially dependent.
Moreover, making $\bmu_{jt}$ and $\beta_{jt}$ spatially dependent is expected to bring in much computational burden, while no inferential gain is expected.

Also note that since unlike space, time is dynamic in nature, 
postulation of dynamic structures for $\bmu_{jt}$ and $\beta_{jt}$ is indispensable for imparting a dynamic structure to our model, but for spatial dependence
no more structure is necessary given spatial dependence of the kernel.

There is also an important issue regarding the temporal dynamics. In many spatio-temporal datasets, the numbers of temporal data points for the spatial
locations are not only very different, but a large number of locations usually contain only a few temporal data points, even as small as just $2$ or even $1$.
These are common issues in pollutant datasets such as PM10 or PM2.5 (see, for example, \ctn{Das20}). To such data, application of our L\'{e}vy-dynamic process
with its temporal dynamics, would not be sensible. However, in these situations we can set $J_t=J$, $\bmu_{jt}=\bmu_j$ and $\beta_{jt}=\beta_j$. 
We may then also set $\bSigma=\bSigma_j$, $\tau=\tau_j$ and $\xi=\xi_j$.
That is, in the aforementioned situation, we propose modification of (\ref{eq:levy6}) to
\begin{equation}
	f(\bs,t)=\sum_{0\leq j< J}K(\bM(\bs)-\bmu_{j},t-\tau_j|\bSigma_j,\xi_j)\beta_{j}.
	\label{eq:levy_nondynamic}
\end{equation}
The temporal component of the spatio-temporal process will be taken care of in the $t-\tau_j$ part of the kernel in (\ref{eq:levy_nondynamic}).
This modified model is of course very much applicable to the above kinds of spatio-temporal data.
With the exponential kernel form $K(\bs,t|\bSigma,\xi)=\exp\left\{-\frac{1}{2}\bs^T\bSigma\bs-\xi |t|\right\}$, the covariance structure will be separable
in this case, but if desired, nonseparability can be easily enforced by slightly modifying the kernel to
$\tilde K(\bs,t|\bSigma)=\exp\left\{-\frac{1}{2}\tilde\bs^T\bSigma\tilde\bs\right\}$, where $\tilde\bs=(\bs^T,t)^T$, 
and where $\bSigma$ is a positive definite matrix with non-zero off-diagonal elements.

Note that our ideas are applicable to purely spatial context as well, by simply replacing (\ref{eq:levy_nondynamic}) with
\begin{equation}
	f(\bs)=\sum_{0\leq j< J}K(\bM(\bs)-\bmu_{j}|\bSigma_j)\beta_{j}.
	\label{eq:levy_spatial}
\end{equation}

\section{Comparison with approaches based on mixtures of Dirichlet processes}
\label{sec:comparison}

First note that the L\'{e}vy measure $\nu(d\mu,d\beta)=\beta^{-1}\exp(-\beta\eta)I_{\{\beta>0\}}d\beta \gamma(d\mu)$, for some $\sigma$-finite measure
$\gamma(d\beta)$ yields $\mathcal L(\bA)\sim \mathcal G(\gamma(\bA),\eta)$, the Gamma distribution with mean $\gamma(\bA)/\eta$, for Borel measurable $\bA$ with
$\gamma(\bA)<\infty$. Since $\nu$ is an infinite measure, this entails $J=\infty$, almost surely. This gives rise to the Gamma random field, of the form (see, for instance,
\ctn{Tu11})
\begin{equation}
	\mathcal L\left[g(x)\right]=\sum_{j=0}^{\infty}g(x,\mu_j)\beta_j,
	\label{eq:grf1}
\end{equation}
where $\beta_j>0$ for all $j$. 
Now consider the L\'{e}vy measure $\nu(d\mu,d\beta)=\alpha\beta^{-1}\exp(-\beta)I_{\{\beta>0\}}d\beta G_0(d\mu)$ where $\alpha>0$ and $G_0$ is a probability distribution.
Then replacing $\beta_j$ in (\ref{eq:grf1}) with $\beta_j/\sum_{r=0}^{\infty}\beta_r$ is equivalent to the form
\begin{align}
	\tilde{\mathcal L}\left[g(x)\right]
	&=\frac{\sum_{j=1}^{\infty}g(x,\mu_j)\beta_j}{\sum_{r=1}^{\infty}\beta_r}\notag\\
	&=\frac{\int g(x,\mu)\mathcal L(d\mu)}{\int \mathcal L(d\mu)}\notag\\
	&=\int g(x,\mu)G(d\mu),
	\label{eq:dp}
\end{align}
where, for the above L\'{e}vy measure leading to Gamma random field $\mathcal L(d\mu)$, 
$\mathcal L(d\mu)/\int \mathcal L(d\mu)\equiv G\sim DP(\alpha G_0)$, the Dirichlet process (\ctn{Ferguson73}) with mean distribution $G_0$ and precision parameter $\alpha$.

\ctn{Das20} generalize form (\ref{eq:dp}) to a spatio-temporal process with desirable properties, by replacing $g(x,\mu)$ with a kernel $K(\bx,\btheta)$ 
where $\bx=(\bs^T,t)^T$ consists of the space-time co-ordinates, and the Dirichlet process $G$ with the order-based dependent Dirichlet process (ODDP) $G_{\bx}$.
ODDP can be briefly described as follows.

\ctn{Griffin06} modify the nonparametric stick-breaking construction of \ctn{Sethuraman94}
in the following way:
for each point $\bx\in D$, where $D$ is some specified domain, they define the following, in the sense of equivalence in distribution:
\begin{equation}
G_{\bx}\equiv\sum_{i=1}^{\infty}p_i(\bx)\delta_{\btheta_{\pi_i(\bx)}},
\label{eq:odpp1}
\end{equation}
where 
\begin{equation}
p_i(\bx)=V_{\pi_i(\bx)}\prod_{j<i}(1-V_{\pi_j(\bx)}).
\label{eq:odpp2}
\end{equation}
In (\ref{eq:odpp1}) and (\ref{eq:odpp2}), $\bi{\pi}(\bx)=(\pi_1(\bx),\pi_2(\bx),\ldots)$
denotes the ordering at $\bx$, where $\pi_i(\bx)\in\{1,2,\ldots\}$ and $\pi_i(\bx)=\pi_j(\bx)$
if and only if $i=j$. The ordering at each $\bx$ is random and is induced by a stationary Poisson point process, creating spatio-temporal dependence.  
For $j=1,2,\ldots$, $\btheta_j\stackrel{iid}{\sim} G_0$ and $V_j\stackrel{iid}{\sim}\mathcal B(1,\alpha)$, where $\mathcal B(a,b)$ denotes the Beta
distribution with parameters $a>0$ and $b>0$.
The process associated with specification (\ref{eq:odpp1}) is the ODDP.
Observe that ODDP reduces to Dirichlet process at all $\bx$ if $\pi_i(\bx)=i$ for each $\bx$ and $i$.

Using kernel convolution of ODDP, \ctn{Das20} considered the following form for their spatio-temporal model: 
\begin{equation}
f(\bx)=\int K(\bx,\btheta)dG_{\bx}(\btheta)= \sum_{i=1}^{\infty}K(\bx,\btheta_{\pi_{i}(\bx)})p_{i}(\bx).
\label{eq:kernel_convolution}
\end{equation}
For nonstationary kernels $K(\bx,\btheta)$, \ctn{Das20} showed that the spatio-temporal process $f(\cdot)$ is nonstationary with respect to space and time, 
yet the spatio-temporal correlation converges to zero as the lag tends to infinity. The correlation structure is nonseparable in general, but separability
can be enforced if desired. Thus, (\ref{eq:kernel_convolution}) qualifies as a very flexible and realistic spatio-temporal process, and is of course superior to
ODDP itself, since the latter is always stationary and nonseparable.

As already mentioned in Section \ref{subsec:other_approaches}, a possible drawback of (\ref{eq:kernel_convolution}) is its non-dynamic 
nature with respect to the temporal component. Dynamism can be imparted to it by letting the ODDP orderings depend only upon the spatial locations $\bs$, while
$G_0$, $\alpha$ and the Poisson point process may be allowed to be time-variant, with explicit temporal dependence structures. However, it is unclear as of now if the strategy
will ensure decay of the spatio-temporal correlations to zero with lag tending to infinity. 
Furthermore, (\ref{eq:kernel_convolution}) is not even continuous, and this may be a limitation where smooth processes are desirable. 

On the other hand, in Section \ref{sec:theoretical_properties}, we show that our L\'{e}vy-dynamic process satisfies all the desirable properties, despite being
temporally dynamic. 
Hence, at least in the current scenario, it seems that our L\'{e}vy-dynamic process is more flexible and useful compared to the Dirichlet process approaches.

\section{Theoretical properties of the L\'{e}vy-dynamic process}
\label{sec:theoretical_properties}

In the spatial context associated with the form (\ref{eq:levy_spatial}) with $\bM(\bs)=\bs$, \ctn{Clyde07} provide several examples to 
illustrate that various isotropic covariance functions  
can be obtained by suitably choosing the kernel and the L\'{e}vy measure. In fact, it follows from the Radon transform based treatment considered in Section 2.5
of \ctn{Chiles99} that almost all of the common isotropic covariance functions correspond to some specific kernel and L\'{e}vy measure associated with 
(\ref{eq:levy_spatial}) with $\bM(\bs)=\bs$. Thus, our general spatio-temporal structure (\ref{eq:levy6}) includes nearly all of the isotropic geostatistical covariance functions
as special cases, and goes on to provide far richer class of covariance functions, almost surely including nonstationary and nonseparable ones when $\bM(\cdot)$ is random. 

Indeed, as iterated several times, in keeping with reality, it is important to ensure nonstationarity of the proposed spatio-temporal process. It is also desirable that
the spatio-temporal correlations converge to zero as the lag tends to infinity, which is expected of the underlying real phenomenon. 
For the L\'{e}vy-dynamic process, we establish both the properties in Section \ref{subsec:cov_properties}. Nonseparability of the covariance structure 
of our process with respect to space and time is evident from the covariance form that we provide with respect to these results, and hence we do not provide
any separate result on nonseparability. 
Furthermore, we investigate continuity and smoothness properties of our process in Sections \ref{subsec:continuity} and \ref{subsec:smoothness}, respectively.

\subsection{Covariance properties}
\label{subsec:cov_properties}
The purpose of this subsection is two-fold. First, we show that the covariance structure of the L\'{e}vy-dynamic process is nonstationary in the sense that 
it does not depend upon the locations and times only through their differences (Theorem \ref{theorem:nonstationary_cov}). 
Thus, the process is not even weakly stationary, implying strong nonstationarity.
Then we show that the lagged covariance structure, in spite of nonstationarity, converges to zero as
the spatio-temporal lag tends to infinity (Theorem \ref{theorem:zero_cov}).
Theorems \ref{theorem:nonstationary_cov} and \ref{theorem:zero_cov} thus establish the realistically desirable properties of the L\'{e}vy-dynamic process.
\begin{theorem}
	\label{theorem:nonstationary_cov}
	The covariance structure with respect to the function $f(\cdot,\cdot)$ given by (\ref{eq:levy6}) is nonstationary, satisfying the following properties:
	\begin{itemize}
		\item[(i)] Given any $t$, $Cov(f(\bs_1,t),f(\bs_2,t))$ does not depend upon $\bs_1$ and $\bs_2$ only through $\bs_1-\bs_2$.
		\item[(ii)] Given any $\bs\in\mathbb R^p$, $Cov(f(\bs,t_1),f(\bs,t_2))$ does not depend upon $t_1$ and $t_2$ only through $t_1-t_2$.
		\item[(ii)] $Cov(f(\bs_1,t_1),f(\bs_2,t_2))$ does not depend upon $\bs_1$, $\bs_2$, $t_1$ and $t_2$ only through $\bs_1-\bs_2$ and $t_1-t_2$.
	\end{itemize}
\end{theorem}

\begin{theorem}
	\label{theorem:zero_cov}
	Assume that $K(\cdot,\cdot)$ is uniformly bounded and
	$K(\bs-\bmu,t-\tau|\bSigma,\xi)\rightarrow 0$ if either $s^{(\ell)}\rightarrow \infty$ for at least one $\ell\in\{1,\ldots,p\}$ or if $t\rightarrow\infty$ or both.
	Then,
	\begin{equation}
	Cov(f(\bs_1,t_1),f(\bs_2,t_2))\rightarrow 0,~\mbox{if either}~|t_1-t_2|\rightarrow\infty~\mbox{or}~\|\bs_1-\bs_2\|\rightarrow\infty~\mbox{or both}.
		\label{eq:cov1}
	\end{equation}
\end{theorem}




\subsection{Continuity properties}
\label{subsec:continuity}

\begin{definition}
A process $\{X(\bi{x}),\bi{x}\in \mathbb{R}^p\}$ is almost surely continuous at $\bi{x}_{0}$ if
$X(\bi{x})\rightarrow X(\bi{x}_0)$ $a.s.$ as $\bi{x}\rightarrow\bi{x_{0}}$.
If the process is almost surely continuous for every $\bi{x_{0}}\in \mathbb{R}^{p}$ then the process is said
to have continuous realizations.
\end{definition}

\begin{definition}
For $r\geq 1$, a process $\{X(\bi{x}),\bi{x}\in \mathbb{R}^p\}$ is $L_r$-continuous at $\bi{x}_0$ if
$$\underset{\bi{x}\rightarrow \bi{x}_0}{\lim} E\left|X(\bi{x})-X(\bi{x}_0)\right|^r =0.$$
\end{definition}

First, it is clear that $f(\bs,t)=\sum_{0\leq j<J_t}K(\bM(\bs)-\bmu_{jt},t-\tau|\bSigma,\xi)\beta_{jt}$ can not be almost surely continuous in 
$(\bs,t)$, since $f(\bs,t)$ is a jump process 
with respect to $t$ (and $J_t$). However, for fixed $t$, $f(\bs,t)$ can be almost surely continuous with respect to $\bs$, as the following result shows.
\begin{theorem}
	\label{theorem:cont1}
	Assume that $\bM(\bs)$ is almost surely continuous in $\bs$ and that $K(\bx-\bmu,t-\tau|\bSigma,\xi)$ is continuous in $\bx$. 
	Then $f(\bs,t)$ is almost surely continuous in $\bs$.
\end{theorem}

\begin{theorem}
	\label{theorem:cont2}
	Assume that $\bM(\bs)$ is almost surely continuous in $\bs$ and that $K(\bx-\bmu,t-\tau|\bSigma,\xi)$ is continuous in $\bx$. 
	Also assume that $K(\cdot,\cdot|\cdot,\cdot)$ is uniformly bounded.
	Then $f(\bs,t)$ is $L_1$-continuous in $\bs$.
\end{theorem}

The following two results show that $f(\bs,t)$ is not $L_2$-continuous even with respect to $\bs$.
\begin{theorem}
	\label{theorem:cont4}
	Assume that $\bM(\bs)$ is almost surely continuous in $\bs$ and that $K(\bx,t)$ is continuous in $(\bx,t)$. 
	Also assume that $K(\cdot,\cdot)$ is uniformly bounded.
	Even then $f(\bs,t)$ is not $L_2$-continuous with respect to $\bs$ for any fixed $t$.
\end{theorem}

The next theorem shows that if only convergence in expectation in considered, then $f(\bs,t)$ converges to $f(\bs_0,t_0)$ in expectation as $(\bs,t)\rightarrow (\bs_0,t_0)$.
\begin{theorem}
	\label{theorem:cont3}
	Assume that $\bM(\bs)$ is almost surely continuous in $\bs$ and that $K(\bx,t)$ is continuous in $(\bx,t)$. 
	Also assume that $K(\cdot,\cdot)$ is uniformly bounded.
	Then, as $(\bs,t)\rightarrow (\bs_0,t_0)$,
	\begin{equation}
		E\left[f(\bs,t)\right]\rightarrow E\left[f(\bs_0,t_0)\right].
		\label{eq:cont2}
	\end{equation}
\end{theorem}

\subsection{Smoothness properties}
\label{subsec:smoothness}
Now we examine differentiability of our spatio-temporal process. 
\begin{definition}
A process $\{X(\bi{x}),\bx\in \mathbb{R}^p\}$ is said to be almost surely differentiable at $\bi{x}_0$ if for any 
direction $\bu$, there exists a process $L_{\bi{x}_0}(\bu)$, linear in $\bu$ such that
\begin{align*}
 X(\bi{x}_0+\bu)=X(\bi{x}_0)+ L_{\bi{x}_0}(\bu) + R(\bi{x}_0,\bu),
	\mbox{ where } \frac{R(\bi{x}_0,\bu)}{\|\bu\|} \stackrel{a.s.}\rightarrow 0,~\mbox{as}~\bu\rightarrow\bzero.
\end{align*}
If the process is almost surely differentiable at all $\bx_0\in\mathbb R^p$, then
it is said to be differentiable almost surely.
\end{definition}
Since $f(\bs,t)$ is not even continuous in $(\bs,t)$, it is certainly not differentiable. However, for fixed $t$, differentiability of our process is given
by the following result.
\begin{theorem}
	\label{theorem:sm1}
	Assume that for almost all paths, all the partial derivatives of the elements of $\bM(\bs)$ with respect to the 
	elements the $\bs$ exist and are continuous. Also assume that all the
	partial derivatives of $K(\bx-\bmu,t-\tau|\bSigma,\xi)$ with respect to the elements of $\bx$ exist and are continuous.
	Then $f(\bs,t)$ is almost surely differentiable with respect to $\bs$.
\end{theorem}

\begin{definition}
For $r\geq 1$, a process $\{X(\bi{x}),\bx\in \mathbb{R}^p\}$ is said to be $L_r$ differentiable at $\bi{x}_0$ if for any 
direction $\bu$, there exists a process $L_{\bi{x}_0}(\bu)$, linear in $\bu$ such that
\begin{align*}
 X(\bi{x}_0+\bu)=X(\bi{x}_0)+ L_{\bi{x}_0}(\bu) + R(\bi{x}_0,\bu),
	\mbox{ where } \frac{R(\bi{x}_0,\bu)}{\|\bu\|} \stackrel{L_r}\rightarrow 0,~\mbox{as}~\bu\rightarrow\bzero.
\end{align*}
\end{definition}

\begin{theorem}
	\label{theorem:sm2}
	Assume the following conditions:
	\begin{enumerate}
		\item[(A1)] For any $t$, $K(\bx-\bmu,t-\tau|\bSigma,\xi)$ has bounded second derivative with respect to $\bx$. 
		\item[(A2)]  $\bM(\cdot)$ has bounded second derivative almost surely.
		\item[(A3)] $E(\left|\beta_t\right|^r)<\infty$, for some $r\geq 1$.
\end{enumerate}
		Then $f(\bs,t)$ is $L_r$-differentiable with respect to $\bs$.
\end{theorem}

\section{Choice of increasing stochastic processes for $\bM$ and stationary processes for $\bmu_{t}$ and $\beta_{t}$}
\label{sec:choices}


\subsection{Smooth increasing stochastic processes for the components of $\bM$}
\label{subsec:smooth}

As valid increasing stochastic processes, the subordinators, which are almost surely increasing L\'{e}vy processes, merit serious consideration. 
Examples of such increasing processes are Poisson processes, $\alpha$-stable subordinators, the L\'{e}vy subordinator, inverse Gaussian subordinators, 
Gamma subordinators, etc. See \ctn{Apple04} for details on subordinators. Since these are L\'{e}vy processes, they have stationary and independent increments.
However, increasing stochastic processes with independent increments must be jump processes; see \ctn{Ferguson72}.
Hence, although subordinators qualify as models for the components of $\bM$, they fail to satisfy smoothness, or even continuity of $f$, as required by 
Theorems \ref{theorem:cont1}, \ref{theorem:cont2}, \ref{theorem:cont3}, \ref{theorem:sm1}, \ref{theorem:sm2}.
This requires us to create new increasing processes that are also smooth. Details follow.

For each $\ell=1,\ldots,p$, let us consider the following stochastic process for $M_\ell$: for $s_1,s_2\in\mathbb R$ such that $s_1>s_2$,
\begin{equation}
	M_\ell(s_1)-M_\ell(s_2)=C_\ell \tilde X_\ell(s_1-s_2)^r,
	\label{eq:mono1}
\end{equation}
where $C_\ell$ is some positive constant and $\tilde X_\ell$ is a positive random variable independent of $s_1$ and $s_2$. We set $r\geq 1$ in (\ref{eq:mono1}).
Almost sure continuity and differentiability of $M_\ell$ is achieved even with $r=1$. 
Note that under (\ref{eq:mono1}), although $M_\ell$ has stationary increments, the increments are not independent, due to the presence of $X_\ell$.

For $\ell=1,\ldots,p$ let $s^{(\ell)}_i$ denote the $\ell$-th component of $\bs_i$, for $i=1,\ldots,n$, and let 
$s^{(\ell)}_{(1)}\leq s^{(\ell)}_{(2)}\leq\ldots s^{(\ell)}_{(n)}$ denote the ordered values of $s^{(\ell)}_i$. Then data-based modeling of $M_{\ell}$ 
corresponding to (\ref{eq:mono1}) reduces to
\begin{equation}
	M_\ell\left(s^{(\ell)}_{(i)}\right)=M_\ell\left(s^{(\ell)}_{(i-1)}\right)+C_\ell \tilde X_\ell\left(s^{(\ell)}_{(i)}-s^{(\ell)}_{(i-1)}\right)^r;~i=2,\ldots,n.
	\label{eq:mono2}
\end{equation}
We shall also set $\tilde X_\ell=|X_\ell|$, where $X_\ell\sim N(\nu_\ell,\omega^2_\ell)$, where $\nu_\ell$, $\omega^2_\ell$ will be treated as unknown. 
The positive constant $C_{\ell}$ will also be treated as unknown in our setup.
We shall set $M_\ell\left(s^{(\ell)}_{(1)}\right)=\tilde C_{\ell}-C_\ell \tilde X_\ell\left|s^{(\ell)}_{(1)}\right|^r$, where $\tilde C_{\ell}>0$ will be treated as unknown.
We shall set $r=2$ in our applications.

A very important advantage of our so-created monotone processes $M_\ell$ is that they are parameterized by only five unknown quantities, namely,
$\tilde C_\ell$, $C_\ell$, $X_\ell$, $\nu_\ell$ and $\omega^2_\ell$, and hence are very amenable to cheap computation. If on the other hand, other processes, such as
the subordinators were employed, then for every $s^{(\ell)}_i$, $M_\ell\left(s^{(\ell)}_{(i)}\right)$ would be unknown, for $i=1,\ldots,n$, 
and for even moderately large $n$, would have led to significant computational burden.

\subsection{Stationary stochastic process models for $\bmu_{t}$ and $\beta_{t}$}
\label{subsec:theta_beta_model}
In Section \ref{sec:levy_dynamic} we have assumed that $\bmu_t$ and $\beta_t$ 
are stationary stochastic processes. This assumption was important in 
proving our theoretical results results. In practice, particularly, for large spatio-temporal data analysis, it is important to keep the forms of the stationary stochastic
processes as simple as possible. Thus, in practice, it is useful to model the components of $\bmu_t$ and $\beta_t$ as independent stationary AR(1) processes.  

In spite of such simplicity, the actual time series $f(\cdot,t)$ 
is nonstationary and has rich enough temporal covariance structure,
borne out by the kernel, which is rendered time-dependent through $\bmu_{jt}$ as well as $t$, and even through $\beta_{jt}$. 
The relevant results on temporal nonstationarity are provided by (ii) and (iii) of Theorem \ref{theorem:nonstationary_cov}, 
and Theorem \ref{theorem:zero_cov} shows that the covariance structure has the desirable asymptotic property.

Of course, if necessary, we can consider any desired stationary stochastic process models for $\bmu_t$ and $\beta_t$, 
with dependence among the components of $\bmu_t$ and $\beta_t$. Theorems \ref{theorem:nonstationary_cov} and \ref{theorem:zero_cov} would continue to hold
in all such situations, and for any almost surely monotonically increasing stochastic processes $M_\ell;~\ell=1,\ldots,p$, 
bringing out the flexibility and generality of our strategies.

In particular, if $t_k-t_{k-1}=1$ for all $k$, then the stationary, first order autoregressive model may be the default choice. For irregularly spaced
time series, we recommend the irregular autoregressive (IAR) model introduced by \ctn{Susana18}, which we briefly review below.

\subsubsection{Irregular autoregressive model}
\label{subsubsec:iar}
For an increasing sequence of observation times $\{t_k:k\geq 1\}$, an IAR process $\{\zeta_{t_k}:k\geq 1\}$ is defined by \ctn{Susana18} as the following:
\begin{equation}
	\zeta_{t_k}=\rho^{t_k-t_{k-1}}_{\zeta}\zeta_{t_{k-1}}+\sigma_{\zeta}\sqrt{1-\rho^{2(t_k-t_{k-1})}_{\zeta}}\epsilon_{t_k},
	\label{eq:iar1}
\end{equation}
where $\epsilon_{t_k}$ are $iid$ random variables with zero mean and unit variance. It is assumed that $\zeta_{t_1}$ is a zero-mean random variable with variance
$\sigma^2_{\zeta}$. 

It can be seen that $E(\zeta_{t_k})=0$ and $Var(\zeta_{t_k})=\sigma^2_{\zeta}$, for $k\geq 1$. For $k_1\geq k_2$, 
the covariance between $\zeta_{t_{k_1}}$ and $\zeta_{t_{k_2}}$ is given by
\begin{equation}
	Cov(\zeta_{t_{k_1}},\zeta_{t_{k_2}})=E(\zeta_{t_{k_1}}\zeta_{t_{k_2}})=\sigma^2_{\zeta}\rho^{t_{k_1}-t_{k_2}}_{\zeta},
	\label{eq:cov_iar}
\end{equation}
implying that for any $t>s$, the autocovariance function can be defined as
$\gamma(t-s)=E(\zeta_t\zeta_s)=\sigma^2_{\zeta}\rho^{t-s}_{\zeta}$, signifying a second-order weakly stationary process.
However, under some mild conditions, strict stationarity and ergodicity can be ensured, as shown in the following result of \ctn{Susana18}:
\begin{theorem}[\ctn{Susana18}]
	\label{theorem:iar}
	Consider the IAR process defined by (\ref{eq:iar1}), and let $0<\rho_{\zeta}<1$. Assume that $t_k-t_{k-n}\geq C\log n$, as $n\rightarrow\infty$, where
	$C$ is a positive constant satisfying $C\log\rho^2_{\zeta}<-1$. Then, there exists a solution to the IAR process and the sequence $\{\zeta_{t_k}:k\geq 1\}$
	is stationary and ergodic.
\end{theorem}
It is noted in \ctn{Susana18} that the case $t_k-t_{k-1}=1$ for all $k\geq 1$ corresponds to regular AR(1), which satisfies the conditions of Theorem \ref{theorem:iar},
since $t_k-t_{k-n}=n>\log n$ and $\rho^2_{\zeta}<1$ is a part of the assumptions regarding stationary AR(1) processes. Observe that regular AR(1) allows $-1<\rho_{\zeta}<1$
for stationarity, while IAR requires $0<\rho_{\zeta}<1$. Indeed, from (\ref{eq:cov_iar}) it is clear that for non-integer positive real values $t_{k_1}-t_{k_2}$,
$\rho^{t_{k_1}-t_{k_2}}_{\zeta}$ will be undefined for negative values of $\rho_{\zeta}$. This is not the case for regular AR(1) since there $t_{k_1}-t_{k_2}$
is always an integer.

Thus, in our applications, we shall model $\beta_t$ and the components of $\bmu_t$ using the regular AR(1) process when the time gap is $1$ 
and with the IAR process otherwise. Henceforth, we shall denote the $\ell$-th component of $\bmu_t$ by $\mu^{(\ell)}_t$, for $\ell=1,\ldots,p$. The corresponding
$\rho_{\zeta}$ and $\sigma^2_{\zeta}$ will be denoted by $\rho_{\ell}$ and $\sigma^2_{\ell}$, respectively. In the case of $\beta_t$, we shall denote
$\rho_{\zeta}$ and $\sigma^2_{\zeta}$ by $\rho_{\beta}$ and $\sigma^2_{\beta}$, respectively.
Although derivation of the important statistical properties of the IAR (or AR(1)) process does not require the normality assumption of the errors, for our applications,
we shall assume normality.

\section{Incorporation of random effects in the L\'{e}vy-dynamic spatio-temporal model}
\label{sec:re}
In practice, the functional form $f(\bs_i,t_k)$ driven by specific choices of the kernel $K$ need not be always sufficient to explain the
underlying spatio-temporal structure in precise details. Hence, we shall attempt to further enhance inference by considering spatio-temporal 
random effects in our model. In other words,
we shall consider the following model for data analysis:
\begin{equation}
	y(\bs_i,t_k)=\alpha+\phi(\bs_i,t_k)+f(\bs_i,t_k)+\epsilon_{ik},
	\label{eq:re_model1}
\end{equation}
where $\alpha$ is the overall effect and $\phi(\bs_i,t_k)$ are the spatio-temporal random effects. We assume that
$\alpha\sim N(\mu_{\alpha},\sigma^2_{\alpha})$ and
\begin{equation}
	\phi(\bs_i,t_k)\sim N(\phi_0(\bs_i,t_k),\sigma^2_\phi),\label{eq:re}
\end{equation}
independently for $i=1,\ldots,n$ and $k=1,\ldots,m$. In the above, $\phi_0(\bs_i,t_k)=y(\bs_{i^*},t_k)$, where $i^*=\arg\min\{\|\bs_i-\bs_j\|:j\neq i\}$. 
In cases where there are multiple minimizers $i^*_1,\ldots,i^*_N$ of  $\{\|\bs_i-\bs_j\|:j\neq i\}$ for some $N>1$, 
we define $\phi_0(\bs_i,t_k)=\sum_{j=1}^Ny(\bs_{i^*_j},t_k)/N$.
In the case of prediction of $y(\tilde\bs,\tilde t)$ at location $\tilde \bs$ and time point $\tilde t$, 
where at least one of $\tilde\bs$ or $\tilde t$ is not in the training dataset,
then, assuming that for $N\geq 1$, $\{(i^*_r,k^*_r):r=1,\ldots,N\}=\arg\min\{\|\tilde\bs-\bs_i\|^2+(\tilde t-t_k)^2:i=1,\ldots,n;~j=1,\ldots,m\}$,
we define $\phi_0(\tilde\bs,\tilde t)=\sum_{r=1}^Ny(\bs_{i^*_r},t_{k^*_r})/N$.

Thus, although the spatio-temporal dependence structure is encapsulated in the dependence among $f(\bs_i,t_k)$, the random effects $\phi(\bs_i,t_k)$,
along with the overall effect $\alpha$, 
are introduced to capture the finer details of the specific spatial location and time point associated with the data and 
enhance inference. In particular, precisions of the predictions at given locations and time points where data are not observed, are likely to
be sharper with these random effects. 
However, given the basic spatio-temporal structure offered by $f(\bs_i,t_k)$ and the way $\phi_0(\bs_i,t_k)$ are constructed, $\alpha$ and $\phi(\bs_i,t_k)$ are not expected to 
have significant variabilities. As such, we shall consider the priors for $\sigma^2_{\alpha}$ and $\sigma^2_\phi$ to reflect the opinion that with relatively high certainty 
they are not much different from zero. Note that $\alpha$ may be viewed as the average of all the $y(\bs_i,t_k)$ and can be set to zero when $y(\bs_i,t_k)$
are standardized to have mean zero and variance one. 

Observe that although there are $nm$ random effects in our model, these can be integrated out from (\ref{eq:re_model1}), so that under the marginalized model, $y(\bs_i,t_k)$
admits the representation
\begin{equation}
	y(\bs_i,t_k)=\alpha+\phi_0(\bs_i,t_k)+f(\bs_i,t_k)+\tilde\epsilon_{ik},
	\label{eq:re_model_marginalized1}
\end{equation}
where $\tilde\epsilon_{ik}\sim N(0,\sigma^2_{\epsilon}+\sigma^2_\phi)$, independently. As argued above, there are reasons to consider a prior for $\sigma^2_\phi$
that concentrates around zero. Thus, it would make sense to deterministically set $\sigma^2_\phi\approx 0$, 
which would also make $\sigma^2_{\epsilon}$ identifiable in the variance
$\sigma^2_{\epsilon}+\sigma^2_\phi$ of $\tilde\epsilon_{ik}$. This entire exercise certainly leads to huge computational savings compared to the original, non-marginalized
version. Importantly, with $\sigma^2_\phi=0$, we shall demonstrate with our simulation experiment that although the non-marginalized version has better MCMC mixing properties, 
the final Bayesian prediction results are remarkably similar for the two versions. Thus, we shall also consider the marginalized version with $\sigma^2_\phi=0$ in the
real data scenario.

It is worth mentioning that ours is not the first spatio-temporal work to incorporate random effects.
Random effects in spatial and spatio-temporal setups have also been considered in \ctn{Kang11} and \ctn{Wu16}; see also Section 4.4.1 of \ctn{Wikle19}. 


\section{Hierarchical form of the L\'{e}vy-dynamic model with prior details}
\label{sec:hier}
For the sake of generality, we assume the time points to be of the form $\{t_k:k\geq 1\}$.
Then our Bayesian L\'{e}vy-dynamic spatio-temporal model admits the following hierarchical form:
for $i=1,\ldots,n$ and $k=1,\ldots,m$,
\begin{align}
	&y(\bs_i,t_k)\sim N\left(\alpha+\phi(\bs_i,t_k)+f(\bs_i,t_k),\sigma^2_{\epsilon}\right);\notag\\ 
	& \alpha\sim N(\mu_\alpha,\sigma^2_{\alpha});~\phi(\bs_i,t_k)\sim N(\phi_0(\bs_i,t_k),\sigma^2_\phi);\notag\\ 
	&f(\bs_i,t_k)=\sum_{0\leq j< J_{t_k}}\exp\left\{-\frac{1}{2}(\bM(\bs_i)-\bmu_{jt_k})^T\bSigma (\bM(\bs_i)-\bmu_{jt_k})-\xi|t_k-\tau|\right\}
	\beta_{jt_k};\notag\\ 
	&J_{t_k}\sim \mathcal P(\lambda);\notag\\ 
	& M_\ell\left(s^{(\ell)}_{(i)}\right)-M_\ell\left(s^{(\ell)}_{(i-1)}\right)=C_\ell \tilde X_\ell\left(s^{(\ell)}_{(i)}-s^{(\ell)}_{(i-1)}\right)^r;~\ell=1,\ldots,p;
	\notag\\ 
	& M_\ell\left(s^{(\ell)}_{(1)}\right)=\tilde C_{\ell}-C_\ell \tilde X_\ell\left|s^{(\ell)}_{(1)}\right|^r;~\ell=1,\ldots,p;\notag\\ 
	&X_{\ell}\sim N\left(\nu_{\ell},\omega^2_{\ell}\right);~\ell=1,\ldots,p;\notag\\ 
	&\beta_{jt_k}\sim N\left(\rho_{\beta}\beta_{j,t_{k-1}},\sigma^2_{\beta}\left(1-\rho^{2(t_k-t_{k-1})}_{\beta}\right)\right);~k=2,\ldots,m,
	~\mbox{where}~0<\rho_{\beta}<1;\notag\\ 
	&\beta_{jt_1}\sim N\left(0,\sigma^2_{\beta}\right);\notag\\ 
	&\mu^{(\ell)}_{jt_k}\sim N\left(\rho_{\ell}\mu^{(\ell)}_{j,t_{k-1}},\sigma^2_{\ell}\left(1-\rho^{2(t_k-t_{k-1})}_{\ell}\right)\right);~k=2,\ldots,m,
	~\mbox{where}~0<\rho_{\ell}<1;~\ell=1,\ldots,p;\notag\\ 
	&\mu^{(\ell)}_{jt_1}\sim N\left(0,\sigma^2_{\ell}\right);~\ell=1,\ldots,p;\notag\\ 
	&(\bSigma,\tau)\sim\pi_{\Sigma}\times\pi_{\tau};\label{eq:prior1}\\
	&(\lambda,\xi,C_1,\ldots,C_p,\tilde C_1,\ldots,\tilde C_p,\nu_1,\ldots,\nu_p,\omega^2_1,\ldots,\omega^2_{\ell},\notag\\
	&\qquad\qquad\rho_{\beta},\sigma^2_{\beta},\rho_1,\ldots,\rho_p,\sigma^2_1,\ldots,\sigma^2_p,\sigma^2_{\epsilon},\sigma^2_{\alpha},\sigma^2_\phi)
	\sim\pi;\label{eq:prior2}
\end{align}
We shall further set $\bSigma$ to be a diagonal matrix with unknown positive diagonal elements $\tilde\sigma^2_1,\ldots,\tilde\sigma^2_p$. 
The specific prior forms for (\ref{eq:prior1}) and (\ref{eq:prior2}) would be the following in our applications:
\begin{align}
	&\tilde\sigma^2_{\ell}\sim \mathcal{IG}(a_{\tilde\sigma^2_\ell},b_{\tilde\sigma^2_\ell});~\ell=1,\ldots,p;\notag\\ 
	&\tau\sim \mathcal{IG}(a_{\tau},b_{\tau});\notag\\ 
	&\lambda\sim \mathcal G(a_{\lambda},b_{\lambda});\notag\\ 
	&\xi\sim \mathcal{IG}(a_{\xi},b_{\xi});\notag\\ 
	&C_{\ell}\sim \mathcal{IG}(a_{C_\ell},b_{C_\ell});~\ell=1,\ldots,p;\notag\\ 
	&\tilde C_{\ell}\sim \mathcal{IG}(a_{\tilde C_\ell},b_{\tilde C_\ell});~\ell=1,\ldots,p;\notag\\ 
	&\nu_{\ell}\sim N(0,\sigma^2_{\nu_\ell});~\ell=1,\ldots,p;\notag\\ 
	&\omega^2_{\ell}\sim \mathcal{IG}(a_{\omega^2_\ell},b_{\omega^2_\ell});~\ell=1,\ldots,p;\notag\\ 
	&\log\left(\frac{\rho_{\beta}}{1-\rho_{\beta}}\right)\sim N(0,\sigma^2_{\rho_{\beta}});\notag\\ 
	&\sigma^2_{\beta}\sim \mathcal{IG}(a_{\sigma^2_\beta},b_{\sigma^2_\beta});\notag\\ 
	&\log\left(\frac{\rho_{\ell}}{1-\rho_{\ell}}\right)\sim N(0,\sigma^2_{\rho_{\ell}});~\ell=1,\ldots,p;\notag\\ 
	&\sigma^2_{\ell}\sim \mathcal{IG}(a_{\sigma^2_\ell},b_{\sigma^2_\ell});~\ell=1,\ldots,p;\notag\\ 
	&\sigma^2_{\epsilon}\sim \mathcal{IG}(a_{\sigma^2_\epsilon},b_{\sigma^2_\epsilon});\notag\\ 
	&\sigma^2_{\alpha}\sim \mathcal{IG}(a_{\sigma^2_{\alpha}},b_{\sigma^2_{\alpha}});\notag\\ 
	&\sigma^2_{\phi}\sim \mathcal{IG}(a_{\sigma^2_\phi},b_{\sigma^2_\phi}).\notag 
\end{align}
In the above, 
$\mathcal {IG}(a,b)$ denotes the inverse Gamma distribution with positive parameters $a$ and $b$ with density 
$h(x)\propto x^{-a-1}\exp(-b/x)$, for $x>0$.
We recommend $a=2.01$ and $b=1.01$ for the inverse gamma priors except those for $\sigma^2_{\alpha}$, $\sigma^2_\phi$ and $\sigma^2_{\epsilon}$.
Thus the means and variances are $b/(a-1)=1$ and $b^2/((a-1)^2(a-2))=100$ in these cases. For $\sigma^2_\alpha$ and $\sigma^2_\phi$ we recommend $a=10^4$ and $b=1$,
so that the means and variances are close to zero, 
to reflect the opinion
that the spatial and temporal random effects do not have significant variabilities in the presence of $f(\bs_i,t_k)$. 
Moreover, in our applications, we fit our Bayesian model after standardizing the space-time datasets, and hence set $\alpha=0$ for model implementation. 
We finally convert our Bayesian predictions to the original locations and scales, for the reporting purpose.

For $\sigma^2_{\epsilon}$, we again recommend $a=10^4$ and $b=1$. Again, this encapsulates our opinion that $\sigma^2_{\epsilon}$ is not much different from zero.
The reason for this opinion about $\sigma^2_{\epsilon}$ is that when $n$ and $m$ are even reasonably large, with very high probability, the 
overall variability of the spatio-temporal data 
is expected to be very large, which would drastically increase the posterior mean and variance
of $\sigma^2_{\epsilon}$, unless its prior means and variances are fixed to be very small. Note that large posterior mean and variance of $\sigma^2_{\epsilon}$
would render predictions at desired spatial locations and time points highly unreliable. Consequently, we set such small values of the mean and variance to obtain
reasonable predictions. For $\lambda$, we set $b_{\lambda}=0.001$ and $a_{\lambda}=10 b_{\lambda}$ in our applications, 
so that the prior mean and variance of $\lambda$ are $10$ and $10^4$, respectively. 

As regards the zero-mean normal priors for $\nu_{\ell}$, $\log\left(\frac{\rho_{\beta}}{1-\rho_{\beta}}\right)$ and $\log\left(\frac{\rho_{\ell}}{1-\rho_{\ell}}\right)$,
we set the variances to be $100$ in our applications.

Recall that when $t_k-t_{k-1}=1$, the IAR model boils down to the regular AR(1) model, so that in such a situation we set $|\rho_{\ell}|<1$ for $\ell=1,\ldots,p$
and $|\rho_{\beta}|<1$ for stationarity. For the priors on $\rho_{\ell}$ and $\rho_{\beta}$, we then set 
$\rho_\ell=-1+\frac{2\exp(\tilde\rho_\ell)}{1+\exp(\tilde\rho_\ell)}$ and $\rho_\beta=-1+\frac{2\exp(\tilde\rho_\beta)}{1+\exp(\tilde\rho_\beta)}$,
with $\tilde\rho_\ell\sim N(0,\sigma^2_{\rho_{\ell}})$ and $\tilde\rho_\beta\sim N(0,\sigma^2_{\rho_{\beta}})$.
To ensure stationarity of the regular AR(1) setup we also set $\beta_{jt_1}\sim N\left(0,\frac{\sigma^2_{\beta}}{1-\rho^2_\beta}\right)$ 
and $\mu^{(\ell)}_{t_1}\sim N\left(0,\frac{\sigma^2_{\ell}}{1-\rho^2_\ell}\right)$.

Now note that the exponential kernel $K(\bM(\bs)-\bmu,t-\tau|\tilde\sigma^2_1,\ldots,\tilde\sigma^2_p,\xi)$ that we use for our purpose will have negligible values
for large values of $\tau$, $\xi$, $\tilde\sigma^2_1,\ldots,\tilde\sigma^2_p$. We reparameterize these non-negative parameters generically by $\exp(\varphi)$, where
$-\infty<\varphi<\infty$, and obtain the prior distribution for $\varphi$ corresponding to the priors for the original non-negative parameters. We then
truncate $\varphi$ on the interval $[-20,5]$. We adopt the same strategy for the non-negative parameters $\sigma^2_{\ell}$, $C_{\ell}$ and $\tilde C_{\ell}$ as well. 
Thus, although we allow small positive values of the original non-negative parameters, large positive values are ruled out.
We also truncate $X_{\ell}$ and $\mu^{(\ell)}_{jt_k}$ on $[-10,10]$. Further, the mixing behaviour of our MCMC is improved by truncating 
the normal distributions associated with $\rho_{\ell}$ and $\rho_{\beta}$ on $[-10,10]$ and adopting the same aforementioned reparameterization and truncation strategy for 
$\sigma^2_{\beta}$.

\section{An overview of our parallel MCMC algorithm}
\label{sec:mcmc_overview}
The form of the joint posterior distribution is provided in Section \ref{sec:form_joint}, using which
the forms of the full conditional distributions of the parameters are detailed in Section \ref{sec:fullcond}.
In our MCMC method, we use Gibbs sampling steps to simulate from most of the standard full conditional distributions. 
We refer to the relevant set of parameters updated using Gibbs steps by $\bzeta$. 
Although the full conditionals of $\sigma^2_1,\ldots,\sigma^2_p$ and $\sigma^2_{\beta}$ are also available in closed forms, instead of using Gibbs steps for these,
we update these parameters simultaneously in a single block consisting of $$\btheta=(X_1,\ldots,X_p,\tilde C_1,\ldots,\tilde C_p,C_1,\ldots,C_p,
	\tilde\sigma^2_1,\ldots,\tilde\sigma^2_p,\tau,\xi,\rho_1,\ldots,\rho_p,\sigma^2_1,\ldots,\sigma^2_p,\rho_{\beta},\sigma^2_{\beta})$$
	using Transformation based Markov Chain Monte Carlo (TMCMC) introduced by \ctn{Dutta14}. The essence of TMCMC is to simultaneously update many
	parameters in a single block using simple deterministic transformations of some low-dimensional (usually, one-dimensional) random variable.
	As can be anticipated, this drastic dimension reduction leads to great improvement in acceptance rates and faster convergence compared to
	traditional MCMC methods. For details on TMCMC, see \ctn{Dutta14}, \ctn{Dey16}, \ctn{Dey17}, \ctn{Dey19}.

	In our case, we consider a mixture of additive and multiplicative TMCMC; 
	such a mixture outperforms both
	additive and multiplicative TMCMC by combining the localised moves of additive TMCMC and the non-local moves of multiplicative TMCMC
	(see \ctn{Dey16} for details). The multiplicative
	transformation has been referred to as ``random dive" by \ctn{Dutta12}.

	After implementing the mixture TMCMC, we supplement this with another deterministic move type consisting of additive and multiplicative 
	transformations to further enhance mixing properties of our methodology.

Now note that due to the Markov property, $(\bU_k,\bbeta_k,J_{t_k})$, for all the odd values of $k\in\{1,\ldots,m\}$ can be updated simultaneously in parallel processors.
Once $(\bU_k,\bbeta_k,J_{t_k})$ are updated for odd values of $k$, those for the even values of $k$ can then be updated simultaneously in parallel processors.

Since $J_{t_k}$ is a random variable, this makes the dimensions of $\bU_k$ and $\bbeta_k$ random, rendering the updating problem
of $(\bU_k,\bbeta_k,J_{t_k})$ a variable-dimensional problem, for every $k=1,\ldots,m$. Since reversible jump MCMC introduced in \ctn{Green95} 
is well-known to be a very inefficient method for handling variable-dimensional problems, \ctn{Das19} came up with a novel and efficient 
alternative to solving variable-dimensional cases, using
appropriate deterministic transformations of fixed and low-dimensional random variables. The method, referred to as Transdimensional Transformation based Markov Chain
Monte Carlo (TTMCMC), is an extension of TMCMC for fixed-dimensional setups to
general variable-dimensional problems.

In our case, for each value of $k$, we update $(\bU_k,\bbeta_k,J_{t_k})$ using TTMCMC in separate parallel processors.
Also, given all other unknowns, we update $\phi(\bs_i,t_k)$ simultaneously in separate parallel processors by sampling from their full conditional distributions.
Note that in order to sample from the full conditional distributions of $\lambda$, $\sigma^2_\phi$, $\sigma^2_{\epsilon}$ and $\alpha$, computations of the sums
$\sum_{k=1}^mJ_k$, $\sum_{i=1}^n\sum_{k=1}^m(\phi(\bs_i,t_k)-\phi_0(\bs_i,t_k))^2$, and $\sum_{i=1}^n\sum_{k=1}^m(y(\bs_i,t_k)-\alpha-\phi(\bs_i,t_k)-f(\bs_i,t_k))^2$ are required. We compute these
by splitting the sums into the available parallel processors, each processor computing only a small part of each sum. The final sum is aggregated into a single processor,
which then updates the relevant parameters. The TMCMC update required in step (\ref{eq:c7}) requires computing  
$$\sum_{i=1}^n\sum_{k=1}^m\log\left[y(\bs_i,t_k)|X_1,\ldots,X_p,C_1,\ldots,C_p,\tilde C_1,\ldots,\tilde C_p,
	\bU_{k},\bbeta_k,J_{t_k},\tilde\sigma^2_1,\ldots,\tilde\sigma^2_p,\tau,\xi,\alpha,\phi(\bs_i,t_k),\sigma^2_{\epsilon}\right],$$ 
	which we again compute by splitting the sum
	into the available parallel processors, finally aggregating the result into a single processor where TMCMC is applied.

For the purpose of prediction of $y(\tilde\bs,\tilde t)$ at any location $\tilde \bs$ and time point $\tilde t$, we substitute the MCMC-simulated realizations
in the distribution associated with (\ref{eq:re_model_marginalized1}), and generate $\tilde y(\tilde\bs,\tilde t)$ from the resultant distribution, which yields
the posterior predictive distribution of $\tilde y(\tilde\bs,\tilde t)$. For multiple locations and time points, for each MCMC realization, we parallelize
our prediction exercise over the required locations and time points, again leading to significant computational savings.

The complete algorithm is provided as Algorithm \ref{algo:ttmcmc} in the supplement.
Note that in the marginalized model where the random effects are integrated out, the algorithm is simplified.

\section{Simulation study}
\label{sec:simstudy}
For our simulation experiment, we generate data from the so-called general quadratic non-linear (GQN) model (\ctn{Wikle10}, \ctn{Cressie11}), 
to which we fit our L\'{e}vy-dynamic 
spatio-temporal model with random effects and make predictions at various locations and time points. Specifically, our data-generating GQN model is of the following form:
for $i=1,\ldots,n$ and $k=1,\ldots,m$, 
\begin{align}
	&y(\bs_i,t_k)=\phi_{1t_k}(\bs_i)+\phi_{2t_k}(\bs_i)\tan(\beta_{t_k}(\bs_i))+\epsilon_{t_k}(\bs_i);\label{eq:gqn1}\\
	&\beta_{t_k}(\bs_i)=\sum_{j=1}^na_{ij}\beta_{t_k-1}(\bs_j)+\sum_{j=1}^n\sum_{l=1}^nb_{ijl}\beta_{t_k-1}(\bs_j)g(\beta_{t_k-1}(\bs_l))+\eta_{t_k}(\bs_i).
	\label{eq:gqn2}
\end{align}
We assume that independently, $\phi_{1t_k}(\cdot),\phi_{2t_k}(\cdot),\eta_{t_k}(\cdot),\epsilon_{t_k}(\cdot),\beta_0(\cdot)\sim GP(0,c(\cdot,\cdot))$, 
a zero-mean Gaussian process with covariance function 
$c(\bs_1,\bs_2)=\exp\left(-\|\bs_1-\bs_2\|\right)$, for any $\bs_1,\bs_2\in\mathbb R^2$, where $\|\cdot\|$ denotes the Euclidean norm.
We further assume that independently, for $i=1,\ldots,n$, $j=1,\ldots,n$, $l=1,\ldots,n$, $a_{ij}\sim N(0,0.001^2)$ and $b_{ijl}\sim N(0,0.001^2)$.
We set $g(\beta_{t_k-1}(\bs_l))=\beta^2_{t_k-1}(\bs_l)$ and for $i=1,\ldots,n$, independently simulate $\bs_i\sim U(0,1)\times U(0,1)$.

We generate $y(\bs_i,t_k)$ for $n=120$ spatial locations and $m=50$ time points from the above model defined by (\ref{eq:gqn1}) and (\ref{eq:gqn2})
and the associated distributions. For the purpose of prediction, we set aside $20$ spatial locations and the associated $50$ times points for each of the locations.
Our main goal in this simulation experiment is to make reliable predictions at the set-aside locations and time points.
As mentioned earlier, we first standardize the dataset, to which we apply our model and methods, and make Bayesian predictions. Finally, we transform the predictions
to the original location and scale for reporting. Thus, for our purpose, we set $\alpha=0$ and consequently, updating $\alpha$ and $\sigma^2_{\alpha}$ are not required.

\subsection{Implementation details}
\label{subsec:implementation}

We implement MCMC Algorithm \ref{algo:ttmcmc} (wth $p=2$) for the non-marginalized Bayesian L\'{e}vy-dynamic model, 
written in C in conjunction with the message passing interface (MPI) protocol, 
on $25$ parallel processors in our 2 TB memory VMWare (each core has about 2.8 GHz CPU speed). Generation of $11\times 10^4$ MCMC realizations took 
$1$ hour $27$ minutes in our VMWare. 
However, on $80$ parallel processors (the maximum number of cores on our VMWare), the time taken is $1$ hour $47$ minutes. This is a consequence of slower communications
among much larger number of processors with a relatively small amount of data. 
That is, the computational overhead with $80$ processors is less than the communication overhead among the $80$ processors. 
Also note that the number of time points is only $50$ and only $25$ processors can be used
at a time to update the associated parameters corresponding to even or odd indices, making the other processors redundant for these parameters.
More experimentations led to the conclusion that $25$ processors provide the most efficiency for this problem.

To prevent extreme propositions in the multiplicative moves of our algorithm brought about by dividing the current realization
by $\epsilon$ too close to zero, we set $\epsilon\sim U(-1,1)$ subject to $|\epsilon|>0.01$. This is of course a theoretically valid step and irreducibility of our algorithm
is preserved by the additive transformation. For details, see \ctn{Dey16}.
We discarded the first $10^4$ realizations as burn-in and stored every $10$-th realization in the next $10^5$ iterations to obtain
$10^4$ MCMC realizations for Bayesian inference. 

\subsection{MCMC convergence}
\label{subsec:mcmc_conv}
Figure \ref{fig:J_plots_simstudy1} of the supplement exhibit the trace plots of $J_{k}$ for different $k$ (for which we display and provide the analyses of
the corresponding spatial predictions) and Figure \ref{fig:parameter_plots_simstudy1} provides the trace plots of some other parameters of our L\'{e}vy-dynamic Bayesian model.
All the trace plots vindicate excellent convergence. Panel (i) of Figure \ref{fig:parameter_plots_simstudy1} shows that even though the prior for $\sigma^2_{\epsilon}$ has
mean and variance close to zero, the posterior distribution still takes on high values occasionally.

That the acceptance rates for the TTMCMC and TMCMC rates are adequate, are evident from the trace plots. Specifically, the average TTMCMC 
acceptance rates of the birth, death and no-change moves over
$50$ time points are approximately $0.09$, $0.726$ and $0.619$, respectively. The average overall TTMCMC acceptance rate is $0.415$.
The fixed-dimensional parameters that are updated using TMCMC have acceptance rate $0.921$, and the corresponding acceptance rate associated with the mixing-enhancement
step is $0.656$.
All these acceptance rates are calculated with respect to the entire set of $11\times 10^4$ MCMC realizations of our algorithm.

\subsection{Results}
\label{susbec:results_simstudy}
\begin{figure}
	\centering
	\subfigure [Spatial index $1$.]{ \label{fig:spatial1}
	\includegraphics[width=7.5cm,height=5.5cm]{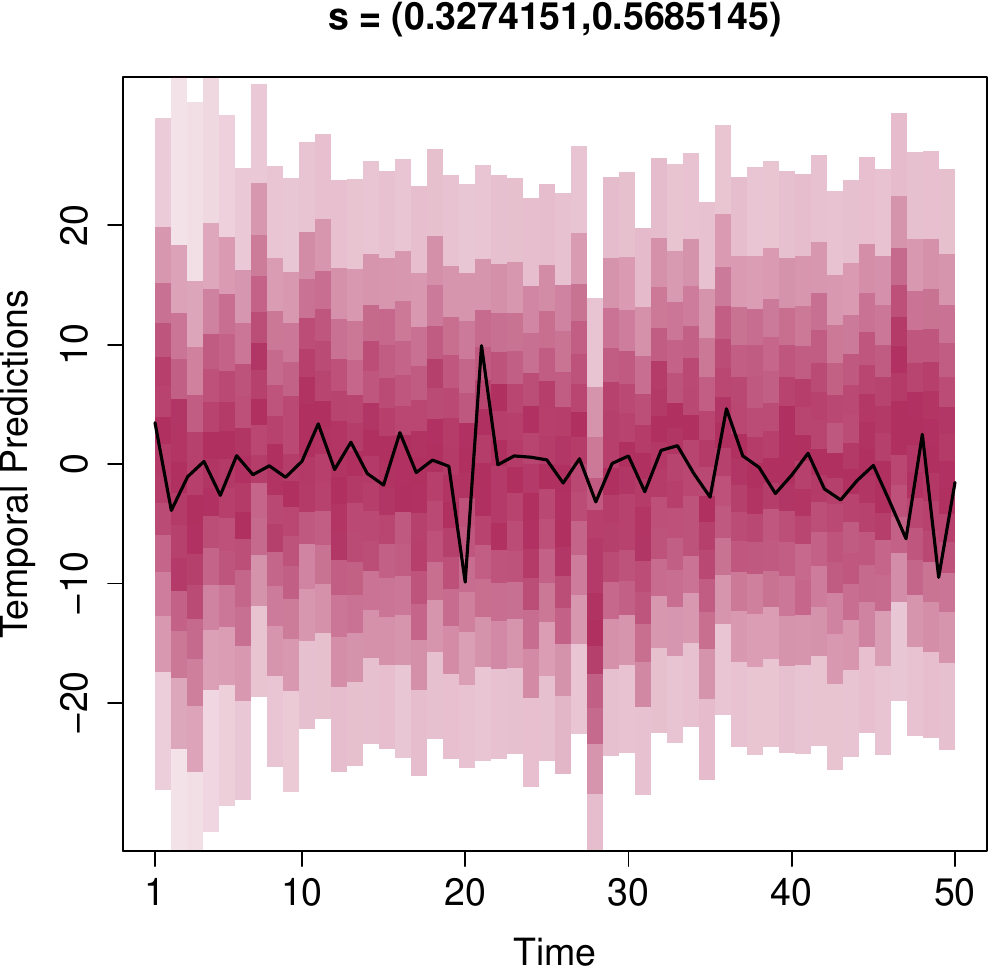}}
	\hspace{2mm}
	\subfigure [Spatial index $5$.]{ \label{fig:spatial5}
	\includegraphics[width=7.5cm,height=5.5cm]{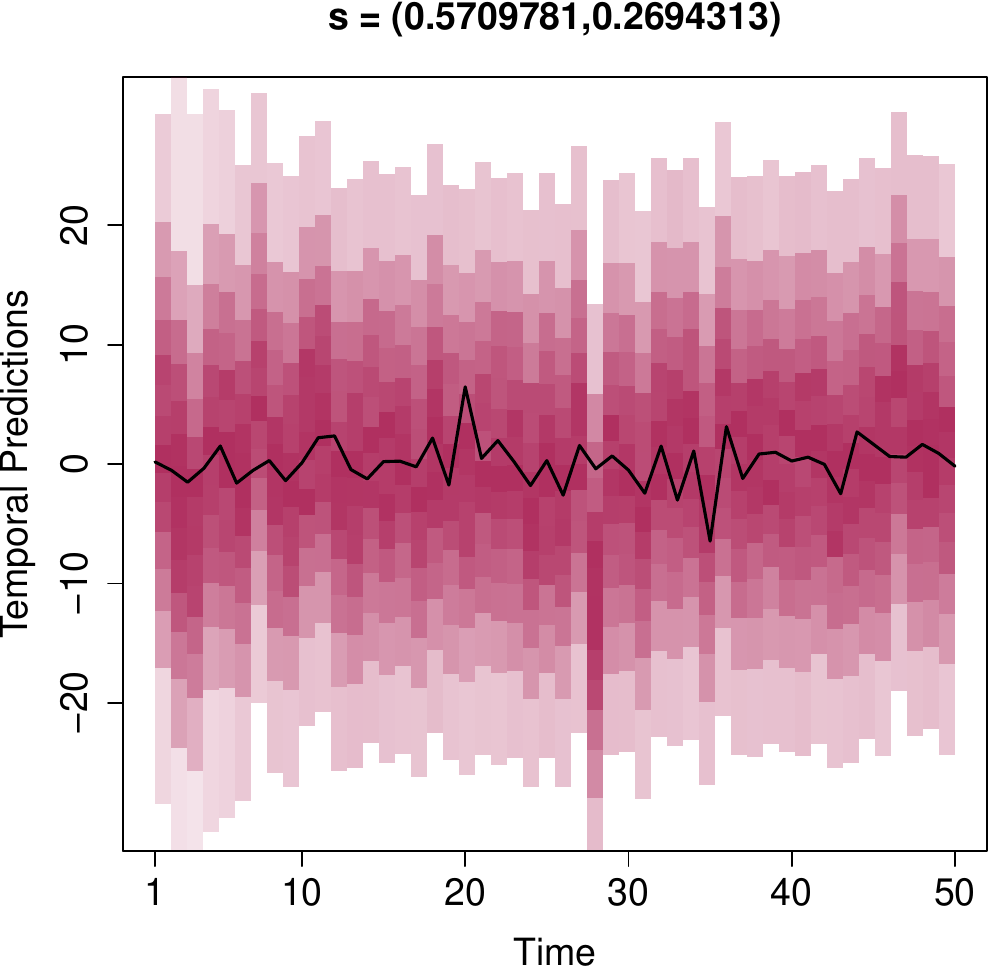}}\\
	\vspace{2mm}
	\subfigure [Spatial index $10$.]{ \label{fig:spatial10}
	\includegraphics[width=7.5cm,height=5.5cm]{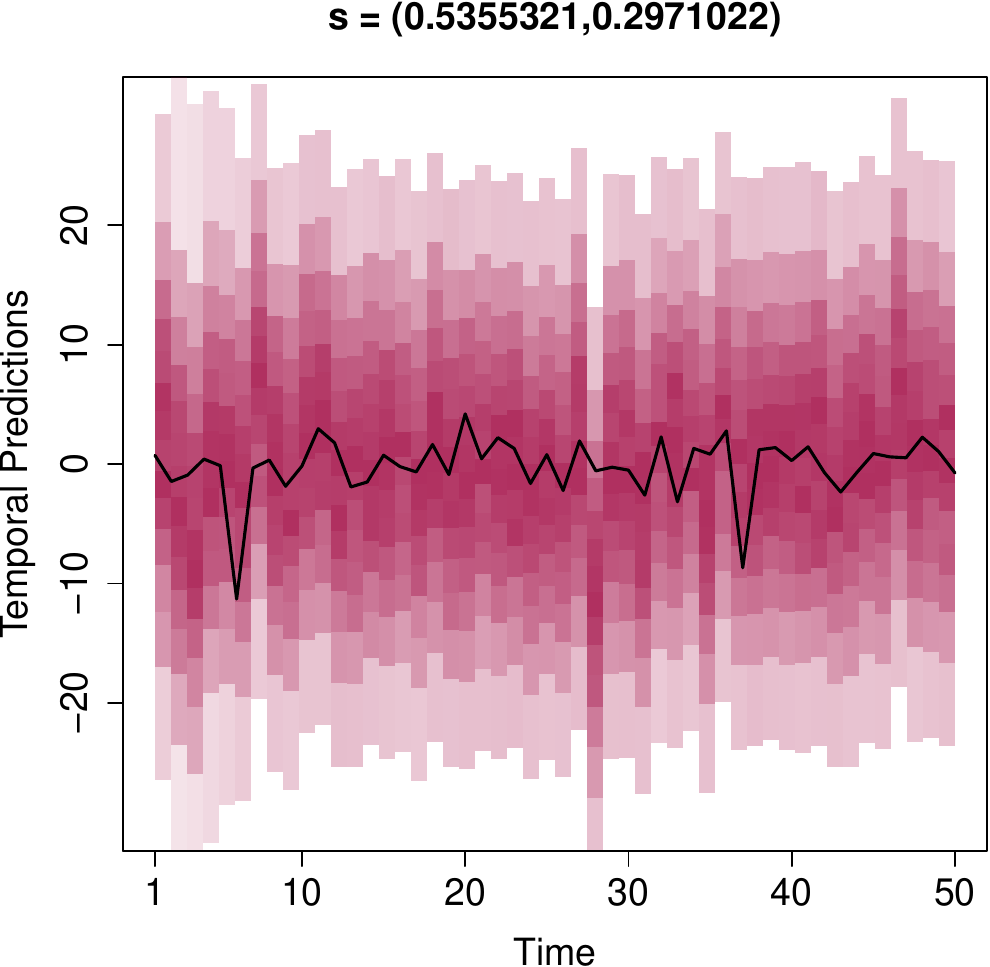}}
	\hspace{2mm}
	\subfigure [Spatial index $15$.]{ \label{fig:spatial15}
	\includegraphics[width=7.5cm,height=5.5cm]{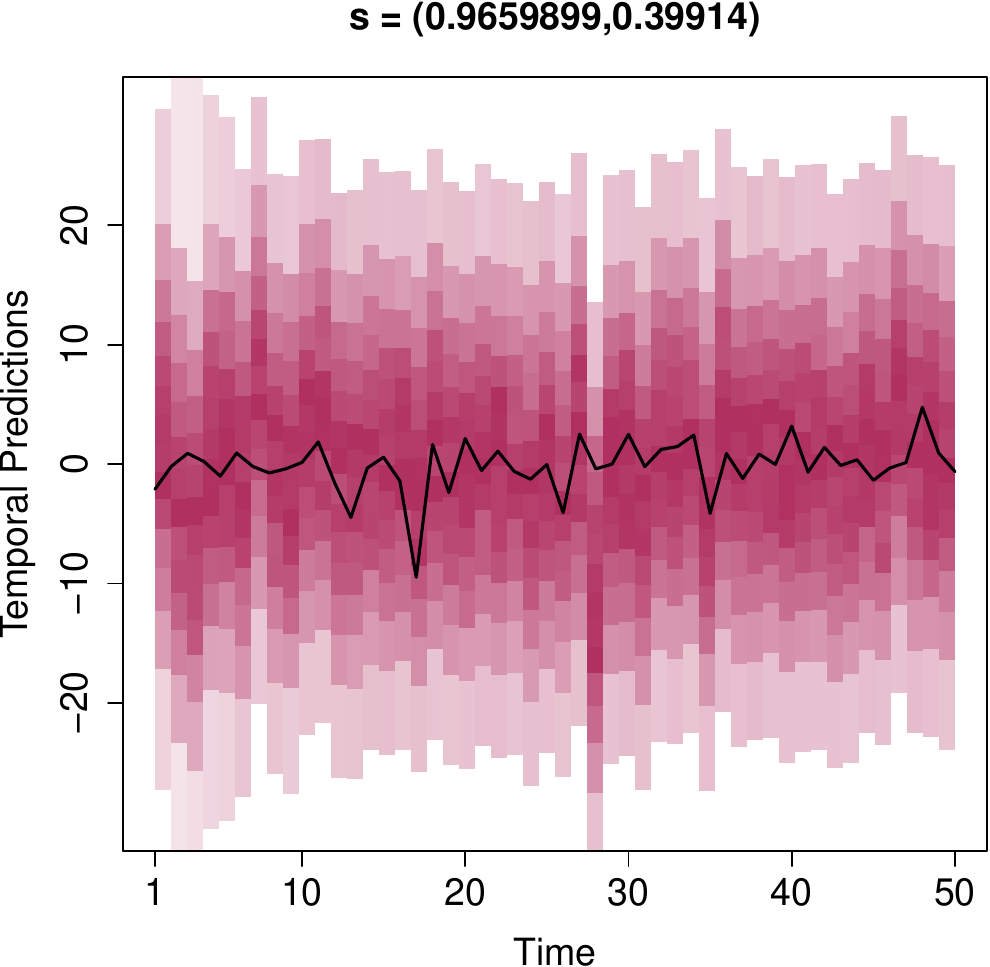}}\\
	\vspace{2mm}
	\subfigure [Spatial index $19$.]{ \label{fig:spatial19}
	\includegraphics[width=7.5cm,height=5.5cm]{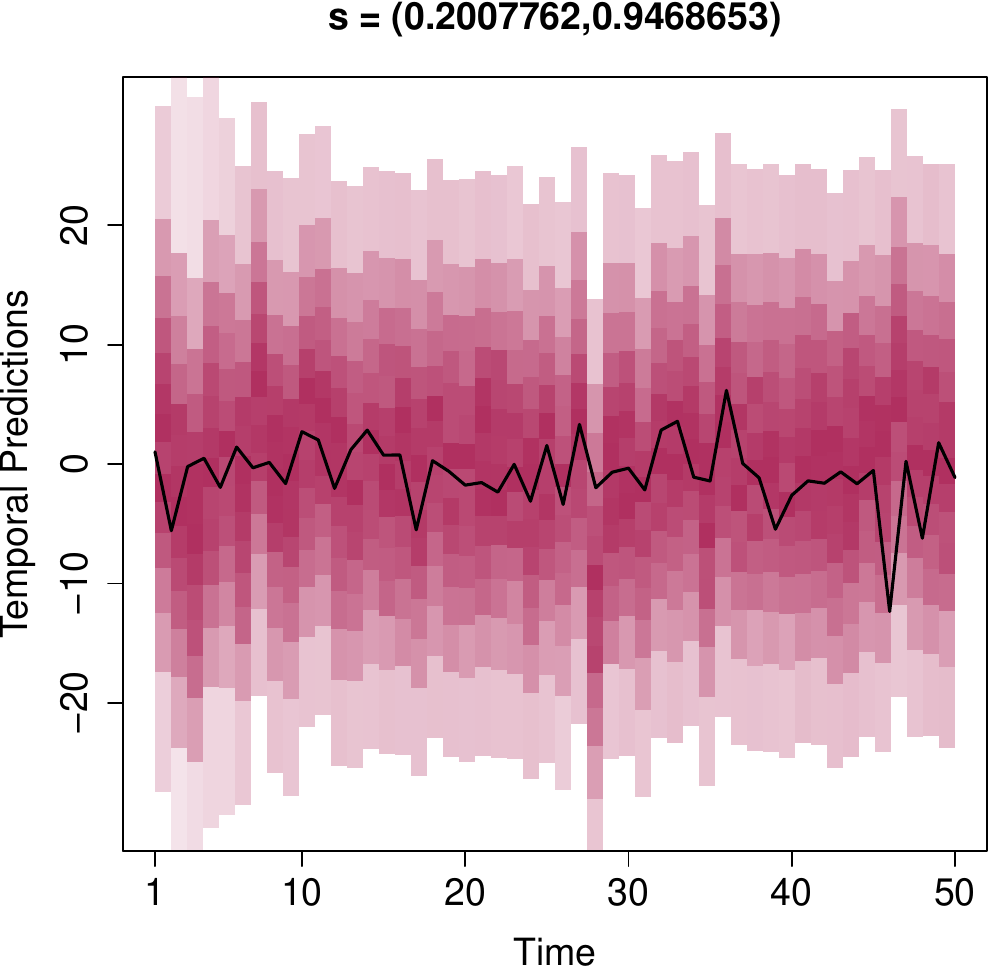}}
	\hspace{2mm}
	\subfigure [Spatial index $20$.]{ \label{fig:spatial20}
	\includegraphics[width=7.5cm,height=5.5cm]{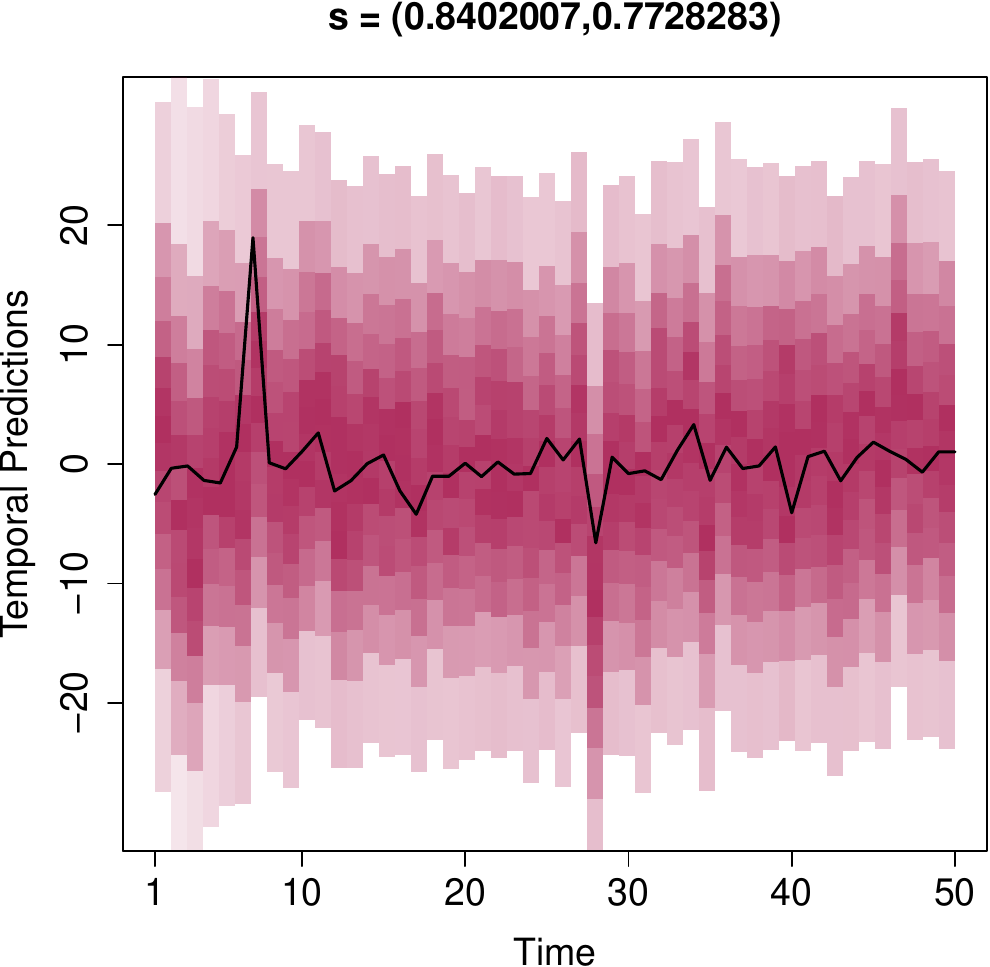}}\\
	\caption{Simulation study: posterior temporal predictions at various spatial locations $\bs$ are shown as colour plots with progressively higher densities 
	depicted by progressively intense colours.}
	\label{fig:temporal_plots_simstudy1}
\end{figure}

Figure \ref{fig:temporal_plots_simstudy1} depicts the densities of the temporal predictions at various spatial locations as color plots. In the color plots
progressively intense colors correspond to higher densities associated with $16$ quantiles dividing the density support (that is, the minimum and the maximum quantiles
undertaken for the color plots are $1/16=0.0625$ and $15/16=0.9375$). The true time series at the spatial locations are denoted by the thick, black line. 
Notice that in almost all the cases, the entire true time series is included in the high density regions of our Bayesian-predicted time series using our
L\'{e}vy-dynamic spatio-temporal process. However, in a few cases, where the actual temporal data points take on very highly positive or negative values,
our predictions have not been adequate (not shown).

\begin{figure}
	\centering
	\subfigure [Temporal index $5$.]{ \label{fig:temporal5}
	\includegraphics[width=7.5cm,height=5.5cm]{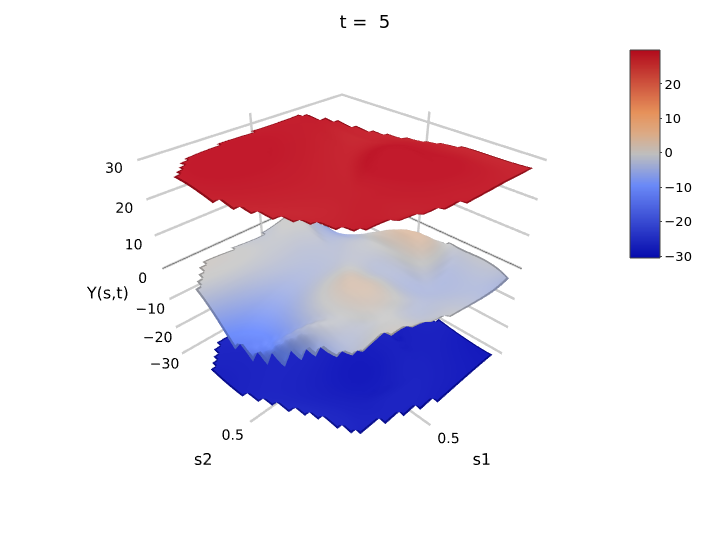}}
	\hspace{2mm}
	\subfigure [Temporal index $10$.]{ \label{fig:temporal10}
	\includegraphics[width=7.5cm,height=5.5cm]{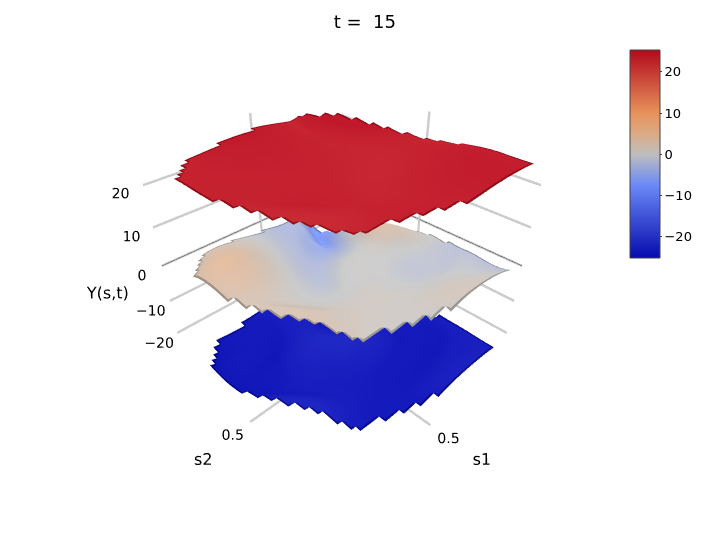}}\\
	\vspace{2mm}
	\subfigure [Temporal index $20$.]{ \label{fig:temporal20}
	\includegraphics[width=7.5cm,height=5.5cm]{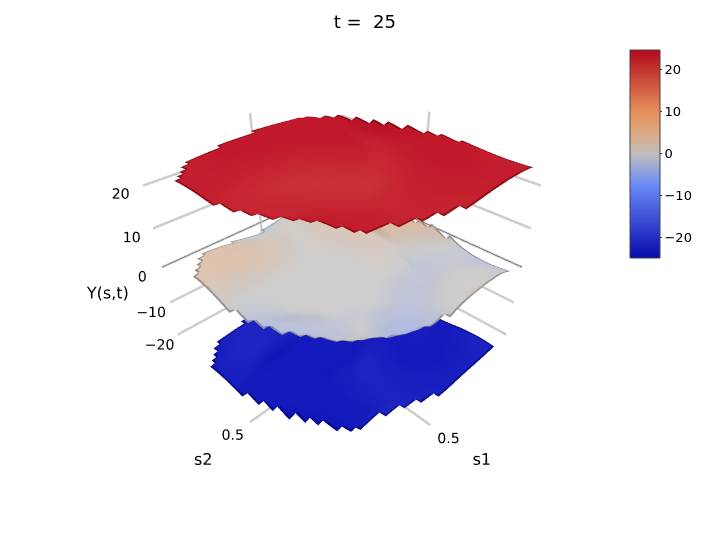}}
	\hspace{2mm}
	\subfigure [Temporal index $30$.]{ \label{fig:temporal30}
	\includegraphics[width=7.5cm,height=5.5cm]{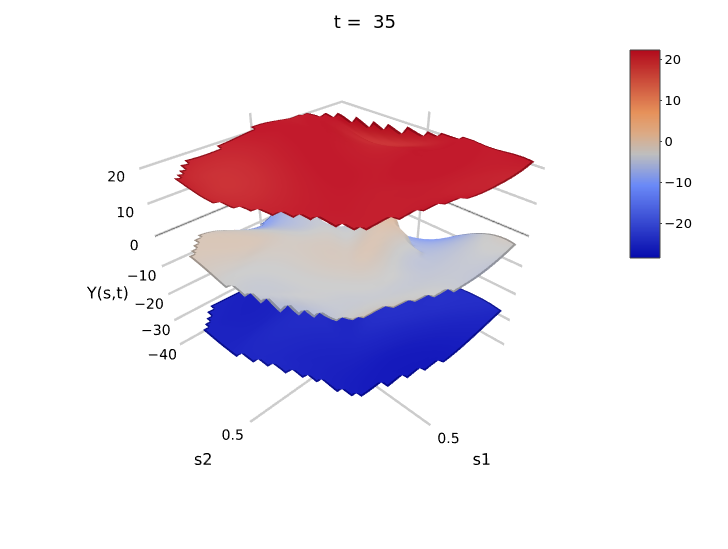}}\\
	\vspace{2mm}
	\subfigure [Temporal index $40$.]{ \label{fig:temporal40}
	\includegraphics[width=7.5cm,height=5.5cm]{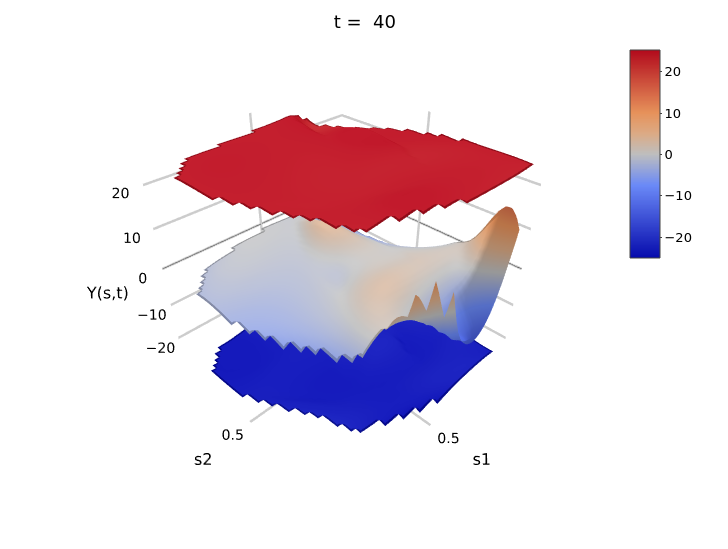}}
	\hspace{2mm}
	\subfigure [Temporal index $50$.]{ \label{fig:temporal50}
	\includegraphics[width=7.5cm,height=5.5cm]{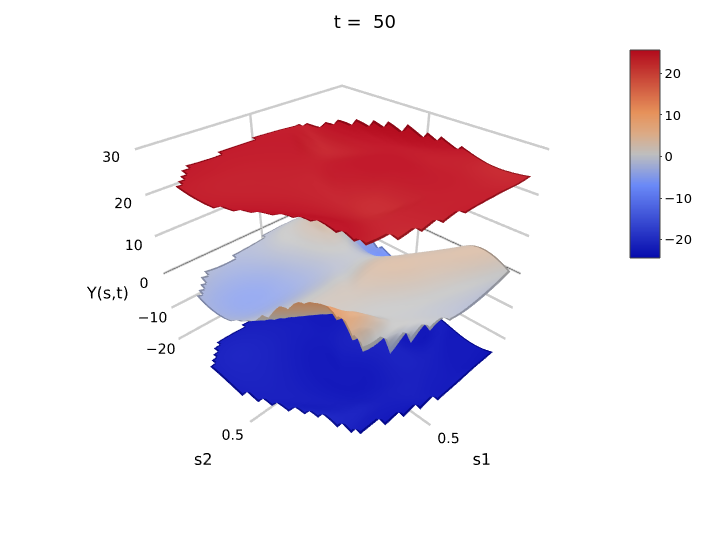}}\\
	\caption{Simulation study: posterior spatial predictions at various time points $t$. The middle surface is the actual spatial data, while the lower and upper surfaces
	are the lower and upper bounds of the $0.875$ credible region.}
	\label{fig:spatial_plots_simstudy1}
\end{figure}

\begin{figure}
	\centering
	\subfigure [Temporal index $5$.]{ \label{fig:temp5}
	\includegraphics[width=7.5cm,height=5.5cm]{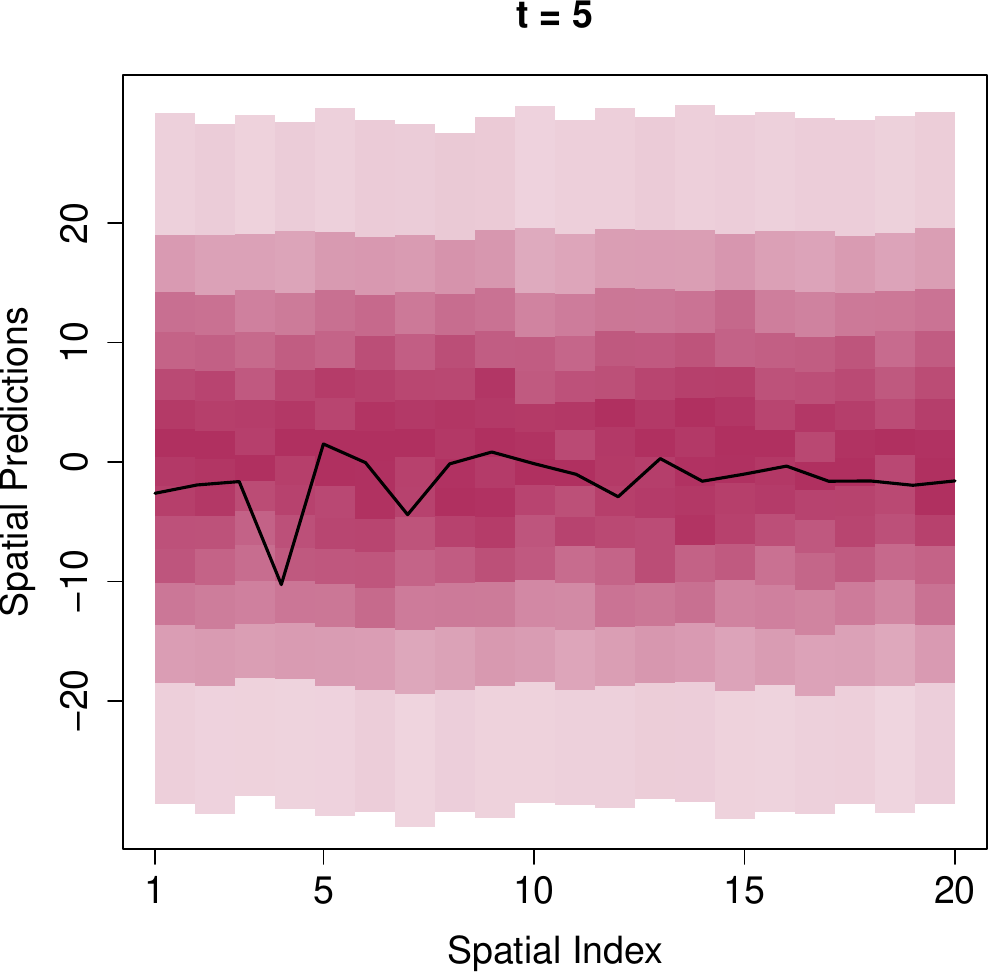}}
	\hspace{2mm}
	\subfigure [Temporal index $15$.]{ \label{fig:temp15}
	\includegraphics[width=7.5cm,height=5.5cm]{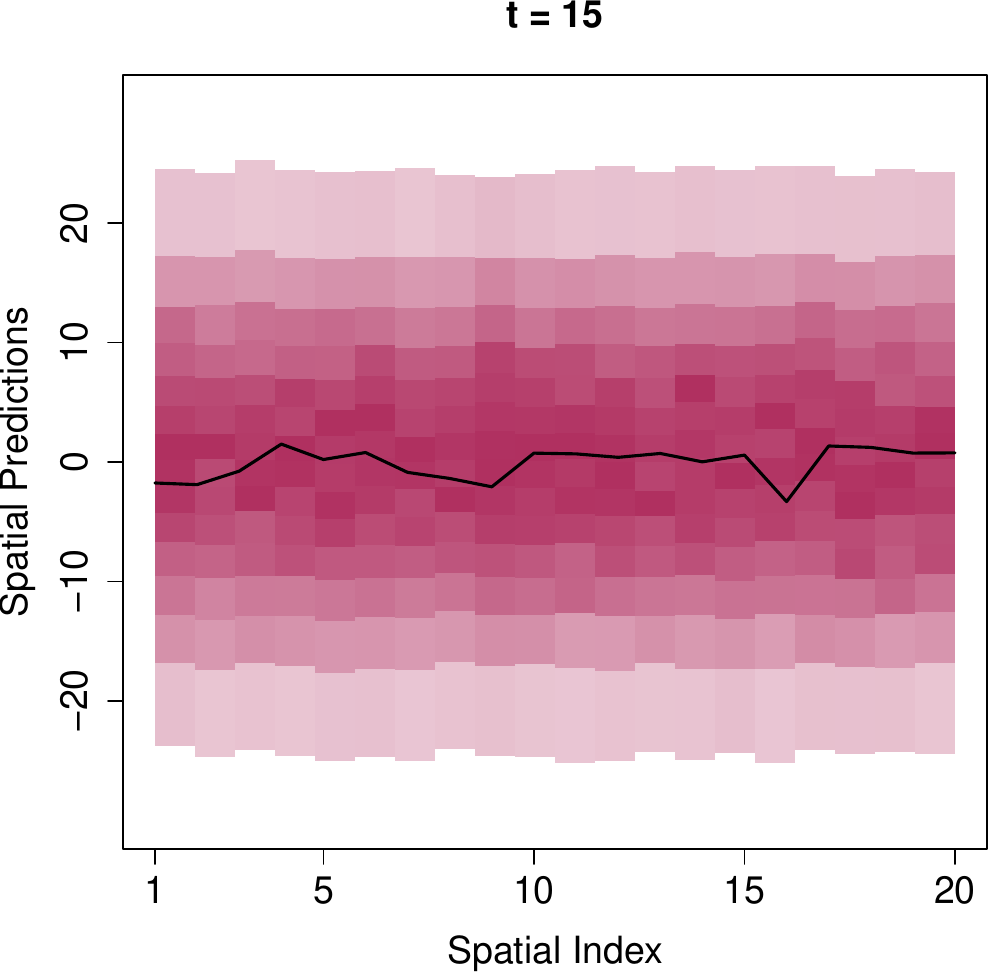}}\\
	\vspace{2mm}
	\subfigure [Temporal index $25$.]{ \label{fig:temp25}
	\includegraphics[width=7.5cm,height=5.5cm]{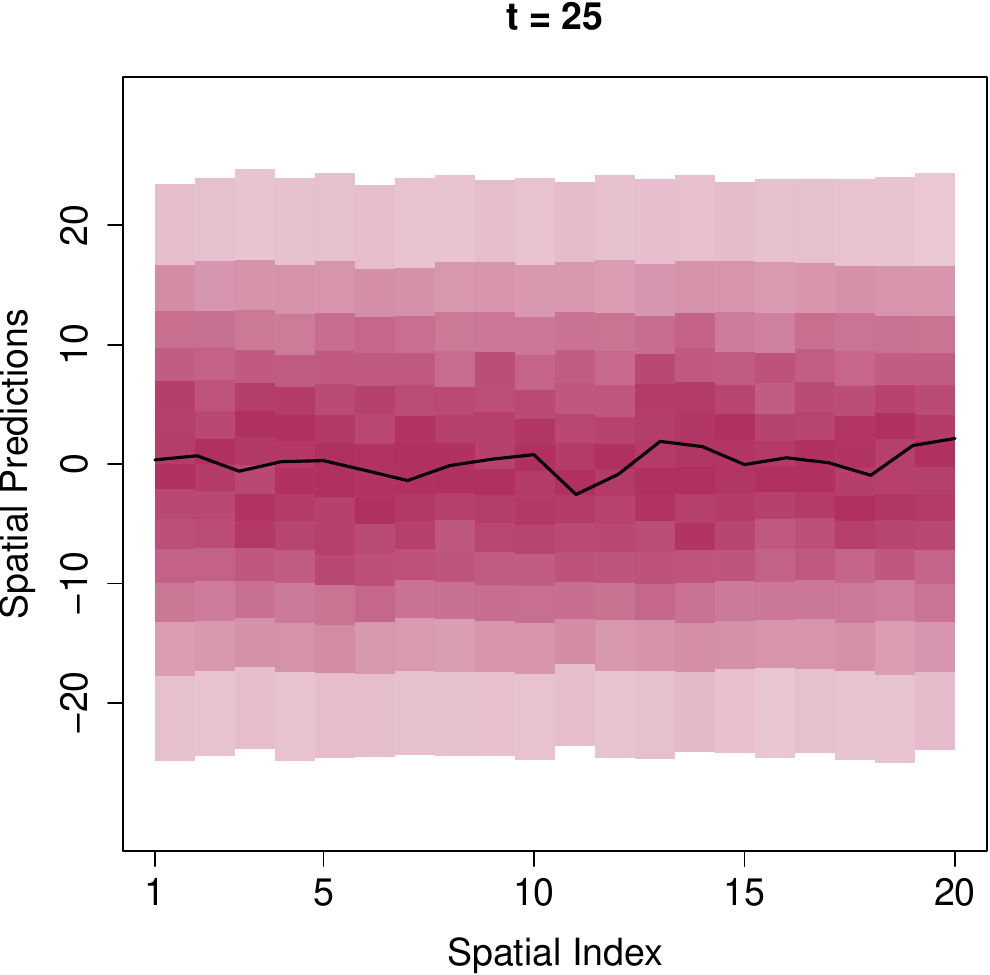}}
	\hspace{2mm}
	\subfigure [Temporal index $35$.]{ \label{fig:temp35}
	\includegraphics[width=7.5cm,height=5.5cm]{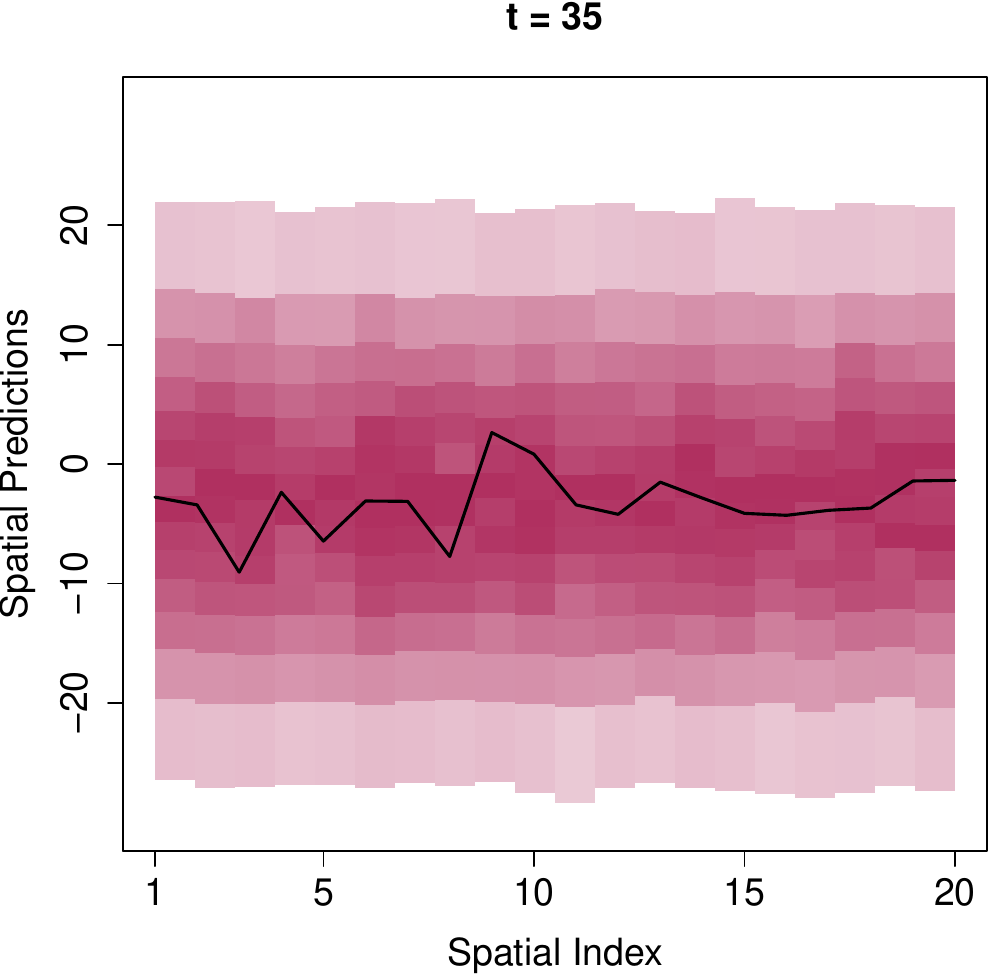}}\\
	\vspace{2mm}
	\subfigure [Temporal index $40$.]{ \label{fig:temp40}
	\includegraphics[width=7.5cm,height=5.5cm]{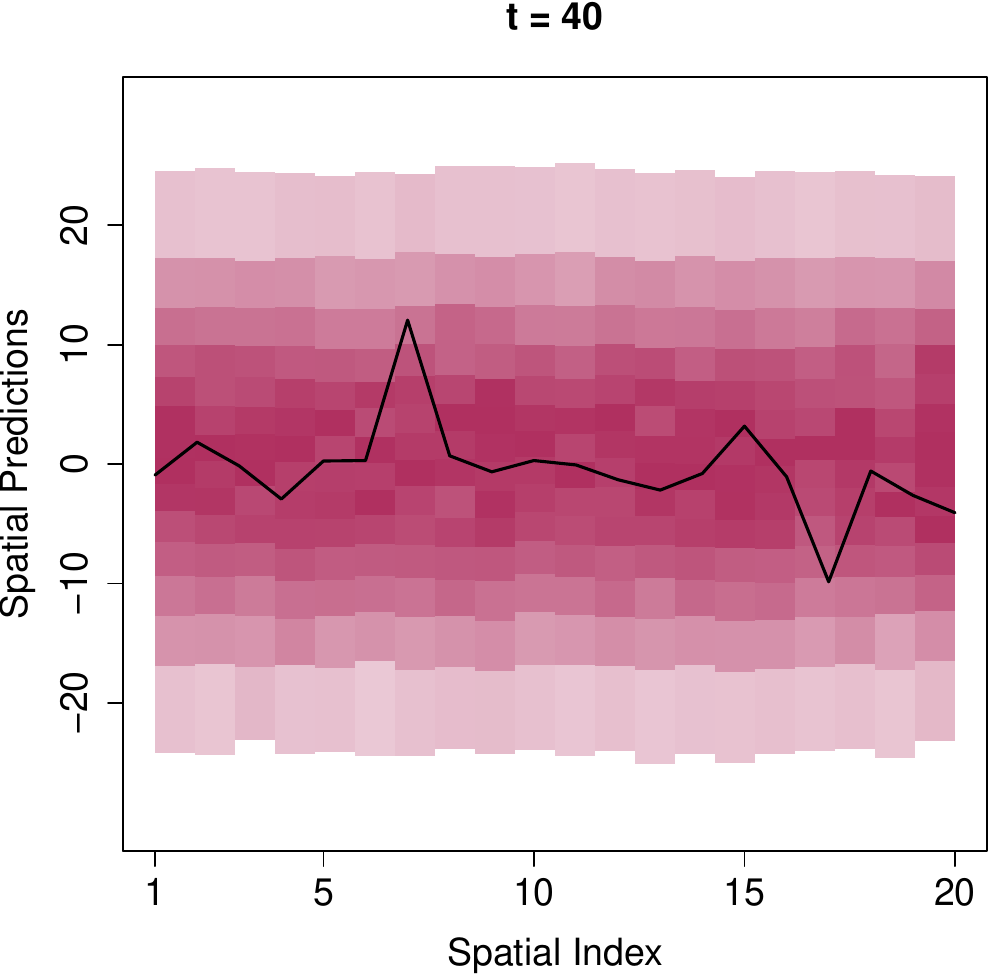}}
	\hspace{2mm}
	\subfigure [Temporal index $50$.]{ \label{fig:temp50}
	\includegraphics[width=7.5cm,height=5.5cm]{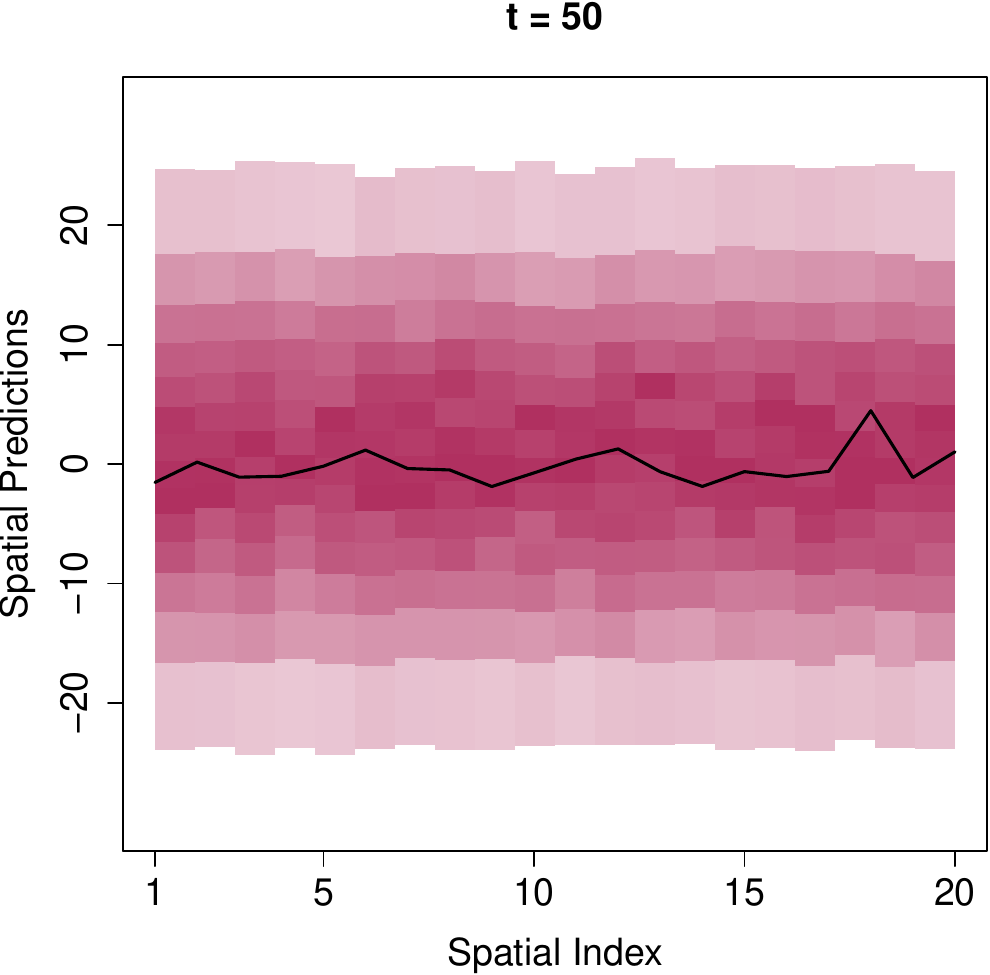}}\\
	\caption{Simulation study: posterior spatial predictions with respect to spatial indices at the time points $t$ as in Figure \ref{fig:spatial_plots_simstudy1} 
	are shown as colour plots with progressively higher densities depicted by progressively intense colours.}
	\label{fig:spatial_plots2_simstudy1}
\end{figure}

Figure \ref{fig:spatial_plots_simstudy1} shows the spatial surface plots at various time points. The middle surface is the actual spatial surface (supplemented
by spline-based interpolations), while the lower and upper surfaces (again, supplemented by spline-based interpolations) are the lower and upper bounds, respectively,
of the $0.875$ credible region. In other words, the lower and upper surfaces correspond to $1/16$-th and $15/16$-th quantiles of the respective posterior predictive 
distributions. The detailed spatial posterior predictive densities are shown in Figure \ref{fig:spatial_plots2_simstudy1}, as color plots akin to the temporal color plots
of Figure \ref{fig:temporal_plots_simstudy1}, with spatial indices replacing the time indices. Unlike the time points, no ordering is intended with respect to the spatial
indices. That is, we simply refer to $i=1,\ldots,20$, as the spatial indices associated with $\bs_i$. Thus, it is evident from the figures that almost all the actual spatial
data fall within the high-density regions of the corresponding posterior predictive distributions. However, again for some time points, there are a few spatial data points  
that are highly positive or negative, and our predictions failed in such cases (not shown).

\subsection{Faster implementation after integrating out the random effects}
\label{subsec:ingore_re_uncertainty}
Note that simulation of $\bphi$, although parallelised, requires many parallel processors for efficiency if either $n$ or $m$ is even moderately large. 
As already reported in Section \ref{subsec:implementation}, this constitutes significant communication overhead 
among the processors and slows down implementation. Redundancy of the cores is also a consequence in the case of moderately large datasets. On the other hand, 
for significantly large datasets with a large number of time points, a large number of parallel processors are necessary for efficiency.
However, such large number of processors are usually not available. 

To get around this problem, for $i=1,\ldots,n$ and $k=1,\ldots,m$, we invoke the marginalized model (\ref{eq:re_model_marginalized1}) and the discussion thereafter and set 
$\phi(\bs_i,t_k)=\phi_0(\bs_i,t_k)$, provided the posterior
uncertainties in $\phi(\bs_i,t_k)$ are negligible. Since the prior distributions of $\phi(\bs_i,t_k)$ are concentrated around 
$\phi_0(\bs_i,t_k)$ (recall that $\sigma^2_{\phi}$ is concentrated around zero {\it a priori}), our experiments reveal that this is indeed the case (not shown for brevity).

To justify our standpoint, we conduct a further experiment by setting $\phi(\bs_i,t_k)=\phi_0(\bs_i,t_k)$ in our model. Not only does the time taken reduce to 
just $24$ minutes from $1$ hour $27$ minutes with $25$ cores, but the prediction results with this setup (see Figures \ref{fig:temporal_plots_simstudy2},
\ref{fig:spatial_plots_simstudy2} and \ref{fig:spatial_plots2_simstudy2} of the supplement) are remarkably similar to those in the previous non-marginalized implementation
(Figures \ref{fig:temporal_plots_simstudy1}, \ref{fig:spatial_plots_simstudy1} and \ref{fig:spatial_plots2_simstudy1}).

In this case, the average TTMCMC 
acceptance rates of the birth, death and no-change moves over
$50$ time points are approximately $0.103$, $0.673$ and $0.617$, respectively, and the average overall TTMCMC acceptance rate is $0.420$.
The TMCMC step has acceptance rate $0.017$, and that of the mixing-enhancement
step is $0.152$.
Thus, compared to the non-marginalized model, here the acceptance rates of the TMCMC and the mixing-enhancement steps
have significantly decreased, but this is not a strong enough reason for concern since both the models yield remarkably similar performances with respect to 
our main goal, Bayesian prediction.

\section{Analysis of sea surface temperature data}
\label{sec:realdata}
We now consider analysis of a real, sea surface temperature dataset available at \url{http://iridl.ldeo.columbia.edu/SOURCES/.CAC/} in the netCDF file format.
The data pertains to monthly sea surface temperatures during January $1970$ -- December $2003$ at the tropical Pacific Ocean region 
covering $124\degree E-70\degree W$ and $30\degree S-30\degree N$, gridded at a $2\degree$ by $2\degree$ resolution. 
Analysis of a somewhat similar (and much smaller) dataset has been reported in \ctn{Cressie11}, on the basis of some simple linear and non-linear dynamic state-space models,
but that data represented monthly temperature anomalies from the normal, rather than the actual temperatures. 
The models of \ctn{Cressie11} are not intended to cover enough grounds like ours, namely, weak and strong nonstationarity, non-separability, 
nonparametric non-Gaussianity and convergence of the lagged correlations to zero. 
Nevertheless, the simplicity of their models enabled them to perform
simple Gibbs sampling based Bayesian analysis, for both of their linear and nonlinear dynamic models. However, the samplers are run for only $6000$ MCMC iterations
(the first $1000$ discarded as burn-in).
\ctn{Wikle19} also consider the anomalies dataset for some simplistic spatio-temporal analyses.

\begin{figure}
	\centering
	\subfigure [La Ni\~{n}a (colder than normal temperatures).]{ \label{fig:sst1}
	\includegraphics[width=12cm,height=5.5cm]{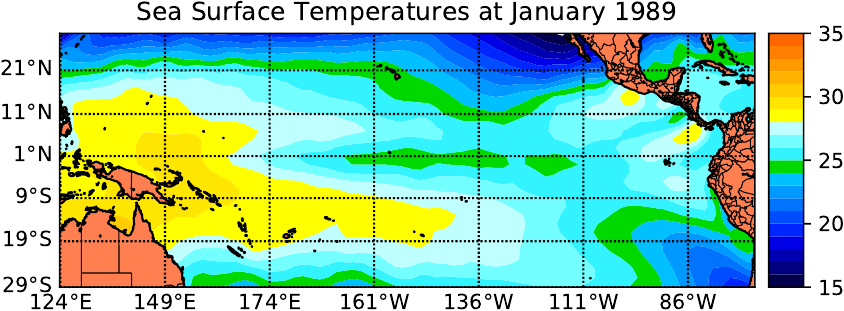}}\\
	\vspace{2mm}
	\subfigure [Normal temperatures.]{ \label{fig:sst2}
	\includegraphics[width=12cm,height=5.5cm]{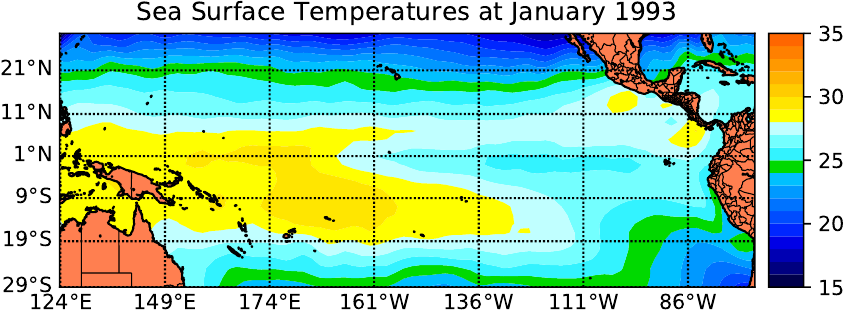}}\\
	\vspace{2mm}
	\subfigure [El Ni\~{n}o (warmer than normal temperatures).]{ \label{fig:sst3}
	\includegraphics[width=12cm,height=5.5cm]{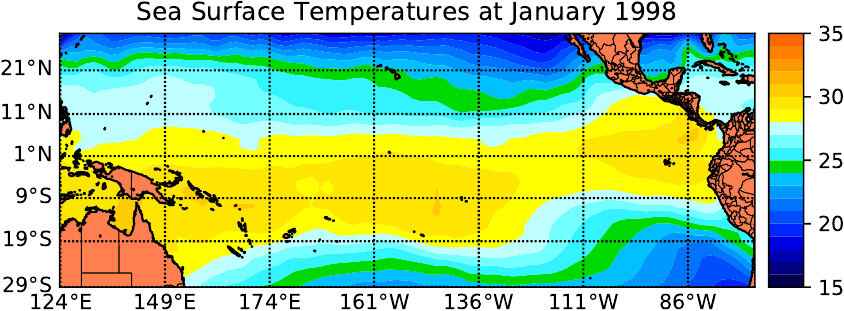}}\\
	\caption{Sea surface temperature plots in January 1898, 1993 and 1998.} 
	\label{fig:sst_plots}
\end{figure}

Our dataset consists of space-time data at $n=2520$ spatial locations, for each of $m=398$ time points. That is, the size of our dataset is
$nm=10,02,960$. Figure \ref{fig:sst_plots} displays the sea surface temperature plots during January $1989$, $1993$ and $1998$ to exhibit the
effects of La Ni\~{n}a (colder than normal temperatures), normal temperatures and El Ni\~{n}o (warmer than normal temperatures), respectively.
\ctn{Wikle19} provide similar plots in essence with their anomalies data using different colouring schemes in the R package,
as opposed to ours in Python, in the context of the actual temperatures.

That this sea surface temperature data arose from a spatio-temporal process that is strictly nonstationary and not even covariance (weak) stationary,
is established in Section \ref{sec:nonstationary} of the supplement, using formal Bayesian methods introduced by \ctn{Roy20}.
Convergence of the lagged spatio-temporal empirical correlations to zero based on this data, in spite of nonstationarity, is detailed in Section \ref{sec:zero_correlation},
while non-Gaussianity of the underlying process is argued in Section \ref{sec:non_gaussian} of the supplement.

\subsection{Bayesian L\'{e}vy-dynamic model implementation and results}
\label{subsec:realdata_results}
The current computational resources at Indian Statistical Institute are certainly not adequate for analysing the entire sea surface temperature dataset within a
reasonable time frame. Hence, we randomly chose $300$ spatial locations and the entire time series associated with each of them. Thus, our selected subsample consists
of $119,400$ spatio-temporal observations, which is not a small dataset with respect to our sophisticated Bayesian hierarchical modeling framework with complex
dependence structures. Indeed, we are not aware of application of realistically sophisticated Bayesian hierarchical models to spatio-temporal datasets as large.  
We further randomly choose another set of $50$ locations from the remaining set of locations and the corresponding time series data of size $398$ for each location 
for evaluation of the 
predictive performance of our model. As before, we standardize the dataset and report the prediction results after transforming them back to the original 
location and scale.

Simplification of the structure induced by the spatio-temporal random effects by setting $\phi(\bs_i,t_k)=\phi_0(\bs_i,t_k)$ for $i=1,\ldots,n=300$ and $k=1,\ldots,m=398$
(for training)
and $\phi(\tilde\bs_i,t_k)=\phi_0(\tilde\bs_i,t_k)$ for $i=1,\ldots,n=50$ and $k=1,\ldots,m=398$ (for prediction)
in our marginalized L\'{e}vy-dyanmic model 
(\ref{eq:re_model_marginalized1}) and the following discussion
brought down the implementation time (on $80$ parallel processors) from
more than $3$ days (estimated) to less than a single day. 
Here the average TTMCMC 
acceptance rates of the birth, death and no-change moves over
$398$ time points are approximately $0.095$, $0.709$ and $0.661$, respectively, and the average overall TTMCMC acceptance rate is $0.439$.
The acceptance rates for TMCMC and the mixing enhancement steps are $0.033$ and $0.03$, respectively. 
Note that these acceptance rates are broadly similar to those reported in the context of the simulation experiment
with the marginalized random effects model (Section \ref{subsec:ingore_re_uncertainty}). Thus, the acceptance rates seem to exhibit a tendency of robustness
with respect to different datasets of varying sizes.

\begin{figure}
	\centering
	\subfigure [Spatial index $1$.]{ \label{fig:spatial1_real}
	\includegraphics[width=7.5cm,height=5.5cm]{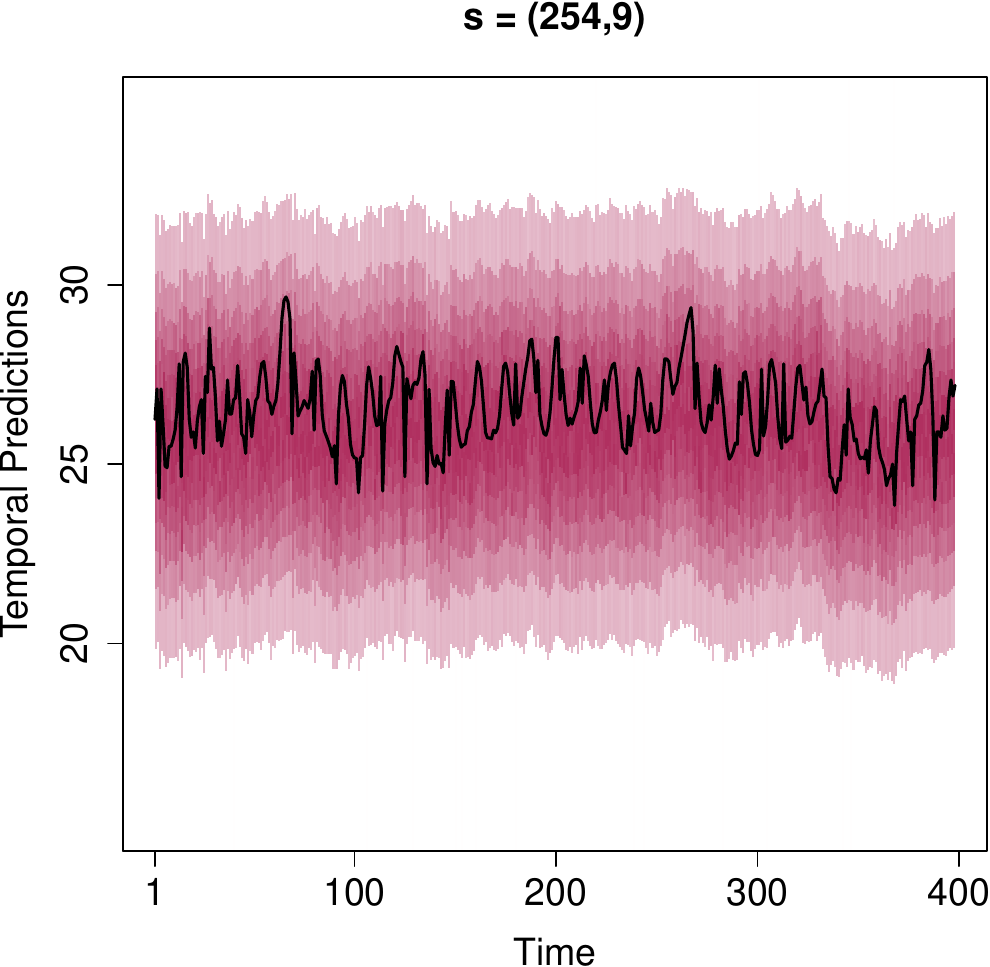}}
	\hspace{2mm}
	\subfigure [Spatial index $15$.]{ \label{fig:spatial15_real}
	\includegraphics[width=7.5cm,height=5.5cm]{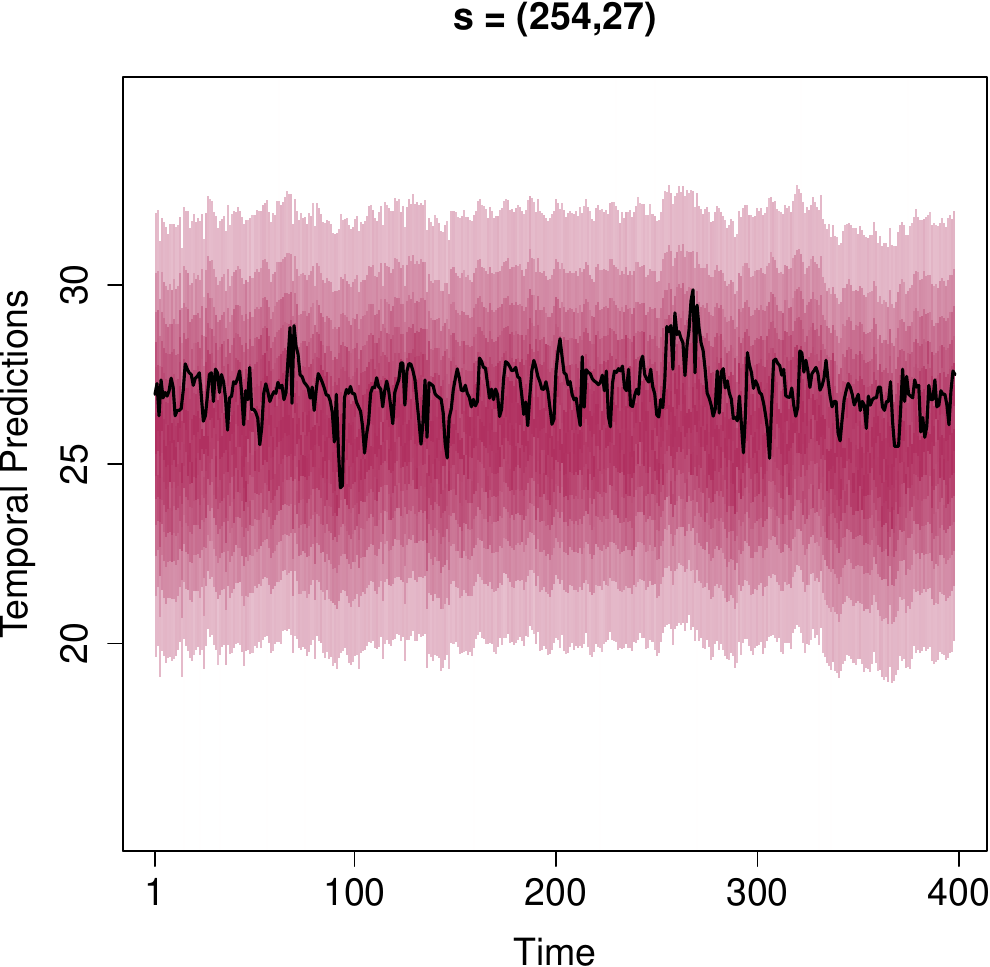}}\\
	\vspace{2mm}
	\subfigure [Spatial index $20$.]{ \label{fig:spatial20_real}
	\includegraphics[width=7.5cm,height=5.5cm]{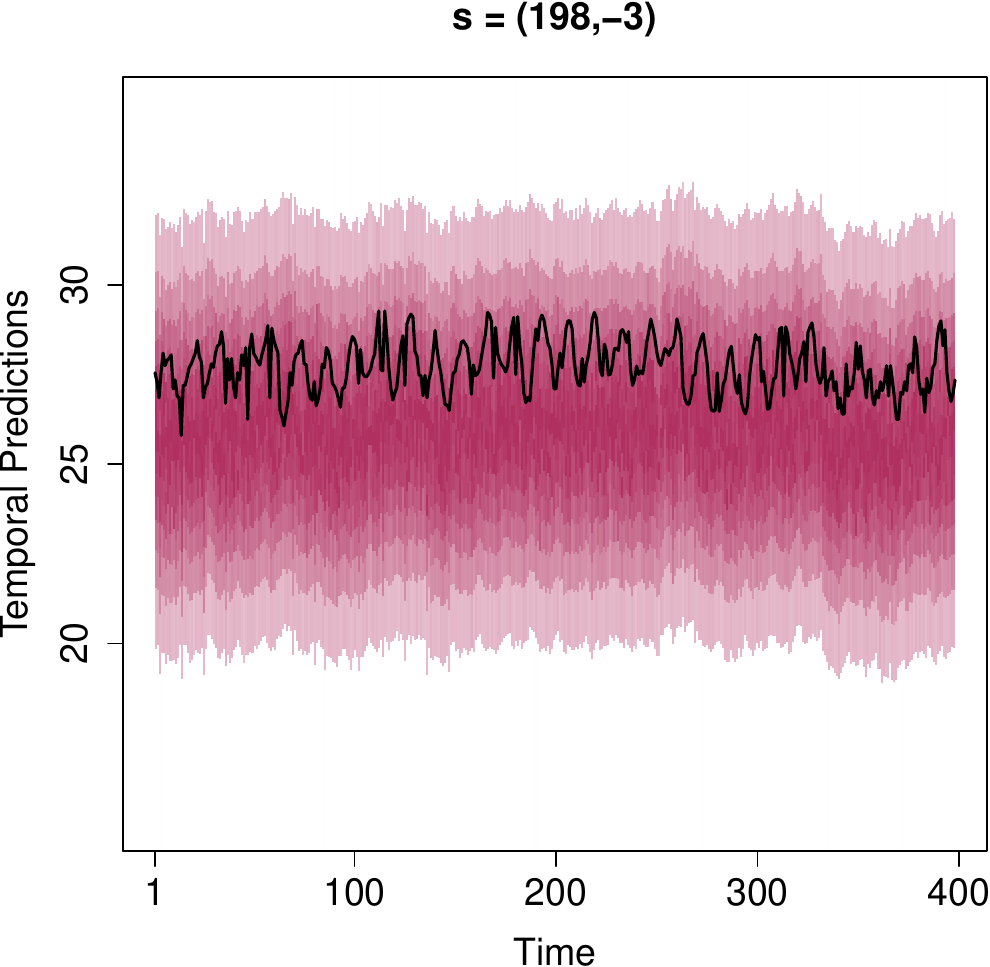}}
	\hspace{2mm}
	\subfigure [Spatial index $30$.]{ \label{fig:spatial30_real}
	\includegraphics[width=7.5cm,height=5.5cm]{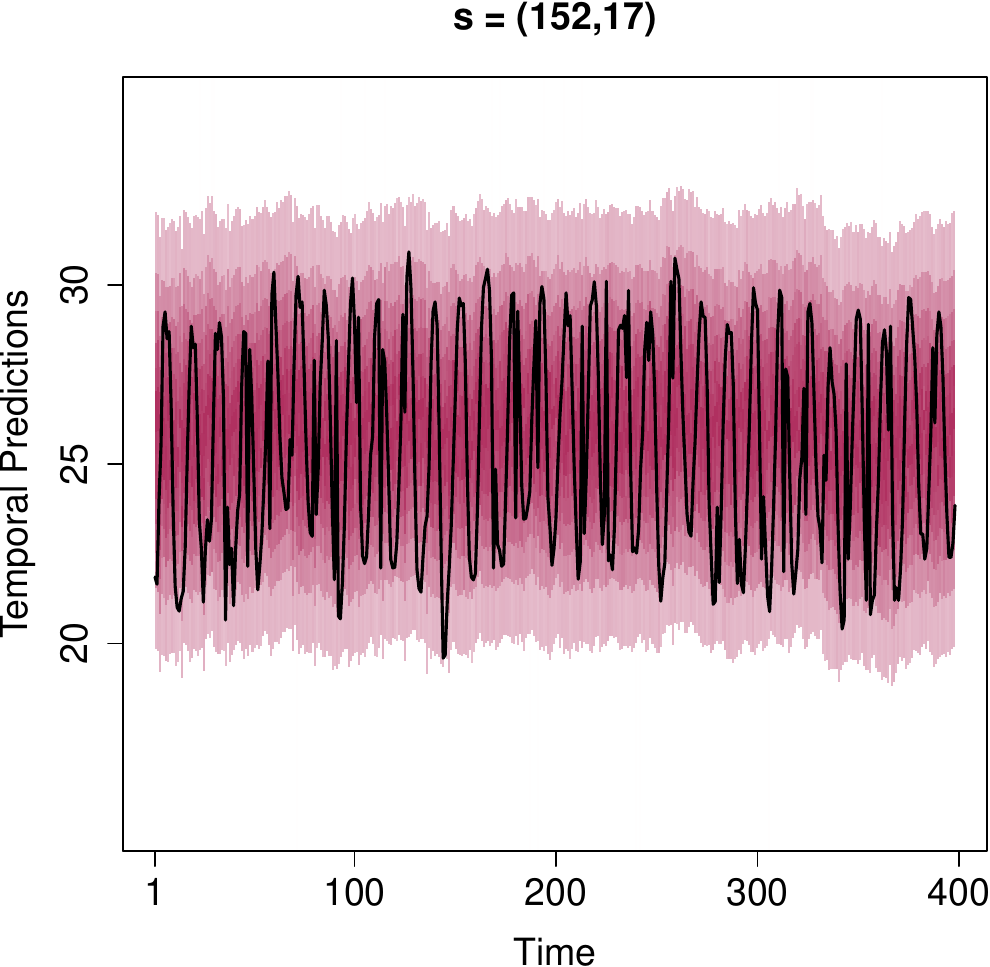}}\\
	\vspace{2mm}
	\subfigure [Spatial index $40$.]{ \label{fig:spatial40_real}
	\includegraphics[width=7.5cm,height=5.5cm]{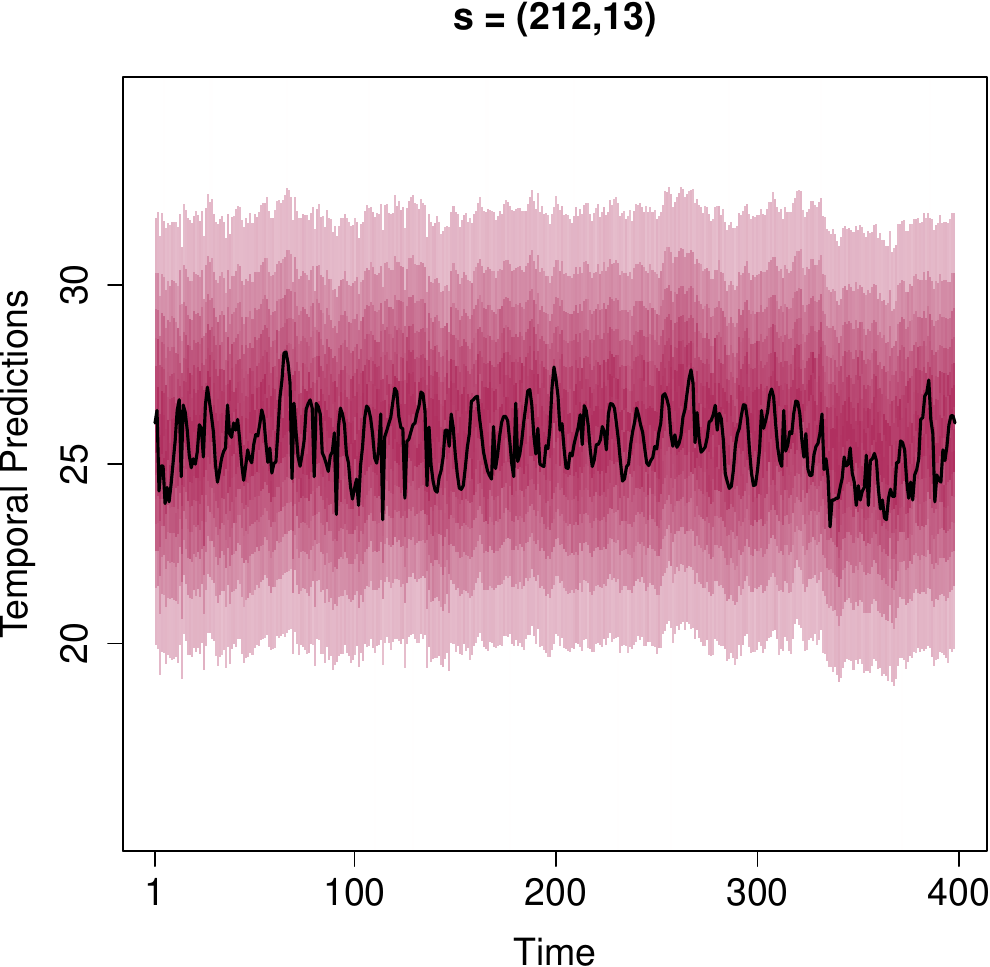}}
	\hspace{2mm}
	\subfigure [Spatial index $50$.]{ \label{fig:spatial50_real}
	\includegraphics[width=7.5cm,height=5.5cm]{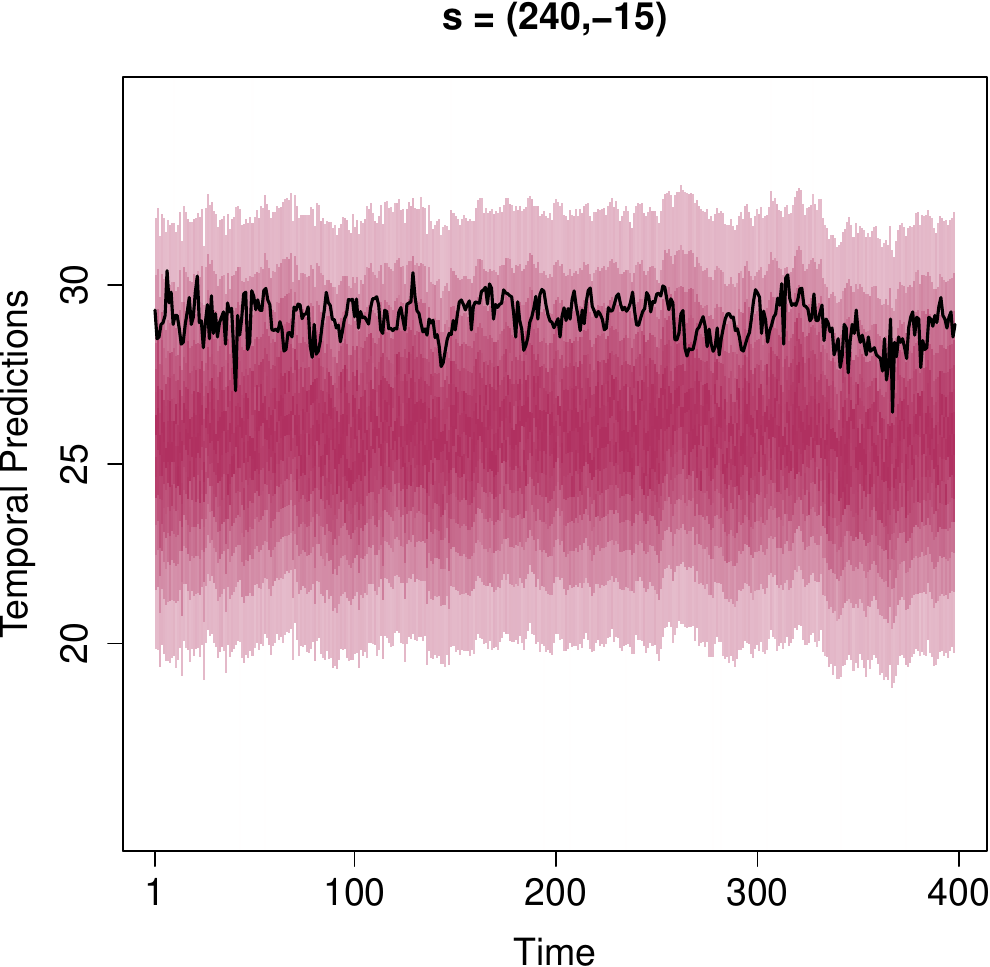}}\\
	\caption{Real data analysis: posterior temporal predictions.}
	\label{fig:temporal_plots_realdata2}
\end{figure}

Figure \ref{fig:temporal_plots_realdata2} shows the time series predictions at a few of the $50$ spatial locations set aside for prediction. Observe that
for each spatial location, the entire true time series falls well within the associated $0.875$ posterior predictive density region. 
Although in panels (c) and in particularly panel (f), the time series do not pass through the highest posterior predictive density regions depicted by the most
intense colours, the trends in these cases seem to be still well-captured by our Bayesian model and methods.

\begin{figure}
	\centering
	\subfigure [Temporal index $10$.]{ \label{fig:temporal10_real}
	\includegraphics[width=7.5cm,height=5.5cm]{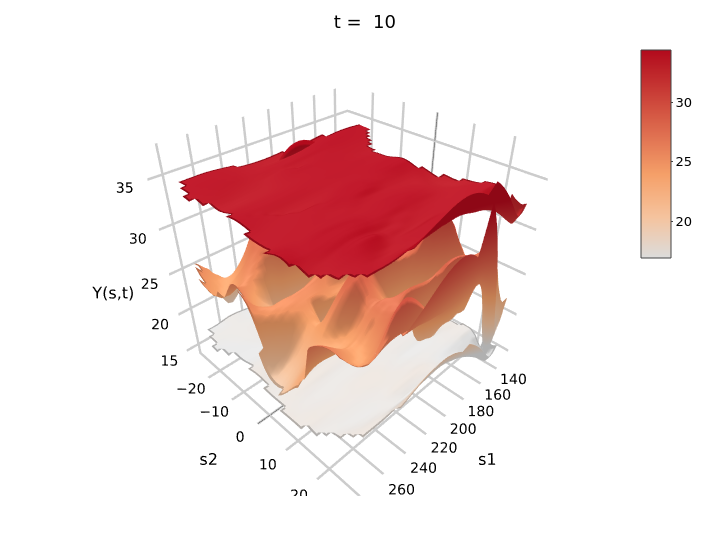}}
	\hspace{2mm}
	\subfigure [Temporal index $50$.]{ \label{fig:temporal50_real}
	\includegraphics[width=7.5cm,height=5.5cm]{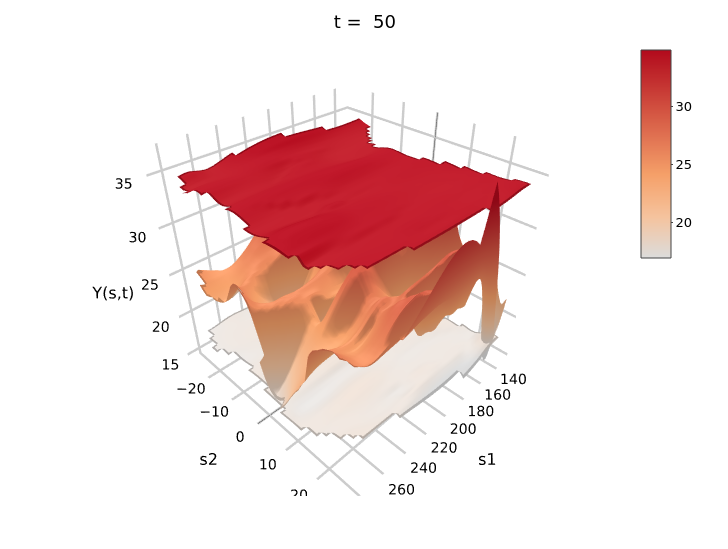}}\\
	\vspace{2mm}
	\subfigure [Temporal index $55$.]{ \label{fig:temporal55_real}
	\includegraphics[width=7.5cm,height=5.5cm]{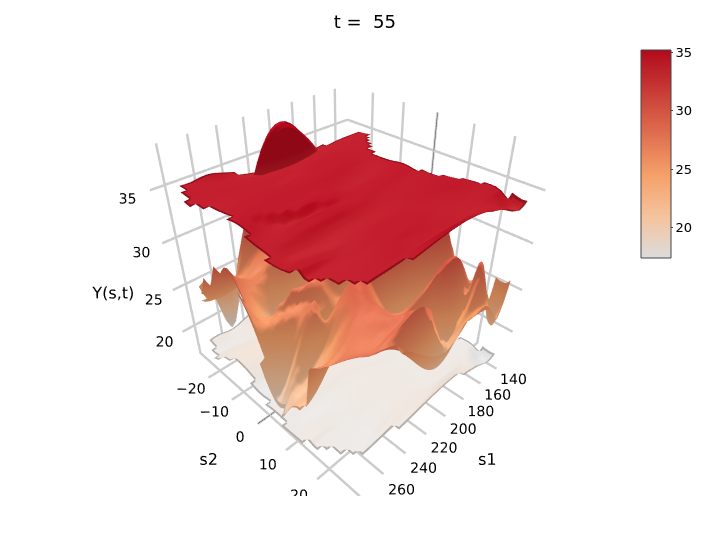}}
	\hspace{2mm}
	\subfigure [Temporal index $155$.]{ \label{fig:temporal155_real}
	\includegraphics[width=7.5cm,height=5.5cm]{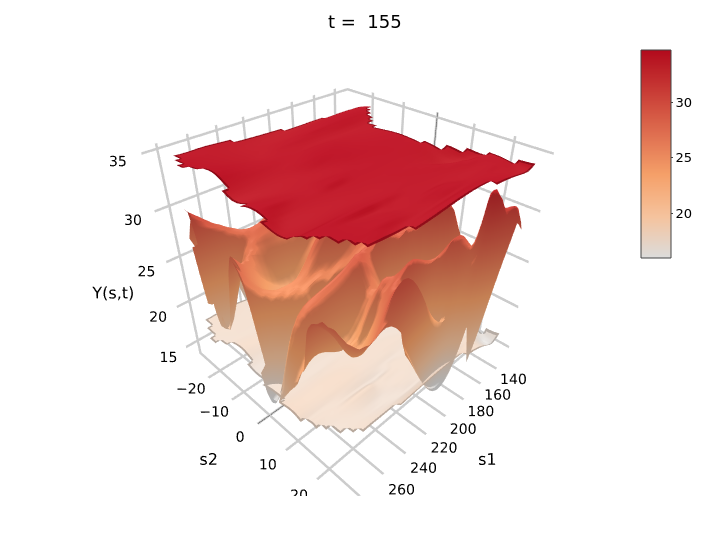}}\\
	\vspace{2mm}
	\subfigure [Temporal index $255$.]{ \label{fig:temporal255_real}
	\includegraphics[width=7.5cm,height=5.5cm]{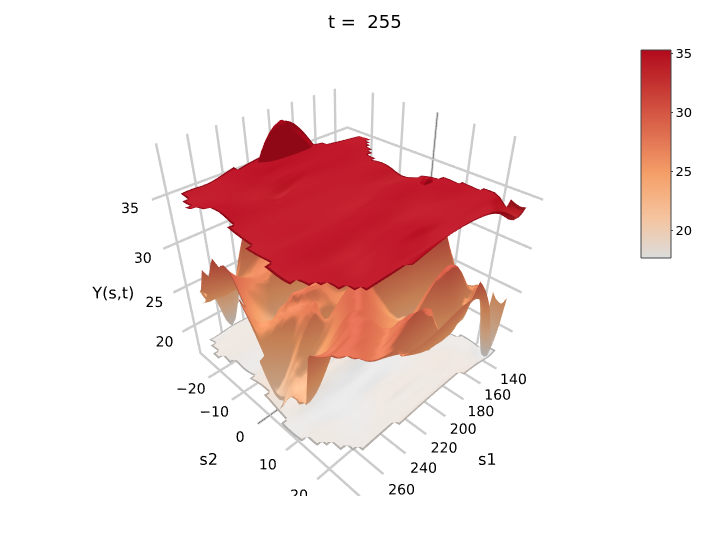}}
	\hspace{2mm}
	\subfigure [Temporal index $398$.]{ \label{fig:temporal398_real}
	\includegraphics[width=7.5cm,height=5.5cm]{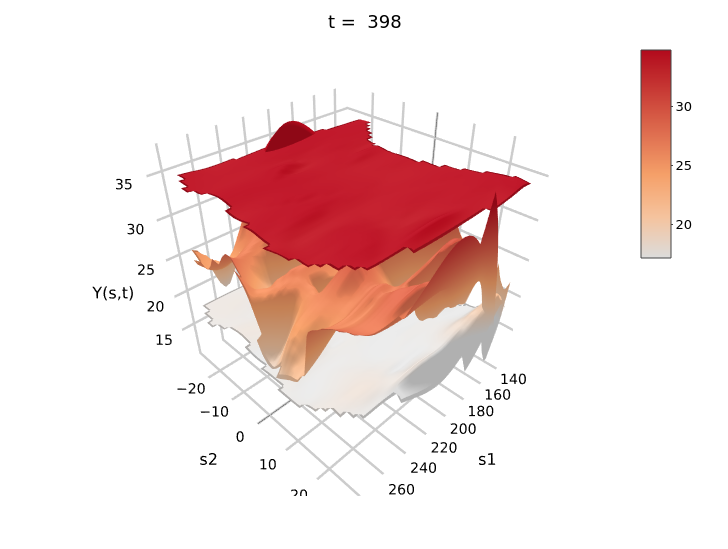}}\\
	\caption{Real data analysis: posterior spatial predictions at various time points $t$.} 
	\label{fig:spatial_plots_realdata2}
\end{figure}

Figure \ref{fig:spatial_plots_realdata2} displays the spatial predictions at a few time points, in the forms of $95\%$ lower and upper Bayesian spatial prediction surfaces,
while the middle, true spatial surface corresponds to the $50$ spatial locations meant for prediction. The reason for choosing $95\%$ surfaces, rather than $0.875$,
as in the simulation experiments is that for many time points, the $0.875\%$ surfaces failed to capture several true spatial data points. Even the $95\%$ surfaces
failed to satisfactorily contain either one or two spatial data points, for several time points (not shown for brevity). 
Setting $\sigma^2_\phi$ to some non-negligible positive value might have solved the issue, but would have increased the Bayesian prediction intervals for most of the
other data points which are already well-captured. 

\begin{figure}
	\centering
	\subfigure [Temporal index $10$.]{ \label{fig:temp10_real}
	\includegraphics[width=7.5cm,height=5.5cm]{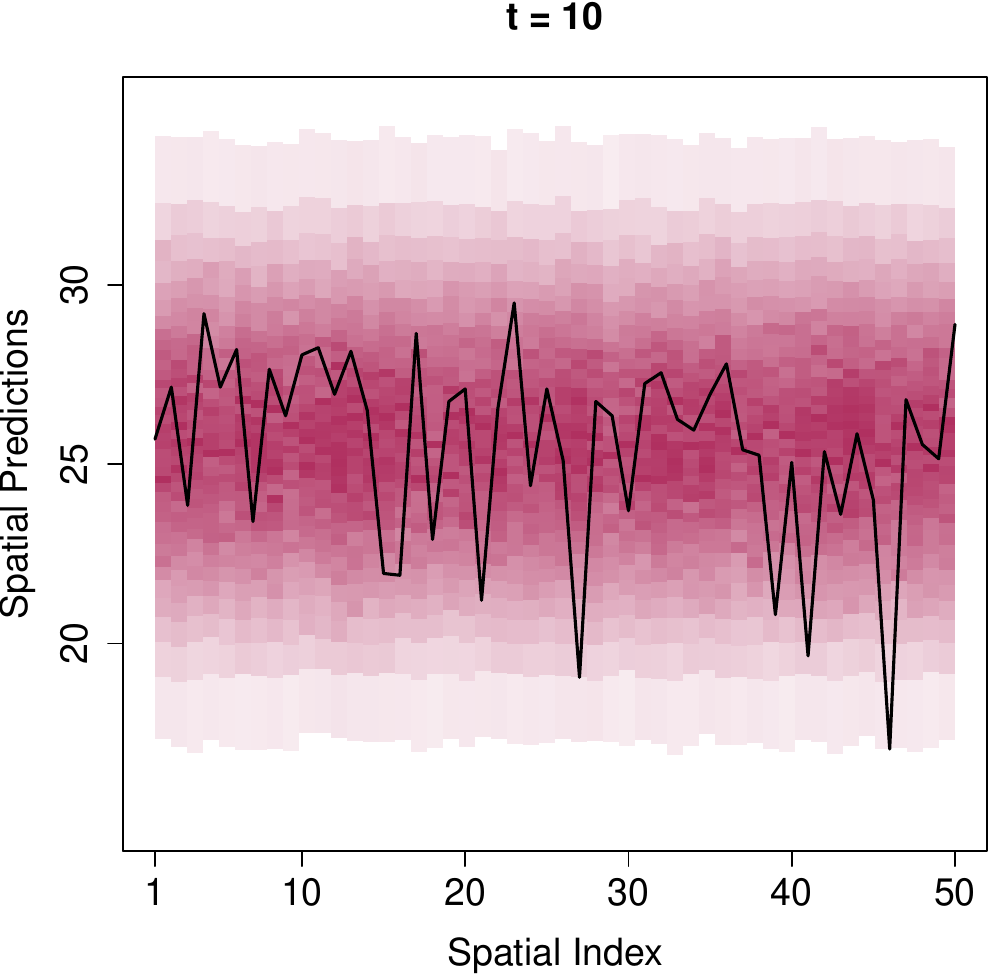}}
	\hspace{2mm}
	\subfigure [Temporal index $50$.]{ \label{fig:temp50_real}
	\includegraphics[width=7.5cm,height=5.5cm]{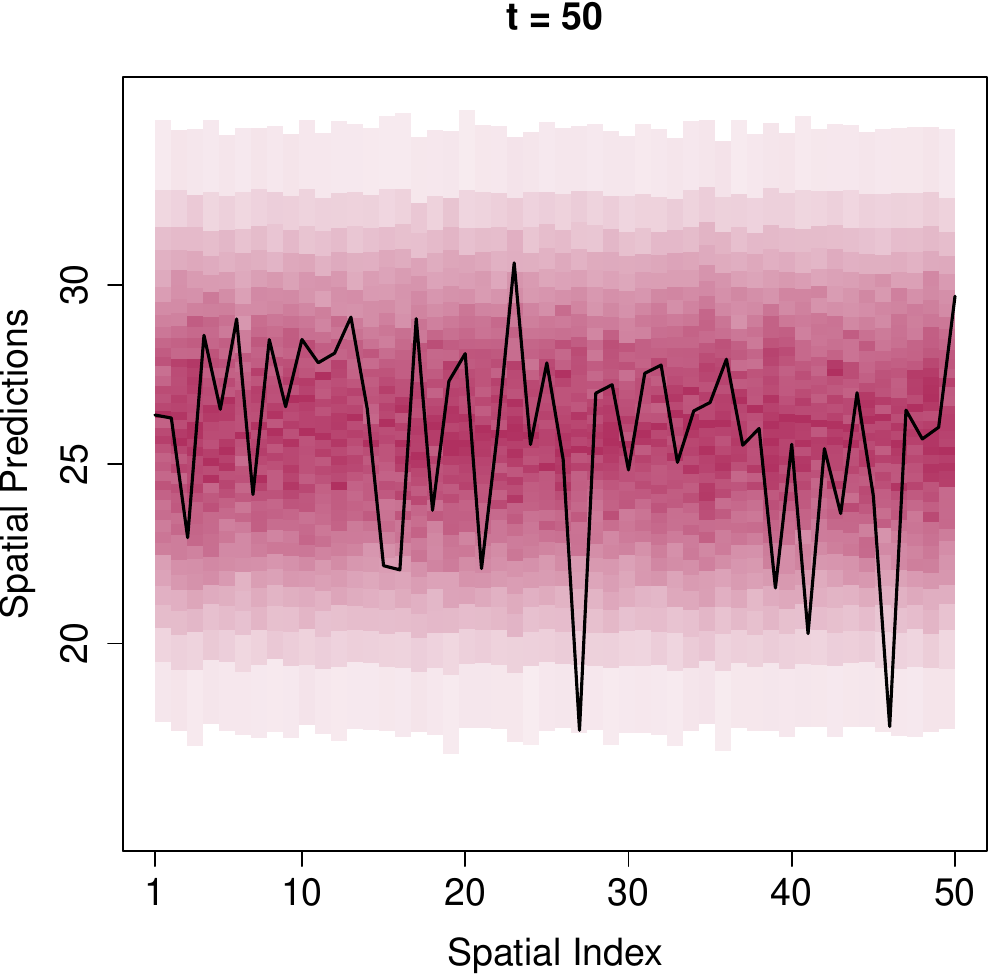}}\\
	\vspace{2mm}
	\subfigure [Temporal index $55$.]{ \label{fig:temp55_real}
	\includegraphics[width=7.5cm,height=5.5cm]{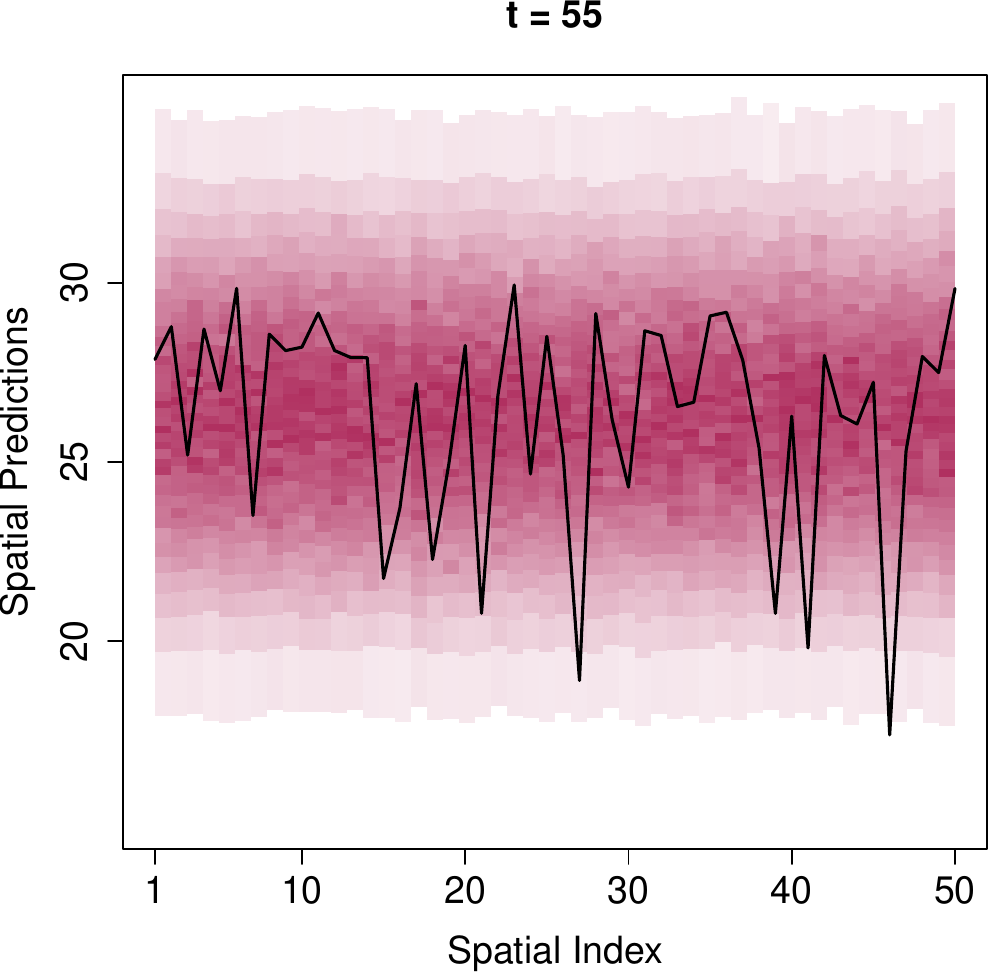}}
	\hspace{2mm}
	\subfigure [Temporal index $155$.]{ \label{fig:temp155_real}
	\includegraphics[width=7.5cm,height=5.5cm]{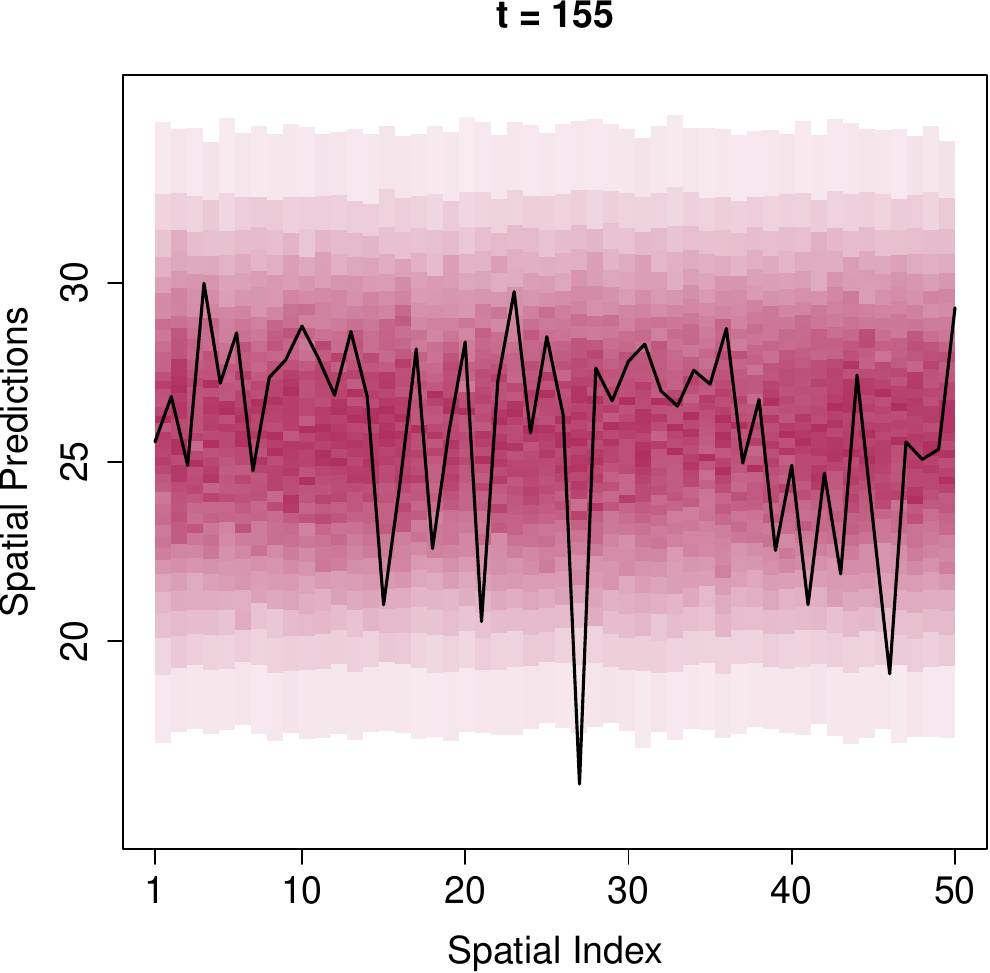}}\\
	\vspace{2mm}
	\subfigure [Temporal index $255$.]{ \label{fig:temp255_real}
	\includegraphics[width=7.5cm,height=5.5cm]{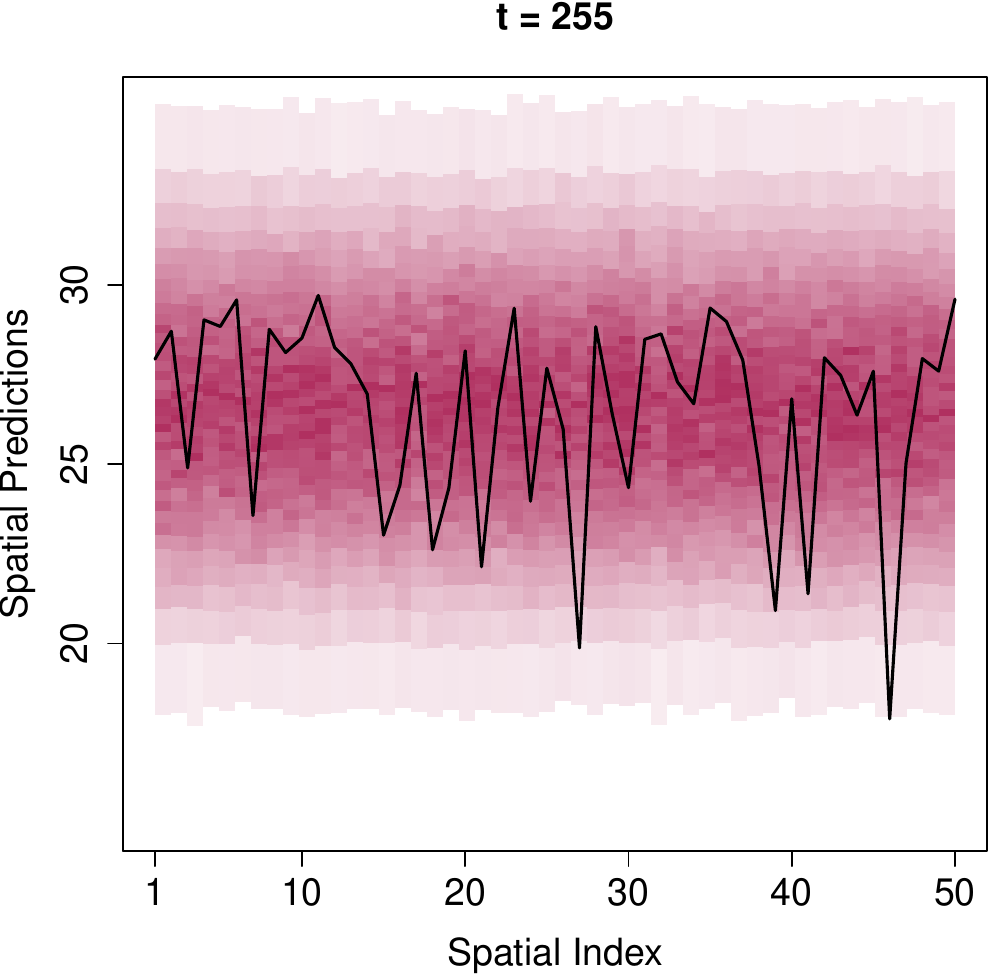}}
	\hspace{2mm}
	\subfigure [Temporal index $398$.]{ \label{fig:temp398_real}
	\includegraphics[width=7.5cm,height=5.5cm]{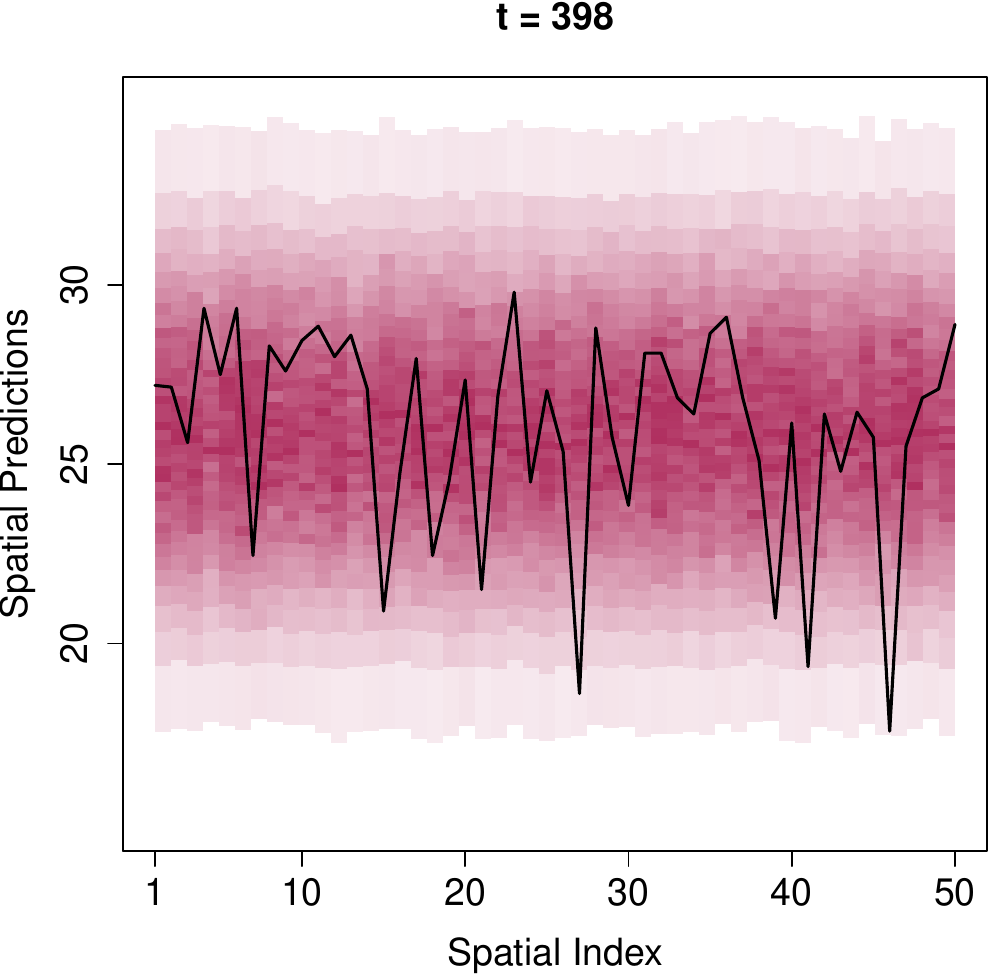}}\\
	\caption{Real data analysis: posterior spatial predictions with respect to spatial indices at the time points $t$ as in Figure \ref{fig:spatial_plots_realdata2}.} 
	\label{fig:spatial_plots2_realdata2}
\end{figure}

Figure \ref{fig:spatial_plots2_realdata2} exhibits the density based colour plots of the spatial predictions corresponding to Figure \ref{fig:spatial_plots_realdata2}.
In this case we use $40$ percentiles to make the figures correspond to $95\%$ credible regions, for comparability with Figure \ref{fig:spatial_plots_realdata2}.

\section{Summary and conclusion}
\label{sec:conclusion}
The Gaussian process is overused in the spatial/spatio-temporal literature, particularly in large data scenarios. The reality issues such as non-Gaussianity,
nonstationarity, nonseparability and properties of the lagged correlations are often relegated to the background in favour of convenience, even in general data-fitting
scenarios. These seem to induce some inflexibility in the current state-of-the-art spatial/spatio-temporal statistics, as even a plethora of existing Gaussian process
based methods and a competition among them failed to yield analysis of any big data in the order of terabytes. The key impediment in the implementation of 
all such methods is matrix-based computation, which may be ameliorated, but can not be avoided. In the zest for simplifying computations in Gaussian processes, the realistic
issues are often forgotten, as mentioned above.

Thus, there is a need to develop realistic spatial and spatio-temporal models and methods that satisfy the realistic properties, and are also amenable to fast and
efficient matrix-free computation to meet the challenges of large data. In this regard, we introduce our Bayesian L\'{e}vy-dynamic spatio-temporal model based upon L\'{e}vy
random fields, and show that it satisfies the desirable realistic properties. 
As we have shown, the approach is flexible enough for modeling space-time data with weak temporal dynamics or even purely spatial data.

For capturing micro-scale spatio-temporal variations, we introduce spatio-temporal random effects, which are amenable to marginalization that enormously simplify
computations.
The model is completely matrix-free, but is variable-dimensional with respect to each of the
time indices. We handle the variable-dimensional parameters using TTMCMC and the fixed-dimensional parameters using TMCMC, all embedded in a novel parallel MCMC
algorithm, which we code in C in the MPI paradigm for parallelism. The model structure allows us to update the variable-dimensional parameters for all the even (odd) 
time indices in parallel, followed by updating those for all odd (even) time indices. Even for fixed-dimensional updates, we compute the acceptance ratios and several
other quantities in parallel. Thus, in conjunction with integrating out the random effects, our parallel MCMC algorithm leads to huge computational savings. However,
the mixing properties are enhanced when the random effects are not integrated out. Despite this, the Bayesian predictions are almost unaffected by the issue of
marginalization of the random effects as borne out by our simulation experiment, providing the green signal to consider the marginalized model for analysis of large datasets. 

We indeed analyse a relatively large sea surface temperature dataset consisting of $139,300$ space-time observations using our marginalized Bayesian L\'{e}vy-dynamic model;
the time taken being less than $24$ hours on our VMWare with $80$ cores. The encouraging results suggest that with more powerful and well-maintained computing
facilities, we may ambitiously begin analyzing ``big data" in the order of terabytes, without any compromise whatsoever on the theoretical properties with respect to
data realism.

\section*{Acknowledgment}
We are sincerely grateful to the Editor and the referee whose comments have led to improved presentation of our manuscript.

\newpage

\renewcommand\thefigure{S-\arabic{figure}}
\renewcommand\thetable{S-\arabic{table}}
\renewcommand\thesection{S-\arabic{section}}
\renewcommand\theequation{S-\arabic{equation}}
\renewcommand\thealgo{S-\arabic{algo}}





\begin{center}
	{\bf\LARGE Supplementary Material}
\end{center}
	
\section{Proof of Theorem \ref{theorem:nonstationary_cov}}
\label{sec:proof_nonstationary_cov}
\begin{proof}
	Let us first prove (i). Note that given $\bM(\bs_1)$ and $\bM(\bs_2)$, the covariance between $f(\bs_1,t)$ and $f(\bs_2,t)$ is given by
	\begin{align}
		&Cov\left(f(\bs_1,t),f(\bs_2,t)\bigg |\bM(\bs_1),\bM(\bs_2)\right)\notag\\
		&=Cov\left(\sum_{0\leq j< J_{t}}K(\bM(\bs_1)-\bmu_{jt},t-\tau|\bSigma,\xi)\beta_{jt},\right.\notag\\
		&\qquad\qquad\left.\sum_{0\leq j< J_{t}}K(\bM(\bs_2)-\bmu_{jt_2},t-\tau|\bSigma,\xi)\beta_{jt}\bigg |\bM(\bs_1),\bM(\bs_2)\right)\notag\\
		&=E\left[Cov\left(\sum_{0\leq j< J_{t}}K(\bM(\bs_1)-\bmu_{jt},t-\tau|\bSigma,\xi)\beta_{jt},\right.\right.\notag\\
		&\qquad\qquad\left.\left.\sum_{0\leq j< J_{t}}K(\bM(\bs_2)-\bmu_{jt},t-\tau|\bSigma,\xi)\beta_{jt}\bigg |\bM(\bs_1),\bM(\bs_2),J_{t}\right)\right]\label{eq:cov2_non}\\
		&\qquad+
		Cov\left(E\left[\sum_{0\leq j< J_{t}}K(\bM(\bs_1)-\bmu_{jt},t-\tau|\bSigma,\xi)\beta_{jt}\bigg |J_{t}\right],\right.\notag\\
		&\qquad\qquad\qquad\qquad\qquad\left.
		E\left[\sum_{0\leq j< J_{t}}K(\bM(\bs_2)-\bmu_{jt},t-\tau|\bSigma,\xi)\beta_{jt}\bigg |J_{t}\right]\Bigg |\bM(\bs_1),\bM(\bs_2)\right).\label{eq:cov3_non}
	\end{align}
	Now the inner covariance structure in (\ref{eq:cov2_non}) has the following form:
	\begin{align}
		&Cov\left(\sum_{0\leq j< J_{t}}K(\bM(\bs_1)-\bmu_{jt},t-\tau|\bSigma,\xi)\beta_{jt},\right.\notag\\
		&\qquad\qquad\left.\sum_{0\leq j< J_{t}}K(\bM(\bs_2)-\bmu_{jt},t-\tau|\bSigma,\xi)\beta_{jt}\bigg |\bM(\bs_1),\bM(\bs_2),J_{t}\right)\notag\\
		&=Cov\left(\bone^T_{J_t}\bX,\bone^T_{J_t}\bY\bigg |\bM(\bs_1),\bM(\bs_2),J_{t}\right)\notag\\
		&=\bone^T_{J_t}Cov\left(\bX,\bY\bigg |\bM(\bs_1),\bM(\bs_2),J_{t}\right)\bone_{J_t},\notag 
	\end{align}
	where $\bone_{J_t}$ is the vector consisting of $J_t$ elements, each element being $1$, 
	$\bX=(K(\bM(\bs_1)-\bmu_{jt},t-\tau|\bSigma,\xi)\beta_{jt},j=0,1,\ldots,J_{t}-1)^T$ and 
	$\bY=(K(\bM(\bs_2)-\bmu_{jt},t-\tau|\bSigma,\xi)\beta_{jt},j=0,1,\ldots,J_{t}-1)^T$.
	Since 
	\begin{align}
		&Cov(\bX,\bY\bigg |\bM(\bs_1),\bM(\bs_2),J_t)\notag\\
		&=Cov\left(K(\bM(\bs_1)-\bmu_{t},t-\tau|\bSigma,\xi)\beta_{t},K(\bM(\bs_2)-\bmu_{t},t-\tau|\bSigma,\xi)\beta_{t}\bigg |\bM(\bs_1),\bM(\bs_2)\right)\bI_{J_t},\notag
	\end{align}
	where $\bI_{J_t}$ is the $J_t\times J_t$ identity matrix,
	\begin{align}
		&\bone^T_{J_t}Cov\left(\bX,\bY\bigg |\bM(\bs_1),\bM(\bs_2),J_{t}\right)\bone_{J_t}\notag\\
		&=J_tCov\left(K(\bM(\bs_1)-\bmu_{t},t-\tau|\bSigma,\xi)\beta_{t},
		K(\bM(\bs_2)-\bmu_{t},t-\tau|\bSigma,\xi)\beta_{t}\bigg |\bM(\bs_1),\bM(\bs_2),J_{t}\right).\notag
	\end{align}
	Hence, (\ref{eq:cov2_non}) is equal to
	\begin{align}
		&E\left[\bone^T_{J_t}Cov\left(\bX,\bY\bigg |\bM(\bs_1),\bM(\bs_2),J_{t}\right)\bone_{J_t}\right]\notag\\
		&=\lambda Cov\left(K(\bM(\bs_1)-\bmu_{t},t-\tau|\bSigma,\xi)\beta_{t},
		K(\bM(\bs_2)-\bmu_{t},t-\tau|\bSigma,\xi)\beta_{t}\bigg |\bM(\bs_1),\bM(\bs_2)\right)\notag\\
		&=\lambda E\left[K(\bM(\bs_1)-\bmu_{t},t)K(\bM(\bs_2)-\bmu_{t},t-\tau|\bSigma,\xi)\beta^2_{t}\bigg |\bM(\bs_1),\bM(\bs_2)\right]\notag\\
		&\qquad-\lambda E\left[K(\bM(\bs_1)-\bmu_{t},t-\tau|\bSigma,\xi)\beta_t\bigg |\bM(\bs_1)\right]
		E\left[K(\bM(\bs_2)-\bmu_{t},t-\tau|\bSigma,\xi)\beta_t
		\bigg |\bM(\bs_2)\right].
		\label{eq:cov5_non}
	\end{align}
       Let us now consider the term (\ref{eq:cov3_non}). Note that for any $\bs\in\mathbb R^p$,
	\begin{equation}
		E\left[\sum_{0\leq j< J_{t}}K(\bM(\bs)-\bmu_{jt},t-\tau|\bSigma,\xi)\beta_{jt}\bigg |\bM(\bs),J_{t}\right]
		=J_tE\left[K(\bM(\bs)-\bmu_{t},t-\tau|\bSigma,\xi)\beta_t\bigg |\bM(\bs)\right].
		\label{eq:cov7_non}
	\end{equation}
	Hence, (\ref{eq:cov3_non}) is given by
	\begin{align}
		&Cov\left(J_tE\left[K(\bM(\bs_1)-\bmu_{t},t-\tau|\bSigma,\xi)\beta_t\right],
		J_tE\left[K(\bM(\bs_2)-\bmu_{t},t-\tau|\bSigma,\xi)\beta_t \right]
		\bigg |\bM(\bs_1),\bM(\bs_2)\right)\notag\\
		&=\lambda E\left[K(\bM(\bs_1)-\bmu_{t},t-\tau|\bSigma,\xi)\beta_t\bigg |\bM(\bs_1)\right]
		E\left[K(\bM(\bs_2)-\bmu_{t},t-\tau|\bSigma,\xi)\beta_t\bigg |\bM(\bs_2)\right].
		\label{eq:cov8_non}
	\end{align}
	From (\ref{eq:cov5_non}) and (\ref{eq:cov8_non}) we obtain 
	\begin{align}
		&Cov\left(f(\bs_1,t),f(\bs_2,t)\bigg |\bM(\bs_1),\bM(\bs_2)\right)\notag\\
		&=\lambda E\left[K(\bM(\bs_1)-\bmu_{t},t-\tau|\bSigma,\xi)K(\bM(\bs_2)-\bmu_{t},t-\tau|\bSigma,\xi)\beta^2_{t}\bigg |\bM(\bs_1),\bM(\bs_2)\right].
		\label{eq:cov9_non}
	\end{align}
	Applying change-of-variable $\bmu_t\mapsto\bM(\bs_2)-\bmu_t$ 
	to the expectation (\ref{eq:cov9_non}) 
	shows that the expectation is a function of $\bM(\bs_1)-\bM(\bs_2)$. If the elements of $\bM(\cdot)$ are L\'{e}vy subordinators, 
	the distribution of $\bM(\bs_1)-\bM(\bs_2)$
	is the same as that of $\bM(\bs_1-\bs_2)$. In general, we allow the distribution of $\bM(\bs_1)-\bM(\bs_2)$ to depend upon 
	$\bs_1-\bs_2$, so that 
	\begin{equation}
		E\left[Cov\left(f(\bs_1,t),f(\bs_2,t)\bigg |\bM(\bs_1),\bM(\bs_2)\right)\right]=g(\bs_1-\bs_2),
		\label{eq:cov10_non}
	\end{equation}
	for some function $g(\cdot)$.

	Now, taking expectation of both sides of (\ref{eq:cov7_non}) yields 
	$$E\left[f(\bs,t)\bigg |\bM(\bs)\right]=\lambda E\left[K(\bM(\bs)-\bmu_{t},t-\tau|\bSigma,\xi)\beta_t\bigg |\bM(\bs)\right],$$ for any $\bs\in\mathbb R^p$.
	Hence,
	\begin{align}
	&Cov\left(E\left[f(\bs_1,t)\bigg |\bM(\bs_1)\right],E\left[f(\bs_2,t)\bigg |\bM(\bs_2)\right]\right)\notag\\
	&=\lambda^2E\left[E\left(K(\bM(\bs_1)-\bmu_{t},t-\tau|\bSigma,\xi)\beta_t\bigg |\bM(\bs_1)\right)\right.\notag\\
		&\qquad\qquad \left. \times E\left(K(\bM(\bs_2)-\bmu_{t},t-\tau|\bSigma,\xi)\beta_t\bigg |\bM(\bs_2)\right)\right]\notag\\
		&\qquad\qquad-\lambda^2E\left[K(\bM(\bs_1)-\bmu_{t},t-\tau|\bSigma,\xi)\beta_t\right]
		E\left[K(\bM(\bs_2)-\bmu_{t},t-\tau|\bSigma,\xi)\beta_t\right].
		\label{eq:cov11_non}
	\end{align}
At least the second term of (\ref{eq:cov11_non}) is not a function of $\bs_1-\bs_2$. Hence, it follows from (\ref{eq:cov9_non}), (\ref{eq:cov10_non}) and
(\ref{eq:cov11_non}) that
\begin{align}
	&Cov\left(f(\bs_1,t),f(\bs_2,t)\right)\notag\\
	&=E\left[Cov\left(f(\bs_1,t),f(\bs_2,t)\bigg |\bM(\bs_1),\bM(\bs_2)\right)\right]\notag\\
	&\qquad\qquad+Cov\left(E\left[f(\bs_1,t)\bigg |\bM(\bs_1)\right],E\left[f(\bs_2,t)\bigg |\bM(\bs_2)\right]\right)\notag
\end{align}
does not depend upon $\bs_1$ and $\bs_2$ only through $\bs_1-\bs_2$.
This proves (i).

To prove (ii), note that for any $\bs\in\mathbb R^p$,
\begin{align}
	&Cov\left(f(\bs,t_1),f(\bs,t_2)\bigg |\bM(\bs)\right)\notag\\
		&=Cov\left(\sum_{0\leq j< J_{t_1}}K(\bM(\bs)-\bmu_{jt_1},t_1-\tau|\bSigma,\xi)\beta_{jt_1},\right.\notag\\
		&\qquad\qquad\left.\sum_{0\leq j< J_{t_2}}K(\bM(\bs)-\bmu_{jt_2},t_2-\tau|\bSigma,\xi)\beta_{jt_2}\bigg |\bM(\bs)\right)\notag\\
		&=E\left[Cov\left(\sum_{0\leq j< J_{t_1}}K(\bM(\bs)-\bmu_{jt_1},t_1-\tau|\bSigma,\xi)\beta_{jt_1},\right.\right.\notag\\
		&\qquad\qquad\left.\left.\sum_{0\leq j< J_{t_2}}K(\bM(\bs)-\bmu_{jt_2},t_2-\tau|\bSigma,\xi)\beta_{jt_2}\bigg |\bM(\bs),J_{t_1},J_{t_2}\right)\right]
		\label{eq:cov2_t}\\
		&\qquad+
		Cov\left(E\left[\sum_{0\leq j< J_{t_1}}K(\bM(\bs)-\bmu_{jt_1},t_1-\tau|\bSigma,\xi)\beta_{jt_1}\bigg |\bM(\bs),J_{t_1}\right],\right.\notag\\
		&\qquad\qquad\qquad\qquad\qquad\left.
		E\left[\sum_{0\leq j< J_{t_2}}K(\bM(\bs)-\bmu_{jt_2},t_2-\tau|\bSigma,\xi)\beta_{jt_2}\bigg |\bM(\bs),J_{t_2}\right]\right).\label{eq:cov3_t}
\end{align}
First note that $$E\left[\sum_{0\leq j< J_{t}}K(\bM(\bs)-\bmu_{jt},t-\tau|\bSigma,\xi)\beta_{jt}\bigg |\bM(\bs),J_{t_1}\right]
=J_tE\left[K(\bM(\bs)-\bmu_{t},t-\tau|\bSigma,\xi)\beta_{t}\right],$$ for any $t$. Hence, the covariance term (\ref{eq:cov3_t}) is 
\begin{equation}
	Cov\left(J_{t_1}E\left[K(\bM(\bs)-\bmu_{t_1},t_1-\tau|\bSigma,\xi)\beta_{t_1}\right],
	J_{t_2}E\left[K(\bM(\bs)-\bmu_{t_2},t_2-\tau|\bSigma,\xi)\beta_{t_2}\right]\right)
	=0,
	\label{eq:cov4_t}
\end{equation}
since $J_{t_1}$ and $J_{t_2}$ are independent.

In (\ref{eq:cov2_t}), the covariance is given by $\bone^T_{J_{t_1}}Cov(\bX,\bY\big |\bM(\bs),J_{t_1},J_{t_2})\bone_{J_{t_2}}$, where 
$$\bX=\left(K(\bM(\bs)-\bmu_{jt_1},t_1-\tau|\bSigma,\xi)\beta_{jt_1},j=0,1,\ldots,J_{t_1}-1\right)$$ and 
$$\bY=\left(K(\bM(\bs)-\bmu_{jt_2},t_2-\tau|\bSigma,\xi)\beta_{jt_2}, j=0,1,\ldots,J_{t_2}-1\right).$$
Hence, $\bone^T_{J_{t_1}}Cov(\bX,\bY\big |\bM(\bs),J_{t_1},J_{t_2})\bone_{J_{t_2}}$ simplifies to 
\begin{align}
&Cov\left(K(\bM(\bs)-\bmu_{t_1},t_1-\tau|\bSigma,\xi)\beta_{t_1},K(\bM(\bs)-\bmu_{t_2},t_2-\tau|\bSigma,\xi)\beta_{t_2}\big |\bM(\bs),J_{t_1},J_{t_2}\right)\notag\\
	&\qquad\qquad\times \min\{J_{t_1},J_{t_2}\}.\notag
\end{align}
Consequently, (\ref{eq:cov2_t}) is given by 
\begin{align}
	&Cov\left(K(\bM(\bs)-\bmu_{t_1},t_1-\tau|\bSigma,\xi)\beta_{t_1},K(\bM(\bs)-\bmu_{t_2},t_2-\tau|\bSigma,\xi)\beta_{t_2}\big |\bM(\bs)\right)\notag\\
	&\qquad\qquad\times E\left[\min\{J_{t_1},J_{t_2}\}\right].\notag
\end{align}
Because of (\ref{eq:cov4_t}), $Cov\left(f(\bs,t_1),f(\bs,t_2))\bigg |\bM(\bs)\right)$ is also the same as the above expression for (\ref{eq:cov2_t}).
Since at least $E\left[\min\{J_{t_1},J_{t_2}\}\right]$ does not depend upon $t_1$ and $t_2$ through $t_1-t_2$, it is clear that the unconditional
covariance structure $Cov\left(f(\bs,t_1),f(\bs,t_2)\right)$ does not depend upon $t_1$ and $t_2$ only through $t_1-t_2$.

The proof of (iii) is similar to that of (ii) with $\bs$ corresponding to $t_1$ and $t_2$ replaced by $\bs_1$ and $\bs_2$, respectively.

\end{proof}

\section{Proof of Theorem \ref{theorem:zero_cov}}
\label{sec:proof_zero_cov}
\begin{proof}
	Let $\bTheta=\{\bmu_{jt_1}:j=0,1,\ldots,J_{t_1}-1\}\cup\{\bmu_{jt_2}:j=0,1,\ldots,J_{t_2}-1\}
	\cup\{\bSigma,\tau,\xi\}$.
	Then
	\begin{align}
		&Cov(f(\bs_1,t_1),f(\bs_2,t_2))\notag\\
		&=Cov\left(\sum_{0\leq j< J_{t_1}}K(\bM(\bs_1)-\bmu_{jt_1},t_1-\tau|\bSigma,\xi)\beta_{jt_1},
		\sum_{0\leq j< J_{t_2}}K(\bM(\bs_2)-\bmu_{jt_2},t_2-\tau|\bSigma,\xi)\beta_{jt_2}\right)\notag\\
		&=E\left[Cov\left(\sum_{0\leq j< J_{t_1}}K(\bM(\bs_1)-\bmu_{jt_1},t_1-\tau|\bSigma,\xi)\beta_{jt_1},\right.\right.\notag\\
		&\qquad\qquad\left.\left.\sum_{0\leq j< J_{t_2}}K(\bM(\bs_2)-\bmu_{jt_2},t_2-\tau|\bSigma,\xi)\beta_{jt_2}
		\bigg |\bM(\bs_1),\bM(\bs_2),J_{t_1},J_{t_2},\bTheta\right)\right]
		\label{eq:cov2}\\
		&\qquad+
		Cov\left(E\left[\sum_{0\leq j< J_{t_1}}K(\bM(\bs_1)-\bmu_{jt_1},t_1-\tau|\bSigma,\xi)\beta_{jt_1}\bigg |\bM(\bs_1),J_{t_1},J_{t_2},\bTheta\right],\right.
		\notag\\
		&\qquad\qquad\qquad\qquad\qquad\left.
		E\left[\sum_{0\leq j< J_{t_2}}K(\bM(\bs_2)-\bmu_{jt_2},t_2-\tau|\bSigma,\xi)\beta_{jt_2}\bigg |\bM(\bs_2),J_{t_1},J_{t_2},\bTheta\right]\right).
		\label{eq:cov3}
	\end{align}
	Now let us consider (\ref{eq:cov2}) in more details. Note that the inner covariance is given by
	\begin{align}
		&Cov\left(\ba'\bX,\bb'\bY\bigg |\bM(\bs_1),\bM(\bs_2),J_{t_1},J_{t_2},\bTheta\right)
		=\ba^TCov\left(\bX,\bY\bigg |\bM(\bs_1),\bM(\bs_2),J_{t_1},J_{t_2},\bTheta\right)\bb,\label{eq:cov4}
	\end{align}
	where $\ba=(K(\bM(\bs_1)-\bmu_{jt_1},t_1-\tau|\bSigma,\xi),j=0,1,\ldots,J_{t_1}-1)^T$, 
	$\bb=(K(\bM(\bs_2)-\bmu_{jt_2},t_2-\tau|\bSigma,\xi),j=0,1,\ldots,J_{t_2}-1)^T$,
	$\bX=(\beta_{jt_1},j=0,1,\ldots,J_{t_1}-1)^T$ and $\bY=(\beta_{jt_2},j=0,1,\ldots,J_{t_2}-1)^T$.

	Under $\pi$, let $Cov(\beta_{t_1},\beta_{t_2})=C_{\beta}(|t_1-t_2|)$, for some function $C_{\beta}(\cdot)$ where $C_{\beta}(|t|)\rightarrow 0$ as $t\rightarrow\infty$.
	Then (\ref{eq:cov4}) reduces to $$C_{\beta}(|t_1-t_2|)\left(\sum_{0\leq j\leq \min\{J_{t_1}-1,J_{t_2}-1\}}K(\bM(\bs_1)-\bmu_{jt_1},t_1-\tau|\bSigma,\xi)
	K(\bM(\bs_2)-\bmu_{jt_2},t_2-\tau|\bSigma,\xi)\right).$$ Hence,
	\begin{align}
		&E\left[Cov\left(\sum_{0\leq j< J_{t_1}}K(\bM(\bs_1)-\bmu_{jt_1},t_1-\tau|\bSigma,\xi)\beta_{jt_1},\right.\right.\notag\\
		&\qquad\qquad\left.\left.\sum_{0\leq j< J_{t_2}}K(\bM(\bs_2)-\bmu_{jt_2},t_2-\tau|\bSigma,\xi)\beta_{jt_2}
		\bigg |\bM(\bs_1),\bM(\bs_2),J_{t_1},J_{t_2},\bTheta\right)\right]
		\notag\\
		&=E\left[C_{\beta}(|t_1-t_2|)\right.\notag\\
		&\quad\left.\times\left(\sum_{0\leq j\leq \min\{J_{t_1}-1,J_{t_2}-1\}}K(\bM(\bs_1)-\bmu_{jt_1},t_1-\tau|\bSigma,\xi)
		K(\bM(\bs_2)-\bmu_{jt_2},t_2-\tau|\bSigma,\xi)\right)\right]\notag\\
		&\leq |C_{\beta}(|t_1-t_2|)|\notag\\
		&\quad \times E\left[\left(\sum_{0\leq j\leq \min\{J_{t_1}-1,J_{t_2}-1\}}\left|K(\bM(\bs_1)-\bmu_{jt_1},t_1-\tau|\bSigma,\xi)\right|
		\left|K(\bM(\bs_2)-\bmu_{jt_2},t_2-\tau|\bSigma,\xi)\right|\right)\right]\notag\\
		&\leq |C_{\beta}(|t_1-t_2|)|E\left[\left(\sum_{0\leq j< J_{t_1}}\left|K(\bM(\bs_1)-\bmu_{jt_1},t_1-\tau|\bSigma,\xi)\right|
		\left|K(\bM(\bs_2)-\bmu_{jt_2},t_2-\tau|\bSigma,\xi)\right|\right)\right]\notag\\
		&\leq |C_{\beta}(|t_1-t_2|)|C^2_K\lambda\notag\\
		&\rightarrow 0,~\mbox{as}~|t_1-t_2|\rightarrow\infty.
		\label{eq:cov_first}
	\end{align}
	where $C_K$ is the upper bound for $|K(\cdot,\cdot|\cdot,\cdot)|$.

	The treatment of the term (\ref{eq:cov3}) is as follows.
	\begin{align}
		&Cov\left(E\left[\sum_{0\leq j< J_{t_1}}K(\bM(\bs_1)-\bmu_{jt_1},t_1-\tau|\bSigma,\xi)\beta_{jt_1}\bigg |\bM(\bs_1),J_{t_1},J_{t_2},\bTheta\right],
		\right.\notag\\
		&\qquad\qquad\qquad\qquad\qquad\left.
		E\left[\sum_{0\leq j< J_{t_2}}K(\bM(\bs_2)-\bmu_{jt_2},t_2-\tau|\bSigma,\xi)\beta_{jt_2}\bigg |\bM(\bs_2),J_{t_1},J_{t_2},\bTheta\right]\right)\notag\\
		&=Cov\left(\sum_{0\leq j< J_{t_1}}K(\bM(\bs_1)-\bmu_{jt_1},t_1-\tau|\bSigma,\xi)E\left(\beta_{jt_1}\right),\right.\notag\\
		&\qquad\qquad\qquad\qquad\left.\sum_{0\leq j< J_{t_2}}K(\bM(\bs_2)-\bmu_{jt_2},t_2-\tau|\bSigma,\xi)E\left(\beta_{jt_2}\right)\right)\notag\\
		&=E\left[Cov\left(\sum_{0\leq j< J_{t_1}}K(\bM(\bs_1)-\bmu_{jt_1},t_1-\tau|\bSigma,\xi)E\left(\beta_{jt_1}\right),\right.\right.\notag\\
		&\qquad\qquad\qquad\qquad
		\left.\left.\sum_{0\leq j< J_{t_2}}K(\bM(\bs_2)-\bmu_{jt_2},t_2)E\left(\beta_{jt_2}\right)\bigg |J_{t_1},J_{t_2}\right)\right]\label{eq:cov5}\\
	&\qquad+Cov\left(E\left[\sum_{0\leq j< J_{t_1}}K(\bM(\bs_1)-\bmu_{jt_1},t_1-\tau|\bSigma,\xi)E\left(\beta_{jt_1}\right)\bigg |J_{t_1}\right],\right.\notag\\
		&\qquad\qquad\qquad\qquad\left. E\left[\sum_{0\leq j< J_{t_2}}K(\bM(\bs_2)-\bmu_{jt_2},t_2-\tau|\bSigma,\xi)E\left(\beta_{jt_2}\right)\bigg |J_{t_2}\right]\right)\label{eq:cov6}
	\end{align}
	First note that (\ref{eq:cov6}) is equal to
	\begin{align}
		&Cov\left(J_{t_1}E\left[K(\bM(\bs_1)-\bmu_{t_1},t_1-\tau|\bSigma,\xi)\right]E\left(\beta_{t_1}\right),\right.\notag\\
		&\qquad\qquad\qquad\qquad\left.J_{t_2}E\left[K(\bM(\bs_2)-\bmu_{t_2},t_2-\tau|\bSigma,\xi)\right]E\left(\beta_{t_2}\right)\right)
		=0,\label{eq:cov7}
	\end{align}
	since $J_{t_1}$ and $J_{t_2}$ are independent.

        Now let $\bX_2=(K(\bM(\bs_1)-\bmu_{jt_1},t_1-\tau|\bSigma,\xi),j=0,1,\ldots,J_{t_1}-1)^T$, 
	$\bY_2=(K(\bM(\bs_2)-\bmu_{jt_2},t_2-\tau|\bSigma,\xi),j=0,1,\ldots,J_{t_2}-1)^T$,
	$\ba_2=(E(\beta_{jt_1}),j=0,1,\ldots,J_{t_1}-1)^T$ and $\bb_2=(E(\beta_{jt_2}),j=0,1,\ldots,J_{t_2}-1)^T$. 
	Let $$C_{12}=Cov\left(K(\bM(\bs_1)-\bmu_{t_1},t_1-\tau|\bSigma,\xi),K(\bM(\bs_2)-\bmu_{t_2},t_2-\tau|\bSigma,\xi)\right).$$ Then (\ref{eq:cov5}) boils down to
	\begin{align}
		E\left[\ba^T_2Cov(\bX_2,\bY_2|J_{t_1},J_{t_2})\bb_2\right]&= E\left[C_{12}\left(\sum_{j=0}^{\min\{J_{t_1}-1,J_{t_2}-1\}}E(\beta_{jt_1})
		E(\beta_{jt_2})\right)\right]\notag\\
		&\leq\left|C_{12}\right|\left|E(\beta_{t_1})\right|\left|E(\beta_{t_2})\right|\left|E(J_{t_1})\right|.\notag
	\end{align}
	Since $E(J_{t_1})=\lambda<\infty$ and 
	$E(\beta_t)<\infty$ for any $t$, we need to show that $C_{12}\rightarrow 0$ if either $\|\bs_1-\bs_2\|\rightarrow\infty$ or
	$|t_1-t_2|\rightarrow\infty$, or both. In this regard, let us write
	\begin{align}
		C_{12}&=E\left[K(\bM(\bs_1)-\bmu_{t_1},t_1-\tau|\bSigma,\xi)K(\bM(\bs_2)-\bmu_{t_2},t_2-\tau|\bSigma,\xi)\right]\notag\\
		&\qquad\qquad -E\left[K(\bM(\bs_1)-\bmu_{t_1},t_1-\tau|\bSigma,\xi)\right]E\left[K(\bM(\bs_2)-\bmu_{t_2},t_2-\tau|\bSigma,\xi)\right].
		\label{eq:cov9}
	\end{align}
	Now, $K(\bM(\bs)-\bmu_{t},t-\tau|\bSigma,\xi)$ has the same distribution as $K(\bM(\bs)-\bmu_{t_0},t-\tau|\bSigma,\xi)$, 
	for any fixed $t_0$, since $\bmu_t$ is a stationary process. Also, since by hypothesis, 
	for $\ell\in\{1,\ldots,p\}$, $M_\ell(s^{(\ell)})\rightarrow\infty$ almost surely as $s^{(\ell)}\rightarrow\infty$, 
	$K(\bM(\bs)-\bmu_{t_0},t-\tau|\bSigma,\xi)\rightarrow 0$, almost surely, if at least one $s^{(\ell)}\rightarrow\infty$ for $\ell\in\{1,\ldots,p\}$, 
	or $t\rightarrow\infty$, or both.
	Since $K(\cdot,\cdot|\cdot,\cdot)$ is bounded by hypothesis, it follows by applying the dominated convergence theorem to (\ref{eq:cov9}), 
	that $C_{12}\rightarrow 0$, and hence
	(\ref{eq:cov5}) tends to zero if either $\|\bs_1-\bs_2\|\rightarrow\infty$ or $|t_1-t_2|\rightarrow\infty$, or both. 
	Combining this result with (\ref{eq:cov7}) and (\ref{eq:cov_first}), result (\ref{eq:cov1}) of Theorem \ref{theorem:zero_cov} is seen to hold. 

\end{proof}


\section{Proof of Theorem \ref{theorem:cont1}}
\label{sec:proof_cont1}
\begin{proof}
	Let us fix $t=t_0$, for some $t_0$. Given $\bM(\cdot)$, $J_{t_0}$, $\bmu_{jt_0}$, $\beta_{jt_0}$, $\tau$, $\bSigma$, 
	for $j=0,1,\ldots,J_{t_0}-1$, and $\xi$, arising from the respective
	non-null sets, $f(\bs,t_0)$ is clearly continuous by the assumptions. Hence, $f(\bs,t)$ is almost surely continuous in $\bs$, given $t$.
\end{proof}

\section{Proof of Theorem \ref{theorem:cont2}}
\label{sec:proof_cont2}
\begin{proof}
	Fix $t=t_0$.
	\begin{align}
	&E\left|f(\bs,t_0)-f(\bs_0,t_0)\right|\notag\\
	&\qquad=E\left|\sum_{0\leq j<J_{t_0}}\beta_{jt_0}\left(K\left(\bM(\bs)-\bmu_{jt_0},t_0-\tau|\bSigma,\xi\right)
		-K\left(\bM(\bs_0)-\bmu_{jt_0},t_0-\tau|\bSigma,\xi\right)\right)\right|\notag\\
	&\qquad\leq E\sum_{0\leq j<J_{t_0}}\left|\beta_{jt_0}\right|\bigg|K\left(\bM(\bs)-\bmu_{jt_0},t_0-\tau|\bSigma,\xi\right)
		-K\left(\bM(\bs_0)-\bmu_{jt_0},t_0-\tau|\bSigma,\xi\right)\bigg|\notag\\
		&\qquad=\lambda E\left|\beta_{t_0}\right|E\bigg|K\left(\bM(\bs)-\bmu_{t_0},t_0-\tau|\bSigma,\xi\right)
		-K\left(\bM(\bs_0)-\bmu_{t_0},t_0-\tau|\bSigma,\xi\right)\bigg|,
		\label{eq:cont1}
	\end{align}
	where both $\lambda$ and $E\left|\beta_{t_0}\right|$ are finite.

	Now, as $\bs\rightarrow\bs_0$, $K\left(\bM(\bs)-\bmu_{t_0},t_0-\tau|\bSigma,\xi\right)\rightarrow 
	K\left(\bM(\bs_0)-\bmu_{t_0},t_0-\tau|\bSigma,\xi\right)$, almost surely,
	by the assumptions of continuity of $\bM$ and $K(\cdot-\bmu,t_0-\tau|\bSigma,\xi)$. 
	Then, due to the uniform boundedness assumption of $K(\cdot,\cdot|\cdot,\cdot)$, it follows using the
	dominated convergence theorem that (\ref{eq:cont1}) converges to zero, as $\bs\rightarrow\bs_0$.
\end{proof}


\section{Proof of Theorem \ref{theorem:cont4}}
\label{sec:proof_cont4}
\begin{proof}
	Let us fix $t=t_0$.
First note that for any $\bs\in\mathbb R^p$,
	\begin{align}
		&E\left[f^2(\bs,t_0)|J_{t_0}\right]\notag\\
		&=E\left[\sum_{0\leq j<J_{t_0}}K^2(\bM(\bs)-\bmu_{jt_0},t_0-\tau|\bSigma,\xi)\beta^2_{jt_0}\right.\notag\\
		&\quad\left.	
		+\sum_{0\leq j_1<J_{t_0};0\leq j_2<J_{t_0};j_1\neq j_2}K(\bM(\bs)-\bmu_{j_1t_0},t_0-\tau|\bSigma,\xi)\beta_{j_1t_0}
		K(\bM(\bs)-\bmu_{j_2t_0},t_0-\tau|\bSigma,\xi)\beta_{j_2t_0}\bigg |J_{t_0}\right]\notag\\
		&=J^2_tE\left[K^2(\bM(\bs)-\bmu_{t_0},t_0-\tau|\bSigma,\xi)\beta^2_{t_0}\right]+
		J_t(J_t-1)\left(E\left[K(\bM(\bs)-\bmu_{t_0},t_0-\tau|\bSigma,\xi)\beta_{t_0}\right]\right)^2,\notag
	\end{align}
	so that
	\begin{align}
		&E\left[f^2(\bs,t_0)\right]=E\left[E\left(f^2(\bs,t_0)|J_{t_0}\right)\right]\notag\\
		&=(\lambda^2+\lambda)E\left[K^2(\bM(\bs)-\bmu_{t_0},t_0-\tau|\bSigma,\xi)\beta^2_{t_0}\right]\notag\\
		&\qquad\qquad+\lambda^2\left(E\left[K(\bM(\bs)-\bmu_{t_0},t_0-\tau|\bSigma,\xi)\beta_{t_0}\right]\right)^2.
		\label{eq:cont3}
	\end{align}
	Now 
	\begin{equation}
	E\left[f(\bs,t_0)f(\bs_0,t_0)\right]=Cov\left(f(\bs,t_0),f(\bs_0,t_0)\right)+E\left[f(\bs,t_0)\right]E\left[f(\bs_0,t_0)\right].
		\label{eq:cont4}
        \end{equation}
	Note that
	\begin{align}
		&E\left[Cov\left(f(\bs,t_0),f(\bs_0,t_0)\bigg | J_{t_0}\right)\right]\notag\\
		&=\lambda Cov\left(K(\bM(\bs)-\bmu_{t_0},t_0-\tau|\bSigma,\xi)\beta_{t_0}, 
		K(\bM(\bs_0)-\bmu_{t_0},t_0-\tau|\bSigma,\xi)\beta_{t_0}\right)\notag\\
		&=\lambda E\left[K(\bM(\bs)-\bmu_{t_0},t_0-\tau|\bSigma,\xi)
		K(\bM(\bs_0)-\bmu_{t_0},t_0-\tau|\bSigma,\xi)\beta^2_{t_0}\right]\notag\\
		&\qquad-\lambda E\left[K(\bM(\bs)-\bmu_{t_0},t_0-\tau|\bSigma,\xi)\beta_{t_0}\right]
		E\left[K(\bM(\bs_0)-\bmu_{t_0},t_0-\tau|\bSigma,\xi)\beta_{t_0}\right]
		\label{eq:cont5}
	\end{align}
and
		\begin{align}
		&Cov\left(E\left[f(\bs,t_0)\big |J_{t_0}\right],E\left[f(\bs_0,t_0)\big | J_{t_0}\right]\right)\notag\\
		&\qquad=Cov\left(J_{t_0}E\left[K(\bM(\bs)-\bmu_{t_0},t_0-\tau|\bSigma,\xi)\beta_{t_0}\right],\right.\notag\\
			&\qquad\qquad\qquad\qquad\left.	J_{t_0}E\left[K(\bM(\bs_0)-\bmu_{t_0},t_0-\tau|\bSigma,\xi)\beta_{t_0}\right]\right)\notag\\
		&\qquad = \lambda E\left[K(\bM(\bs)-\bmu_{t_0},t_0-\tau|\bSigma,\xi)\beta_{t_0}\right]
			E\left[K(\bM(\bs_0)-\bmu_{t_0},t_0-\tau|\bSigma,\xi)\beta_{t_0}\right],
			\label{eq:cont6}
	\end{align}
	so that adding up (\ref{eq:cont5}) and (\ref{eq:cont6}) yields
	\begin{align}
		&Cov\left(f(\bs,t_0),f(\bs_0,t_0)\right)\notag\\
		&=\lambda E\left[K(\bM(\bs)-\bmu_{t_0},t_0-\tau|\bSigma,\xi)
		K(\bM(\bs_0)-\bmu_{t_0},t_0-\tau|\bSigma,\xi)\beta^2_{t_0}\right].
		\label{eq:cont7}
	\end{align}
	Since for any $\bs\in\mathbb R^p$,
	\begin{equation}
		E\left[f(\bs,t_0)\right]=\lambda E\left[K(\bM(\bs)-\bmu_{t_0},t_0-\tau|\bSigma,\xi)\beta_{t_0}\right],
		\label{eq:cont8}
	\end{equation}
	it follows from (\ref{eq:cont4}), (\ref{eq:cont7}) and (\ref{eq:cont8}), that
	\begin{align}
		&E\left[f(\bs,t_0)f(\bs_0,t_0)\right]=\lambda E\left[K(\bM(\bs)-\bmu_{t_0},t_0-\tau|\bSigma,\xi)
		K(\bM(\bs_0)-\bmu_{t_0},t_0-\tau|\bSigma,\xi)\beta^2_{t_0}\right]\notag\\
		&\qquad\qquad+\lambda^2 E\left[K(\bM(\bs)-\bmu_{t_0},t_0-\tau|\bSigma,\xi)\beta_{t_0}\right]
		E\left[K(\bM(\bs_0)-\bmu_{t_0},t_0-\tau|\bSigma,\xi)\beta_{t_0}\right].
		\label{eq:cont9}
	\end{align}
	From (\ref{eq:cont3}) and (\ref{eq:cont9}) we obtain
	\begin{align}
		&E\left[f(\bs,t_0)-f(\bs_0,t_0)\right]^2\notag\\
		&=(\lambda^2+\lambda)E\left[K^2(\bM(\bs)-\bmu_{t_0},t_0-\tau|\bSigma,\xi)\beta^2_{t_0}\right]
		+\lambda^2\left(E\left[K(\bM(\bs)-\bmu_{t_0},t_0-\tau|\bSigma,\xi)\beta_{t_0}\right]\right)^2\notag\\
		&~+(\lambda^2+\lambda)E\left[K^2(\bM(\bs_0)-\bmu_{t_0},t_0-\tau|\bSigma,\xi)\beta^2_{t_0}\right]
		+\lambda^2\left(E\left[K(\bM(\bs_0)-\bmu_{t_0},t_0-\tau|\bSigma,\xi)\beta_{t_0}\right]\right)^2\notag\\
		&\qquad-2\lambda E\left[K(\bM(\bs)-\bmu_{t_0},t_0-\tau|\bSigma,\xi)
		K(\bM(\bs_0)-\bmu_{t_0},t_0-\tau|\bSigma,\xi)\beta^2_{t_0}\right]\notag\\
		&\qquad-2\lambda^2 E\left[K(\bM(\bs)-\bmu_{t_0},t_0-\tau|\bSigma,\xi)\beta_{t_0}\right]
		E\left[K(\bM(\bs_0)-\bmu_{t_0},t_0-\tau|\bSigma,\xi)\beta_{t_0}\right].
		\label{eq:cont10}
	\end{align}
	By the assumptions of this theorem and by the applications of the dominated convergence theorem to the terms of 
	the right hand side of (\ref{eq:cont10}) it follows that as $\bs\rightarrow\bs_0$,
	\begin{align}
		&E\left[f(\bs,t_0)-f(\bs_0,t_0)\right]^2\notag\\
		&\rightarrow
		2(\lambda^2+\lambda)E\left[K^2(\bM(\bs_0)-\bmu_{t_0},t_0-\tau|\bSigma,\xi)\beta^2_{t_0}\right]\notag\\
		&\qquad +2\lambda^2\left(E\left[K(\bM(\bs_0)-\bmu_{t_0},t_0-\tau|\bSigma,\xi)\beta_{t_0}\right]\right)^2\notag\\
		&\qquad -2\lambda E\left[K^2(\bM(\bs_0)-\bmu_{t_0},t_0-\tau|\bSigma,\xi)\beta^2_{t_0}\right]\notag\\
		&\qquad-2\lambda^2 \left(E\left[K(\bM(\bs_0)-\bmu_{t_0},t_0-\tau|\bSigma,\xi)\beta_{t_0}\right]\right)^2\notag\\
		&=2\lambda^2E\left[K^2(\bM(\bs_0)-\bmu_{t_0},t_0-\tau|\bSigma,\xi)\beta^2_{t_0}\right]\notag\\
		&>0,\notag
	\end{align}
	showing that $f(\bs,t)$ is not mean square continuous in $\bs$, for fixed $t$.
\end{proof}

\section{Proof of Theorem \ref{theorem:cont3}}
\label{sec:proof_cont3}
\begin{proof}
	Note that $E\left[f(\bs,t)\right]=\lambda E\left[K\left(\bM(\bs)-\bmu_{t},t-\tau|\bSigma,\xi\right)\beta_{t}\right]$. 
	Since $\bmu_t$ and $\beta_t$ are stationary processes,
	$E\left[K\left(\bM(\bs)-\bmu_{t},t-\tau|\bSigma,\xi\right)\beta_{t}\right]
	=E\left[K\left(\bM(\bs)-\bmu_{t_1},t-\tau|\bSigma,\xi\right)\beta_{t_1}\right]$,
	for any $t_1$.
	By the assumptions and by the dominated convergence theorem, 
	$$E\left[K\left(\bM(\bs)-\bmu_{t_1},t-\tau|\bSigma,\xi\right)\beta_{t_1}\right]\rightarrow 
	E\left[K\left(\bM(\bs_0)-\bmu_{t_1},t_0-\tau|\bSigma,\xi\right)\beta_{t_1}\right],$$
	as $(\bs,t)\rightarrow (\bs_0,t_0)$. The proof of (\ref{eq:cont2}) follows by noting that 
	$$E\left[f(\bs_0,t_0)\right]=\lambda E\left[K\left(\bM(\bs_0)-\bmu_{t_1},t_0-\tau|\bSigma,\xi\right)\beta_{t_1}\right].$$
\end{proof}

\section{Proof of Theorem \ref{theorem:sm1}}
\label{sec:proof_sm1}
\begin{proof}
	A sufficient condition for differentiability of (multivariate) functions with multiple arguments is that all the partial derivatives of the vector of 
	(matrix of, for multivariate functions) partial derivatives exist and are continuous. 
	The result for $f(\bs,t)$ then follows by the chain rule of differentiation applied pathwise, to almost all paths of $f(\bs,t)$, for fixed $t$.
\end{proof}

\section{Proof of Theorem \ref{theorem:sm2}}
\label{sec:proof_sm2}

\begin{proof}
By Taylor's series expansion, 
	\begin{align*}
		&K(\bM(\bs_0+\bu)-\bmu,t-\tau|\bSigma,\xi)=K(\bM(\bs_0)-\bmu,t-\tau|\bSigma,\xi)\notag\\
		&\qquad\qquad+\bu'\nabla K(\bM(\bs_0)-\bmu,t-\tau|\bSigma,\xi)+\frac{1}{2}\bu'\nabla\nabla K(\bM(\bs^*)-\bmu,t-\tau|\bSigma,\xi)\bu,
	\end{align*}
	where $\nabla$ denotes gradient and $\bs^*$ lies on the line joining $\bs_0$ and $\bs_0+\bu$. Due to assumptions (A1)and (A2),
	\begin{equation}
		\bu'\nabla\nabla K(\bM(\bs^*)-\bmu,t-\tau|\bSigma,\xi)\bu\leq C\|\bu\|^2,~\mbox{for some}~C>0.
		\label{eq:diff2}
	\end{equation}
	It follows that
	\begin{equation}
		f(\bs_0+\bu,t)=f(\bs_0,t)+\bu'\nabla f(\bs_0,t)+\frac{1}{2}\bu'\nabla\nabla f(\bs^*,t)\bu,
		\label{eq:diff3}
	\end{equation}
	where, using (\ref{eq:diff2}) we obtain
	\begin{align}
		&\|\bu\|^{-r}E\left[\bu'\nabla\nabla f(\bs^*,t)\bu\right]^r
		=\|\bu\|^{-r}E\left[\sum_{0\leq j<J_t}\bu'\nabla\nabla K(\bM(\bs^*)-\bmu_{jt},t-\tau|\bSigma,\xi)\bu\beta_{jt}\right]^r\notag\\
		&\leq C^r\|\bu\|^rE\left(\sum_{0\leq j<J_t}\beta_{jt}\right)^r\notag\\
		&\rightarrow 0,~\mbox{as}~\bu\rightarrow\bzero,\notag
	\end{align}
	since $E\left(\sum_{0\leq j<J_t}\beta_{jt}\right)^r=E\left[E\left\{\left(\sum_{0\leq j<J_t}\beta_{jt}\right)^r\bigg |J_t\right\}\right]<\infty$ due to (A3).
	In other words, $f(\bs,t)$ is $L_r$-differentiable with respect to $\bs$.
\end{proof}

\section{Form of the joint posterior distribution}
\label{sec:form_joint}
Let $\bY_{nm}=\{y(\bs_i,t_k):i=1,\ldots,n;~k=1,\ldots,m\}$ be the observed data.
For $k=1,\ldots,m$, let $\bU_k=\{\bmu_{1t_k},\ldots,\bmu_{J_{t_k}t_k}\}$,
$\bbeta_k=\{\beta_{1t_k},\ldots,\beta_{J_{t_k}t_k}\}$. Thus, the components of $\{(\bU_k,\bbeta_k):k=1,\ldots,m\}$ are Markov-dependent, although
the number of components, $J_{t_k}$, can be different for different $k$.
Let $\bJ=\{J_{t_1},\ldots,J_{t_m}\}$ and $\bM^{(\ell)}_n=\left\{M_{\ell}\left(s^{(\ell)}_1\right),\ldots,M_{\ell}\left(s^{(\ell)}_n\right)\right\}$; $\ell=1,\ldots,p$.
Also, set $\bphi=\{\phi(\bs_i,t_k):~i=1,\ldots,n,~k=1,\ldots,m\}$.

Then the joint posterior distribution of the unknowns is proportional to the following:
\begin{align}
	&\pi\left(\bJ,\bM^{(1)}_n,\ldots,\bM^{(p)}_n,\bU_1,\ldots,\bU_m,\bbeta_1,\ldots,\bbeta_m,\lambda,\tilde\sigma^2_1,\ldots,\tilde\sigma^2_p,\tau,\xi,
	\bA,\bB,\sigma^2_A,\sigma^2_B,X_1,\ldots,X_p,\right. \notag\\
&\qquad\left. \nu_1,\ldots,\nu_p,\omega^2_1,\ldots,\omega^2_p,C_1,\ldots,C_p,\tilde C_1,\ldots,\tilde C_p,\rho_{\beta},\sigma^2_{\beta},
	\rho_1,\ldots,\rho_p,\sigma^2_1,\ldots,\sigma^2_p,
	\sigma^2_{\epsilon}\bigg |\bY_{nm}\right)\notag\\
	&\propto \prod_{i=1}^n\prod_{k=1}^m\left[y(\bs_i,t_k)|X_1,\ldots,X_p,C_1,\ldots,C_p,\tilde C_1,\ldots,\tilde C_p,
	\bU_{k},\bbeta_k,J_{t_k},\tilde\sigma^2_1,\ldots,\tilde\sigma^2_p,\tau,\xi,A_i,B_k,\sigma^2_{\epsilon}\right]\notag\\
&\qquad\times\prod_{k=1}^m[J_{t_k}|\lambda]\times[\bU_1|\sigma^2_1,\ldots,\sigma^2_p]\times\prod_{k=2}^m[\bU_k|\bU_{k-1},\rho_1,\ldots,\rho_p,\sigma^2_1,\ldots,\sigma^2_p]\notag\\
	&\qquad\qquad	\times[\bbeta_1|\sigma^2_{\beta}]\times\prod_{k=2}^m[\bbeta_k|\bbeta_{k-1},\rho_{\beta},\sigma^2_{\beta}]
	\times\prod_{\ell=1}^p[X_{\ell}|\nu_{\ell},\sigma^2_{\ell}]
	\times\prod_{\ell=1}^p[C_{\ell}]\times\prod_{\ell=1}^p[\tilde C_{\ell}]\times\prod_{\ell=1}^p[\tilde\sigma^2_{\ell}]\notag\\
	&\qquad\qquad\times\prod_{\ell=1}^p[\nu_{\ell}]\times\prod_{\ell=1}^p[\omega^2_{\ell}]\times\prod_{\ell=1}^p[\rho_{\ell}]\times\prod_{\ell=1}^p[\sigma^2_{\ell}]
	\times[\alpha]\times[\bphi]\notag\\
	&\qquad\qquad\times[\lambda,\tau,\xi,\rho_{\beta},\sigma^2_{\beta},\sigma^2_\alpha,\sigma^2_\phi,\sigma^2_{\epsilon}].\notag
\end{align}

\section{Full conditional distributions}
\label{sec:fullcond}
\begin{align}
	&\left[\bU_1,\beta_1,J_{t_1}|\cdots\right]\propto [J_{t_1}|\lambda]\times[\bU_1|\sigma^2_1,\ldots,\sigma^2_p]
	\times[\bU_2|\bU_1,\rho_1,\ldots,\rho_p,\sigma^2_1,\ldots,\sigma^2_p]\notag\\
	&\qquad\qquad\times[\bbeta_1|\sigma^2_{\beta}]\times[\bbeta_2|\bbeta_1,\rho_{\beta},\sigma^2_{\beta}]
	\times\prod_{i=1}^n\left[y(\bs_i,t_1)|\bM(\bs_i),\bU_{1},\bbeta_1,J_{t_1},\tilde\sigma^2_1,\ldots,\tilde\sigma^2_p,\tau,\xi,\alpha,\phi(\bs_i,t_k),\sigma^2_{\epsilon}\right];\notag\\ 
	&\left[\bU_k,\bbeta_k,J_{t_k}|\cdots\right]\propto [J_{t_k}|\lambda]\times[\bU_{k+1}|\bU_k,\rho_1,\ldots,\rho_p,\sigma^2_1,\ldots,\sigma^2_p]\notag\\
	&\qquad\qquad\times[\bU_k|\bU_{k-1},\rho_1,\ldots,\rho_p,\sigma^2_1,\ldots,\sigma^2_p]
	\times[\bbeta_{k+1}|\bbeta_k,\rho_{\beta},\sigma^2_{\beta}]\times[\bbeta_k|\bbeta_{k-1},\rho_{\beta},\sigma^2_{\beta}]\notag\\
	&\qquad\qquad\times\prod_{i=1}^n\left[y(\bs_i,t_k)|\bM(\bs_i),\bU_{k},\bbeta_k,J_{t_k},\tilde\sigma^2_1,\ldots,\tilde\sigma^2_p,\tau,\xi,\alpha,\phi(\bs_i,t_k),\sigma^2_{\epsilon}\right];
	~k=2,\ldots,m-1;\notag\\ 
	&\left[\bU_m,\bbeta_m,J_{t_m}|\cdots\right]\propto [J_{t_m}|\lambda]\times[\bU_{m}|\bU_{m-1},\rho_1,\ldots,\rho_p,\sigma^2_1,\ldots,\sigma^2_p]
	\times[\bbeta_{m}|\bbeta_{m-1},\rho_{\beta},\sigma^2_{\beta}]\notag\\
	&\qquad\qquad\times\prod_{i=1}^n\left[y(\bs_i,t_m)|\bM(\bs_i),\bU_{m},\bbeta_m,J_{t_m},\tilde\sigma^2_1,\ldots,\tilde\sigma^2_p,\tau,\xi,\alpha,\phi(\bs_i,t_k),\sigma^2_{\epsilon}\right];
	\notag\\ 
	&[X_1,\ldots,X_p,\tilde C_1,\ldots,\tilde C_p,C_1,\ldots,C_p,
	\tilde\sigma^2_1,\ldots,\tilde\sigma^2_p,\tau,\xi,\rho_1,\ldots,\rho_p,\sigma^2_1,\ldots,\sigma^2_p,\rho_{\beta},\sigma^2_{\beta}|\ldots]\notag\\
	&\quad\propto [X_1,\ldots,X_p,\tilde C_1,\ldots,\tilde C_p,C_1,\ldots,C_p,
	\tilde\sigma^2_1,\ldots,\tilde\sigma^2_p,\tau,\xi,\rho_1,\ldots,\rho_p,\sigma^2_1,\ldots,\sigma^2_p,\rho_{\beta},\sigma^2_{\beta}]\notag\\
	&\quad\times\prod_{i=1}^n\prod_{k=1}^m\left[y(\bs_i,t_k)|X_1,\ldots,X_p,C_1,\ldots,C_p,\tilde C_1,\ldots,\tilde C_p,
	\bU_{k},\bbeta_k,J_{t_k},\tilde\sigma^2_1,\ldots,\tilde\sigma^2_p,\tau,\xi,\alpha,\phi(\bs_i,t_k),\sigma^2_{\epsilon}\right]\notag\\
         &\qquad\times [\bU_1|\sigma^2_1,\ldots,\sigma^2_p]\times\prod_{k=2}^m[\bU_k|\bU_{k-1},\rho_1,\ldots,\rho_p,\sigma^2_1,\ldots,\sigma^2_p]\notag\\
         &\qquad\qquad\times[\bbeta_1|\sigma^2_{\beta}]\times\prod_{k=2}^m[\bbeta_k|\bbeta_{k-1},\rho_{\beta},\sigma^2_{\beta}];\label{eq:c7}\\
	 &[\alpha|\cdots]\propto[\alpha|\sigma^2_{\alpha}]\times\prod_{i=1}^n\prod_{k=1}^m\left[y(\bs_i,t_k)|X_1,\ldots,X_p,C_1,\ldots,C_p,\tilde C_1,\ldots,\tilde C_p,\right.\notag\\
	 &\qquad\qquad\qquad\qquad\left.\bU_{k},\bbeta_k,J_{t_k},\tilde\sigma^2_1,\ldots,\tilde\sigma^2_p,\tau,\xi,\alpha,\phi(\bs_i,t_k),\sigma^2_{\epsilon}\right];
	 \notag\\ 
	 &[\phi(\bs_i,t_k)|\cdots]\propto [\phi(\bs_i,t_k)|\sigma^2_{\phi}]\times
	 \left[y(\bs_i,t_k)|X_1,\ldots,X_p,C_1,\ldots,C_p,\tilde C_1,\ldots,\tilde C_p,\right.\notag\\
	 &\qquad\qquad\qquad\qquad\left.\bU_{k},\bbeta_k,J_{t_k},\tilde\sigma^2_1,\ldots,\tilde\sigma^2_p,\tau,\xi,\alpha,\phi(\bs_i,t_k),\sigma^2_{\epsilon}\right];
	 \notag\\ 
	&[\nu_1,\ldots,\nu_p,\omega^2_1,\ldots,\omega^2_p|\cdots]\propto\prod_{\ell=1}^p[\nu_\ell]\times\prod_{\ell=1}^p[\omega^2_{\ell}]
	\times\prod_{\ell=1}^p[X_{\ell}|\nu_{\ell},\omega^2_{\ell}];\notag\\ 
	&[\lambda|\cdots]\propto [\lambda]\times\prod_{k=1}^m[J_{t_k}|\lambda];\notag\\ 
	&[\sigma^2_{\alpha}|\cdots]\propto [\sigma^2_{\alpha}]\times [\alpha|\sigma^2_{\alpha}];\notag\\ 
	&[\sigma^2_\phi|\cdots]\propto [\sigma^2_\phi]\times\prod_{i=1}^n\prod_{k=1}^m[\phi(\bs_i,t_k)|\sigma^2_\phi];\notag\\ 
	&[\sigma^2_{\epsilon}|\cdots]\propto [\sigma^2_{\epsilon}]\times\prod_{i=1}^n\prod_{k=1}^m
	\left[y(\bs_i,t_k)|\bM(\bs_i),\bU_{k},\bbeta_k,J_{t_k},\tilde\sigma^2_1,\ldots,\tilde\sigma^2_p,\tau,\xi,\alpha,\phi(\bs_i,t_k),\sigma^2_{\epsilon}\right].
	\notag 
\end{align}
The full conditional distributions of $\alpha$, $\phi(\bs_i,t_k)$, $\nu_{\ell}$, $\omega^2_{\ell}$, $\lambda$, $\sigma^2_\alpha$, 
$\sigma^2_\phi$ and $\sigma^2_{\epsilon}$, are available
in closed forms. Specifically, 
\begin{align}
	&[\alpha|\cdots]\equiv N\left(\mu_{\alpha},\tilde\sigma^2_{\alpha}\right),~\mbox{where}\notag\\
	&\mu_{\alpha}=\left(\frac{1}{\sigma^2_\alpha}+\frac{nm}{\sigma^2_{\epsilon}}\right)^{-1}
	\left(\frac{\mu_{\alpha}}{\sigma^2_{\alpha}}+\sum_{i=1}^n\sum_{k=1}^m\frac{(y(\bs_i,t_k)-\phi(\bs_i,t_k)-f(\bs_i,t_k))^2}{\sigma^2_{\epsilon}}\right);\notag\\
	&\tilde\sigma^2_{\alpha}=\left(\frac{1}{\sigma^2_\alpha}+\frac{nm}{\sigma^2_{\epsilon}}\right)^{-1}.\notag\\
	&[\sigma^2_{\alpha}|\cdots]\equiv IG\left(a_{\sigma^2_{\alpha}}+\frac{1}{2},b_{\sigma^2_{\alpha}}+\frac{1}{2}(\alpha-\mu_{\alpha})^2\right).\notag
	\\[10mm]
	&[\phi(\bs_i,t_k)|\cdots]\equiv N\left(\mu_{\phi(\bs_i,t_k)},\tilde\sigma^2_{\phi(\bs_i,t_k)}\right),~\mbox{where}\notag\\
	&\mu_{\phi(\bs_i,t_k)}=\left(\frac{1}{\sigma^2_{\phi}}+\frac{1}{\sigma^2_{\epsilon}}\right)^{-1}
	\left(\frac{\phi_0(\bs_i,t_k)}{\sigma^2_{\alpha}}+\frac{y(\bs_i,t_k)-f(\bs_i,t_k)}{\sigma^2_{\epsilon}}\right);\notag\\
	&\sigma^2_{\phi(\bs_i,t_k)}=\left(\frac{1}{\sigma^2_{\phi}}+\frac{1}{\sigma^2_\epsilon}\right)^{-1}.\notag\\
	&[\sigma^2_\phi|\cdots]\equiv IG\left(a_{\sigma^2_{\phi}}+\frac{1}{2},b_{\sigma^2_{\phi}}+\frac{1}{2}\sum_{i=1}^n\sum_{k=1}^m(\phi(\bs_i,t_k)-\phi_0(\bs_i,t_k))^2\right).
	\notag
	\\[10mm]
	&[\nu_{\ell}|\cdots]\equiv N\left(\tilde\mu_{\nu_{\ell}},\tilde\sigma^2_{\nu_{\ell}}\right),~\mbox{where}\notag\\
	&\tilde\mu_{\nu_{\ell}}=\left(\frac{1}{\omega^2_{\ell}}+\frac{1}{\sigma^2_{\nu_{\ell}}}\right)^{-1}\left(\frac{X_{\ell}}{\omega^2_{\ell}}\right);\notag\\
	&\tilde\sigma^2_{\nu_{\ell}}=\left(\frac{1}{\omega^2_{\ell}}+\frac{1}{\sigma^2_{\nu_{\ell}}}\right)^{-1}.\notag
	\\[10mm]
	&[\omega^2_{\ell}|\cdots]\equiv IG\left(a_{\omega^2_{\ell}}+\frac{1}{2},b_{\omega^2_{\ell}}+\frac{1}{2}(X_\ell-\nu_\ell)^2\right);\notag\\
	&[\lambda|\cdots]\equiv G\left(a_{\lambda}+\sum_{k=1}^mJ_k,b_{\lambda}+m\right);\notag\\
	&[\sigma^2_{\epsilon}|\cdots]\equiv IG\left(a_{\sigma^2_{\epsilon}}+\frac{mn}{2},b_{\sigma^2_{\epsilon}}
	+\frac{1}{2}\sum_{i=1}^n\sum_{k=1}^m(y(\bs_i,t_k)-\alpha-\phi(\bs_i,t_k)-f(\bs_i,t_k))^2\right).\notag
\end{align}
\newpage
\begin{algo}\label{algo:ttmcmc} \topline A parallel MCMC algorithm for L\'{e}vy-dynamic inference.
\botline \normalfont \ttfamily
\begin{itemize}
	\item 
		Let the initial values of $\btheta$ and $\bzeta$ be $\btheta^{(0)}$ and $\bzeta^{(0)}$, respectively. Also, let 
		$\{(\bU^{(0)}_k,\bbeta^{(0)}_k,J^{(0)}_{t_k}):k=1,\ldots,m\}$ denote the initial values of the parameters associated with the variable-dimensional context.
		In $\bU^{(0)}_k$, we denote by $\mu^{(\ell,0)}_{jt_k}$ the initial value of $\mu^{(\ell)}_{jt_k}$, and in general, at the $r$-th iteration, we denote 
		the value of $\mu^{(\ell)}_{jt_k}$ by $\mu^{(\ell,r)}_{jt_k}$.
 \item For $r=0,1,2,\ldots$
\begin{enumerate}
	\item Split the odd values of $k\in\{1,\ldots,m\}$ into separate parallel processors.
\item In any parallel processor, for odd $k$, update $(\bU_k,\bbeta_k,J_{t_k})$ using TTMCMC in the following manner. 
\item Generate $u=(u_1,u_2,u_3)\sim Multinomial (1;w_{b,J^{(r)}_{t_k}},w_{d,J^{(r)}_{t_k}},w_{nc,J^{(r)}_{t_k}})$, where 
	$w_{b,J^{(r)}_{t_k}}$, $w_{d,J^{(r)}_{t_k}}$ and $w_{nc,J^{(r)}_{t_k}}$ are birth, death
		and no-change probabilities, given $J^{(r)}_{t_k}$. Thus, these are non-negative quantities and sum to one. 
		Also, $w_{d,J^{(r)}_{t_k}}=0$ if $J^{(r)}_{t_k}=1$. 
		If a maximum value of $J_{t_k}$ is specified, $J_{\max,k}$, say, then $w_{b,J^{(r)}_{t_k}}=0$ if $J^{(r)}_{t_k}=J_{\max,k}$. 
 \item If $u_1=1$ (increase dimension), generate $U\sim U(0,1)$ and do the following: 
	 \begin{enumerate}
		 \item  If $U\leq \tilde p$, where $\tilde p\in[0,1]$ (use additive transformation for dimension change),
 \begin{enumerate}
 \item Randomly select a value from $\{1,\ldots,J^{(r)}_{t_k}\}$ assuming uniform probability $1/J^{(r)}_{t_k}$. Let $j$ denote the chosen co-ordinate.
 \item Generate $\e_1\sim N(0,1)$, and independently, for $\ell=1,\ldots,p$, $\epsilon^{(\ell)} \sim N(0,1)$. 
 Propose the following birth move: 
		 $$ \bbeta'_k=\left(\beta^{(r)}_{1t_k},\ldots,\beta^{(r)}_{j-1,t_k},\beta^{(r)}_{j,t_k}+a_{\beta,j,t_k}|\e_1|,
		 \beta^{(r)}_{j,t_k}-a_{\beta,j,t_k}|\e_1|,\beta^{(r)}_{j+1,t_k},\ldots,
		 \beta^{(r)}_{J^{(r)}_{t_k},t_k}\right);$$	 
		 $$ \bmu'_{\ell k}=\left(\mu^{(\ell,r)}_{1t_k},\ldots,\mu^{(\ell,r)}_{j-1,t_k},\mu^{(\ell,r)}_{jt_k}+a_{\mu^{(\ell)},j,t_k}|\e^{(\ell)}|,
		 \mu^{(\ell,r)}_{jt_k}-a_{\mu^{(\ell)},j,t_k}|\e^{(\ell)}|,\mu^{(r)}_{j+1,t_k},\ldots,
		 \mu^{(\ell,r)}_{J^{(r)}_{t_k},t_k}\right),$$
		 for $\ell=1,\ldots,p$. 
		 In the above, $a_{\vartheta,j,t_k}$ is a general notation standing for the appropriate positive scaling constant associated with the $j$-th co-ordinate of any 
		 general parameter vector $\bvartheta$ depending upon $t_k$.
	 \item Re-label the elements of $\bbeta'_k$ as $(\beta'_{1t_k},\beta'_{2t_k},\ldots,\beta'_{J'_{t_k}t_k})$, and those of $\bmu'_{\ell k}$ as
		 $(\mu'_{1\ell t_k},\mu'_{2\ell t_k},\ldots,\mu'_{J'_{t_k}\ell t_k})$, for $\ell=1,\ldots,p$, where $J'_{t_k}=J^{(r)}_{t_k}+1$. 
		 Then letting $\bmu'_{jk}=(\mu'_{j 1 k},\ldots,\mu'_{j p k})^T$, set $\bU'_k=\left\{\bmu'_{1k},\ldots,\bmu'_{J'_{t_k}k}\right\}$.
\item The acceptance probability of the birth move is:
 \begin{align}
 a_b &= 
	 \min\left\{1, \frac{1}{J^{(r)}_{t_k}+1}\times\frac{w_{d,J'_{t_k}}}{w_{b,J^{(r)}_{t_k}}}\times 
	 \dfrac{\pi\left(\bU'_k,\bbeta'_k,J'_{t_k}|\cdots\right)}{\pi\left(\bU^{(r)}_k,\bbeta^{(r)}_k,J^{(r)}_{t_k}|\cdots\right)}
	 \times 2^{p+1}	a_{\beta,j,t_k}\times\prod_{\ell=1}^pa_{\mu^{(\ell)},j,t_k}\right\}.\notag 
 \end{align}
\item Set \[ (\bU^{(r+1)}_k,\bbeta^{(r+1)}_k,J^{(r+1)}_{t_k})= \left\{\begin{array}{ccc}
		(\bU'_k,\bbeta'_k,J'_{t_k}) & \mbox{ with probability } & a_b \\
	(\bU^{(r)}_k,\bbeta^{(r)}_k,J^{(r)}_{t_k}) & \mbox{ with probability } & 1 - a_b.
\end{array}\right.\]
\end{enumerate}
	 \end{enumerate}

	 \begin{enumerate}
		 \item[(b)]  If $U> \tilde p$ (use multiplicative transformation for dimension change),
 \begin{enumerate}
 \item Randomly select a value from $\{1,\ldots,J^{(r)}_{t_k}\}$ assuming uniform probability $1/J^{(r)}_{t_k}$. Let $j$ denote the chosen co-ordinate.
 \item Generate $\e_1\sim U(-1,1)$, and independently, for $\ell=1,\ldots,p$, $\epsilon^{(\ell)} \sim U(-1,1)$. 
 Propose the following birth move: 
		 $$ \bbeta'_k=\left(\beta^{(r)}_{1t_k},\ldots,\beta^{(r)}_{j-1,t_k},\beta^{(r)}_{j,t_k}\e_1,
		 \beta^{(r)}_{j,t_k}/\e_1,\beta^{(r)}_{j+1,t_k},\ldots,
		 \beta^{(r)}_{J^{(r)}_{t_k},t_k}\right);$$	 
		 $$ \bmu'_{\ell k}=\left(\mu^{(\ell,r)}_{1t_k},\ldots,\mu^{(\ell,r)}_{j-1,t_k},\mu^{(\ell,r)}_{jt_k}\e^{(\ell)},
		 \mu^{(\ell,r)}_{jt_k}/\e^{(\ell)},\mu^{(r)}_{j+1,t_k},\ldots,
		 \mu^{(\ell,r)}_{J^{(r)}_{t_k},t_k}\right),$$
		 for $\ell=1,\ldots,p$. 
	 \item Re-label the elements of $\bbeta'_k$ as $(\beta'_{1t_k},\beta'_{2t_k},\ldots,\beta'_{J'_{t_k}t_k})$, and those of $\bmu'_{\ell k}$ as
		 $(\mu'_{1\ell t_k},\mu'_{2\ell t_k},\ldots,\mu'_{J'_{t_k}\ell t_k})$, for $\ell=1,\ldots,p$, where $J'_{t_k}=J^{(r)}_{t_k}+1$. 
		 Then letting $\bmu'_{jk}=(\mu'_{j 1 k},\ldots,\mu'_{j p k})^T$, set $\bU'_k=\left\{\bmu'_{1k},\ldots,\bmu'_{J'_{t_k}k}\right\}$.
\item Then the acceptance probability of the birth move is:
 \begin{align}
 a_b &= 
%
	 \min\left\{1, \frac{1}{J^{(r)}_{t_k}+1}\times\frac{w_{d,J'_{t_k}}}{w_{b,J^{(r)}_{t_k}}}\times 
	 \dfrac{\pi\left(\bU'_k,\bbeta'_k,J'_{t_k}|\cdots\right)}{\pi\left(\bU^{(r)}_k,\bbeta^{(r)}_k,J^{(r)}_{t_k}|\cdots\right)}
	 \times \frac{|\beta^{(r)}_{jt_k}|}{|\e_1|}\times\prod_{\ell=1}^p\frac{|\mu^{(\ell,r)}_{jt_k}|}{|\e^{(\ell)}|}\right\}.\notag 
 \end{align}
\item Set \[ (\bU^{(r+1)}_k,\bbeta^{(r+1)}_k,J^{(r+1)}_{t_k})= \left\{\begin{array}{ccc}
		(\bU'_k,\bbeta'_k,J'_{t_k}) & \mbox{ with probability } & a_b \\
	(\bU^{(r)}_k,\bbeta^{(r)}_k,J^{(r)}_{t_k}) & \mbox{ with probability } & 1 - a_b.
\end{array}\right.\]
\end{enumerate}
	 \end{enumerate}

 \item If $u_2=1$ (decrease dimension), generate $U\sim U(0,1)$ and do the following: 
	 \begin{enumerate}
		 \item  If $U\leq \tilde p$ (use additive transformation for dimension change),
 \begin{enumerate}
	 \item Randomly select a co-ordinate $j$ from $\{1,\ldots,J^{(r)}_{t_k}\}$ 
		 assuming uniform probability $1/J^{(r)}_{t_k}$ for each co-ordinate, and randomly select a co-ordinate $j'$ from 
		 $\{1,\ldots,J^{(r)}_{t_k}\}\backslash\{j\}$ 
		 with probability $1/(J^{(r)}_{t_k}-1)$.
		 Assuming $j<j'$, let $\beta^*_{jt_k}=(\beta^{(r)}_{jt_k}+\beta^{(r)}_{j't_k})/2$. Replace $\beta^{(r)}_{jt_k}$ with $\beta^*_{jt_k}$ 
		 and delete $\beta^{(r)}_{j't_k}$. Similarly, for $\ell=1,\ldots,p$, 
		 let $\mu^*_{j\ell t_k}=(\mu^{(\ell,r)}_{jt_k}+\mu^{(\ell,r)}_{j't_k})/2$.
		 Replace $\mu^{(\ell,r)}_{jt_k}$ with $\mu^*_{jt_k}$ and delete $\mu^{(\ell,r)}_{j't_k}$.
 \item Propose the following death move: 
	 $$ \bbeta'_k=(\beta^{(r)}_{1t_k},\ldots,\beta^{(r)}_{j-1,t_k},\beta^*_{jt_k},\beta^{(r)}_{j+1,t_k},\ldots,\beta^{(r)}_{j'-1,t_k},
		 \beta^{(r)}_{j'+1,t_k},\ldots,\beta^{(r)}_{J^{(r)}_{t_k}});$$	 
		 $$ \bmu'_{\ell k}=(\mu^{(\ell,r)}_{1t_k},\ldots,\mu^{(\ell,r)}_{j-1,t_k},\mu^*_{jt_k},\mu^{(\ell,r)}_{j+1,t_k},\ldots,\mu^{(\ell,r)}_{j'-1,t_k},
		 \mu^{(\ell,r)}_{j'+1,t_k},\ldots,\mu^{(\ell,r)}_{J^{(r)}_{t_k}}),$$
		 for $\ell=1,\ldots,p$.
	 \item Re-label the elements of $\bbeta'_k$ as $(\beta'_{1t_k},\beta'_{2t_k},\ldots,\beta'_{J'_{t_k}t_k})$, and those of $\bmu'_{\ell k}$ as
		 $(\mu'_{1\ell t_k},\mu'_{2\ell t_k},\ldots,\mu'_{J'_{t_k}\ell t_k})$, for $\ell=1,\ldots,p$, where $J'_{t_k}=J^{(r)}_{t_k}-1$. 
		 Then letting $\bmu'_{jk}=(\mu'_{j 1 k},\ldots,\mu'_{j p k})^T$, set $\bU'_k=\left\{\bmu'_{1k},\ldots,\bmu'_{J'_{t_k}k}\right\}$.
\item Then the acceptance probability of the death move is:
 \begin{align}
 a_d &= 
	 \min\left\{1, J^{(r)}_{t_k}\times\frac{w_{b,J'_{t_k}}}{w_{d,J^{(r)}_{t_k}}} 
	 \times\dfrac{\pi\left(\bU'_k,\bbeta'_k,J'_{t_k}|\cdots\right)}{\pi\left(\bU^{(r)}_k,\bbeta^{(r)}_k,J^{(r)}_{t_k}|\cdots\right)}
	 \times2^{-p-1}\frac{1}{a_{\beta,j,t_k}}\times\prod_{\ell=1}^p\frac{1}{a_{\mu^{(\ell)},j,t_k}}\right\}.\notag 
 \end{align}

\item Set \[ (\bU^{(r+1)}_k,\bbeta^{(r+1)}_k,J^{(r+1)}_{t_k})= \left\{\begin{array}{ccc}
		(\bU'_k,\bbeta'_k,J'_{t_k}) & \mbox{ with probability } & a_d \\
	(\bU^{(r)}_k,\bbeta^{(r)}_k,J^{(r)}_{t_k}) & \mbox{ with probability } & 1 - a_d.
\end{array}\right.\]
\end{enumerate}
	 \end{enumerate}

	 \begin{enumerate}
		 \item[(b)]  If $U>\tilde p$ (use multiplicative transformation for dimension change),
 \begin{enumerate}
	 \item Randomly select a co-ordinate $j$ from $\{1,\ldots,J^{(r)}_{t_k}\}$ 
		 assuming uniform probability $1/J^{(r)}_{t_k}$ for each co-ordinate, and randomly select a co-ordinate $j'$ from $\{1,\ldots,J^{(r)}_{t_k}\}\backslash\{j\}$ 
		 with probability $1/(J^{(r)}_{t_k}-1)$.
		 Assuming $j<j'$, let $\beta^*_{jt_k}=\sqrt{|\beta^{(r)}_{jt_k}\beta^{(r)}_{j't_k}|}$ with probability $1/2$ and set 
		 $\beta^*_{jt_k}=-\sqrt{|\beta^{(r)}_{jt_k}\beta^{(r)}_{j't_k}|}$ with the remaining probability. 
		 Replace $\beta^{(r)}_{jt_k}$ with $\beta^*_{jt_k}$ and delete $\beta^{(r)}_{j't_k}$.
		 Similarly, for $\ell=1,\ldots,p$, let $\mu^*_{j\ell t_k}=\sqrt{|\mu^{(\ell,r)}_{jt_k}\mu^{(\ell,r)}_{j't_k}|}$ with probability $1/2$
		 and $\mu^*_{j\ell t_k}=-\sqrt{|\mu^{(\ell,r)}_{jt_k}\mu^{(\ell,r)}_{j't_k}|}$ with the remaining probability.
		 Replace $\mu^{(\ell,r)}_{jt_k}$ with $\mu^*_{j\ell t_k}$ and delete $\mu^{(\ell,r)}_{j't_k}$.
 \item Propose the following death move: 
	 $$ \bbeta'_k=(\beta^{(r)}_{1t_k},\ldots,\beta^{(r)}_{j-1,t_k},\beta^*_{jt_k},\beta^{(r)}_{j+1,t_k},\ldots,\beta^{(r)}_{j'-1,t_k},
		 \beta^{(r)}_{j'+1,t_k},\ldots,\beta^{(r)}_{J^{(r)}_{t_k}});$$	 
		 $$ \bmu'_{\ell k}=(\mu^{(\ell,r)}_{1t_k},\ldots,\mu^{(\ell,r)}_{j-1,t_k},\mu^*_{jt_k},\mu^{(\ell,r)}_{j+1,t_k},\ldots,\mu^{(\ell,r)}_{j'-1,t_k},
		 \mu^{(\ell,r)}_{j'+1,t_k},\ldots,\mu^{(\ell,r)}_{J^{(r)}_{t_k}}),$$
		 for $\ell=1,\ldots,p$.
	 \item Re-label the elements of $\bbeta'_k$ as $(\beta'_{1t_k},\beta'_{2t_k},\ldots,\beta'_{J'_{t_k}t_k})$, and those of $\bmu'_{\ell k}$ as
		 $(\mu'_{1\ell t_k},\mu'_{2\ell t_k},\ldots,\mu'_{J'_{t_k}\ell t_k})$, for $\ell=1,\ldots,p$, where $J'_{t_k}=J^{(r)}_{t_k}-1$. 
		 Then letting $\bmu'_{jk}=(\mu'_{j 1 k},\ldots,\mu'_{j p k})^T$, set $\bU'_k=\left\{\bmu'_{1k},\ldots,\bmu'_{J'_{t_k}k}\right\}$.
\item Then the acceptance probability of the death move is:
 \begin{align}
 a_d &= 
	 \min\left\{1, J^{(r)}_{t_k}\times\frac{w_{b,J'_{t_k}}}{w_{d,J^{(r)}_{t_k}}} 
	 \times\dfrac{\pi\left(\bU'_k,\bbeta'_k,J'_{t_k}|\cdots\right)}{\pi\left(\bU^{(r)}_k,\bbeta^{(r)}_k,J^{(r)}_{t_k}|\cdots\right)}
	 \times\frac{1}{\beta^{(r)}_{j',t_k}}\times\prod_{\ell=1}^p\frac{1}{\mu^{(\ell,r)}_{j',t_k}}\right\}.\notag 
 \end{align}

\item Set \[ (\bU^{(r+1)}_k,\bbeta^{(r+1)}_k,J^{(r+1)}_{t_k})= \left\{\begin{array}{ccc}
		(\bU'_k,\bbeta'_k,J'_{t_k}) & \mbox{ with probability } & a_d \\
	(\bU^{(r)}_k,\bbeta^{(r)}_k,J^{(r)}_{t_k}) & \mbox{ with probability } & 1 - a_d.
\end{array}\right.\]

\end{enumerate}
	 \end{enumerate}

\item If $u_3=1$ (dimension remains unchanged), then given that there are $d$ dimensions in the current iteration, 
		generate $U\sim U(0,1)$. 
	\begin{enumerate}
	\item If $U\leq \tilde p$, then do the following:
	\begin{enumerate}
		\item[(i)] Let $d=(2p+1)J^{(r)}_{t_k}$, the number of parameters in $\bV_k=(\bU_k,\bbeta_k)$ given $J^{(r)}_{t_k}$. Let $v_{jk}$ 
			denote the $j$-th element of $\bV_k$. 
			For $j=1,\ldots,d$, set, for some appropriate $c\in (0,1)$, 
			$\tilde a_{v,j,t_k}=ca_{v,j,t_k}$, where $a_{v,j,t_k}$ denotes the positive scaling constant associated with the additive transformation of $v_{jk}$. 
			Thus, the previous scaling constants associated with the birth and death moves are multiplied by $c$ to enhance acceptance rates, since
			$d$ parameters are now updated at once.
		\item[(ii)] Generate $\varepsilon\sim N(0,1)$, $b_j\stackrel{iid}{\sim}U(\{-1,1\})$ for $j=1,\ldots,d$, and set
			$v'_{jk}=v^{(r)}_{jk}+b_j\tilde a_{v,j,t_k}|\varepsilon|$, for $j=1,\ldots,d$, where $v^{(r)}_{jk}$ denotes the value of $v_{jk}$ at the
			$r$-th iteration. Let $\bV'_k=(v'_{1k},\ldots,v'_{dk})$ and $\bV^{(r)}_k=(v^{(r)}_{1k},\ldots,v^{(r)}_{dk})$. 
		\item[(iii)] Evaluate 
                        \begin{equation*}
				\alpha_1=\min\left\{1,\frac{\pi(\bV'_k|\cdots)}{\pi\left(\bV^{(r)}_k|\cdots\right)}\right\}.
                        \end{equation*}
		\item[(iv)] Set $(\bV^{(r+1)}_k,J^{(r+1)}_{t_k})=(\bV'_k,J^{(r)}_{t_k})$ with probability $\alpha_1$, 
			else set $(\bV^{(r+1)},J^{(r+1)}_{t_k})=(\bV^{(r)}_k,J^{(r)}_{t_k})$.
	\end{enumerate}
	\end{enumerate}

      \begin{enumerate}
	      \item[(b)] If $U>\tilde p$, then do the following:
	\begin{enumerate}
		\item[(i)] Generate $\varepsilon\sim U(-1,1)$, $b_j\stackrel{iid}{\sim}U(\{-1,0,1\})$ for $j=1,\ldots,d$, and set
			$v'_{jk}=v^{(r)}_{jk}\varepsilon$ if $b_j=1$, $v'_{jk}=v^{(r)}_{jk}/\varepsilon$ if $b_j=-1$,  
			for $j=1,\ldots,d$. 
			Calculate $|J|=|\varepsilon|^{\sum_{j=1}^db_j}$.
		\item[(ii)] Evaluate 
                        \begin{equation*}
				\alpha_2=\min\left\{1,\frac{\pi(\bV'_k|\cdots)}{\pi(\bV^{(r)}_k|\cdots)}\times|J|\right\}.
                        \end{equation*}
		\item[(iii)] Set $(\bV^{(r+1)}_k,J^{(r+1)}_{t_k})=(\bV'_k,J^{(r)}_{t_k})$ with probability $\alpha_2$, 
			else set $(\bV^{(r+1)}_k,J^{(r+1)}_{t_k})=(\bV^{(r)}_k,J^{(r)}_{t_k})$.
	\end{enumerate}
	\end{enumerate}

\item Repeat steps 1.-- 5. with even values of $k\in\{1,\ldots,m\}$, split into separate processors. 	

\item Send the results of updating in steps 1.-- 7. to processor $0$.	
\item ({\it TMCMC step for updating $\btheta$ in processor $0$})	
		Let $d=6p+10$, the dimension of $\btheta$. Generate $U\sim U(0,1)$. 
	\begin{enumerate}
	\item If $U\leq \tilde p$, then do the following:
	\begin{enumerate}
		\item[(i)] Generate $\varepsilon\sim N(0,1)$, $b_j\stackrel{iid}{\sim}U(\{-1,1\})$ for $j=1,\ldots,d$, and set
			$\theta'_{j}=\theta^{(r)}_{j}+b_ja_{\theta,j}|\varepsilon|$, for $j=1,\ldots,d$, where $\theta^{(r)}_{j}$ denotes the value of $\theta_{j}$ at the
			$r$-th iteration. Let $\btheta'=(\theta'_{1},\ldots,\theta'_{d})$ and $\btheta^{(r)}=(\theta^{(r)}_{1},\ldots,\theta^{(r)}_{d})$. 
		\item[(ii)] Evaluate 
                        \begin{equation*}
				\alpha_1=\min\left\{1,\frac{\pi(\btheta'|\cdots)}{\pi\left(\btheta^{(r)}|\cdots\right)}\right\}.
                        \end{equation*}
		\item[(iii)] Set $\btheta^{(r+1)}=\btheta'$ with probability $\alpha_1$, 
			else set $\btheta^{(r+1)}=\btheta^{(r)}$.
	\end{enumerate}
	\end{enumerate}

      \begin{enumerate}
	      \item[(b)] If $U>\tilde p$, then do the following:
	\begin{enumerate}
		\item[(i)] Generate $\varepsilon\sim U(-1,1)$, $b_j\stackrel{iid}{\sim}U(\{-1,0,1\})$ for $j=1,\ldots,d$, and set
			$\theta'_{j}=\theta^{(r)}_{j}\varepsilon$ if $b_j=1$, $\theta'_{j}=\theta^{(r)}_{j}/\varepsilon$ if $b_j=-1$,  
			for $j=1,\ldots,d$. 
			Calculate $|J|=|\varepsilon|^{\sum_{j=1}^db_j}$.
		\item[(ii)] Evaluate 
                        \begin{equation*}
				\alpha_2=\min\left\{1,\frac{\pi(\btheta'|\cdots)}{\pi(\btheta^{(r)}|\cdots)}\times|J|\right\}.
                        \end{equation*}
		\item[(iii)] Set $\btheta^{(r+1)}=\btheta'$ with probability $\alpha_2$, 
			else set $\btheta^{(r+1)}=\btheta^{(r)}$.
	\end{enumerate}
	\end{enumerate}
		
\item ({\it Mixing-enhancement step at processor $0$}) 
	Let $d=6p+10$. Generate $U\sim U(0,1)$. 
	\begin{enumerate}
		\item If $U\leq \tilde q$, where $\tilde q\in (0,1)$, then do the following
	\begin{enumerate}
		\item[(i)] For $j=1,\ldots,d$, set 
	$\tilde a_{\theta,j}=ca_{\btheta,j}$, for some appropriate $c\in (0,1)$. 
		\item[(ii)] Generate $\tilde U\sim U(0,1)$ and $\varepsilon\sim N(0,1)$. If $\tilde U<1/2$, set
			$\theta''_j=\theta^{(r+1)}_j+\tilde a_{\theta,j}|\varepsilon|$, for $j=1,\ldots,d$; else, 
			set $\theta''_j=\theta^{(r+1)}_j-\tilde a_{\theta,j}|\varepsilon|$, for $j=1,\ldots,d$. 
		\item[(iii)] Letting $\bta''=(\theta''_1,\ldots,\theta''_d)$, evaluate 
                        \begin{equation*}
				\alpha_3=\min\left\{1,\frac{\pi(\bta''|\cdots)}{\pi(\bta^{(r+1)}|\cdots)}\right\}.
                        \end{equation*}
		\item[(iv)] Set $\tilde\bta^{(r+1)}=\bta''$ with probability $\alpha_3$, 
			else set $\tilde\bta^{(r+1)}=\bta^{(r+1)}$. 
	\end{enumerate}
	\end{enumerate}

	\begin{enumerate}
        \item[(b)] If $U> \tilde q$, then
	\begin{enumerate}
		\item[(i)] Generate $\varepsilon\sim U(-1,1)$ and $\tilde U\sim U(0,1)$. If $\tilde U<1/2$, set
			$\theta''_j=\theta^{(t+1)}_j\varepsilon$ for $j=1,\ldots,d$ and $|J|=|\varepsilon|^d$, else set 
			$\theta''_j=\theta^{(t+1)}_j/\varepsilon$ for $j=1,\ldots,d$ and 
			$|J|=|\varepsilon|^{-d}$.
		\item[(ii)] Evaluate 
               \begin{equation*}
		       \alpha_4=\min\left\{1,\frac{\pi(\bta''|\cdots)}{\pi(\bta^{(r+1)}|\cdots)}\times|J|\right\}.
               \end{equation*}
       \item[(iii)] Set $\tilde\bta^{(r+1)}=\bta''$ with probability $\alpha_4$, 
	       else set $\tilde\bta^{(r+1)}=\bta^{(r+1)}$. 
	\end{enumerate}
	\end{enumerate}
\item Broadcast $\tilde\btheta^{(r+1)}$ from processor $0$ to all the processors.
\item Update the random effects $\bphi=\{\phi(\bs_i,t_k):~i=1,\ldots,n;~j=1,\ldots,m\}$ using Gibbs sampling in parallel processors. 
\item Send the updates of the random effects to processor $0$, denote the update by $\bphi^{(r+1)}$ and from processor $0$, 
broadcast $\bphi^{(r+1)}$ to all the processors.
\item Update $\bzeta$ in processor $0$ by Gibbs sampling. In this exercise, parallelize the computations of the relevant sums over the available processors, and aggregate
the final sum in processor $0$, where Gibbs sampling is then performed.
Denote the updated $\bzeta$ vector by $\bzeta^{(r+1)}$.
\item Broadcast $\bzeta^{(r+1)}$ from processor $0$ to all the processors. 
\item For the purpose of prediction of $y(\tilde\bs,\tilde t)$ at any location $\tilde \bs$ and time point $\tilde t$, substitute the MCMC-simulated realizations
	in the distribution associated with (\ref{eq:re_model_marginalized1}), and generate $\tilde y(\tilde\bs,\tilde t)$ from the resultant distribution, which would yield
	the posterior predictive distribution of $\tilde y(\tilde\bs,\tilde t)$. For multiple locations and time points, for each MCMC realization, parallelize
	the prediction exercise over the required locations and time points.
\end{enumerate}
\item End for
\item Instruct processor $0$ to store 
$$\left\{\left\{\left\{\left(\bU^{(r)}_k,\bbeta^{(r)}_k,J^{(r)}_{t_k}\right):k=1,\ldots,m\right\},\btheta^{(r)},\bzeta^{(r)}\right\}:r=0,1,2,\ldots\right\}$$ 
for Bayesian inference.
\item 
\end{itemize}
\botline \rmfamily
\end{algo}
In our applications of Algorithm \ref{algo:ttmcmc}, we set $\tilde p=\tilde q=1/2$. We also set the positive scaling constants associated with the additive transformations
of the parameters to $0.05$ and $c$ to $0.01$. Algorithm \ref{algo:ttmcmc}, along with these choices of the tuning parameters, exhibited adequate mixing properties.

\begin{figure}
	\centering
	\subfigure []{ \label{fig:J5}
	\includegraphics[width=6.5cm,height=5.5cm]{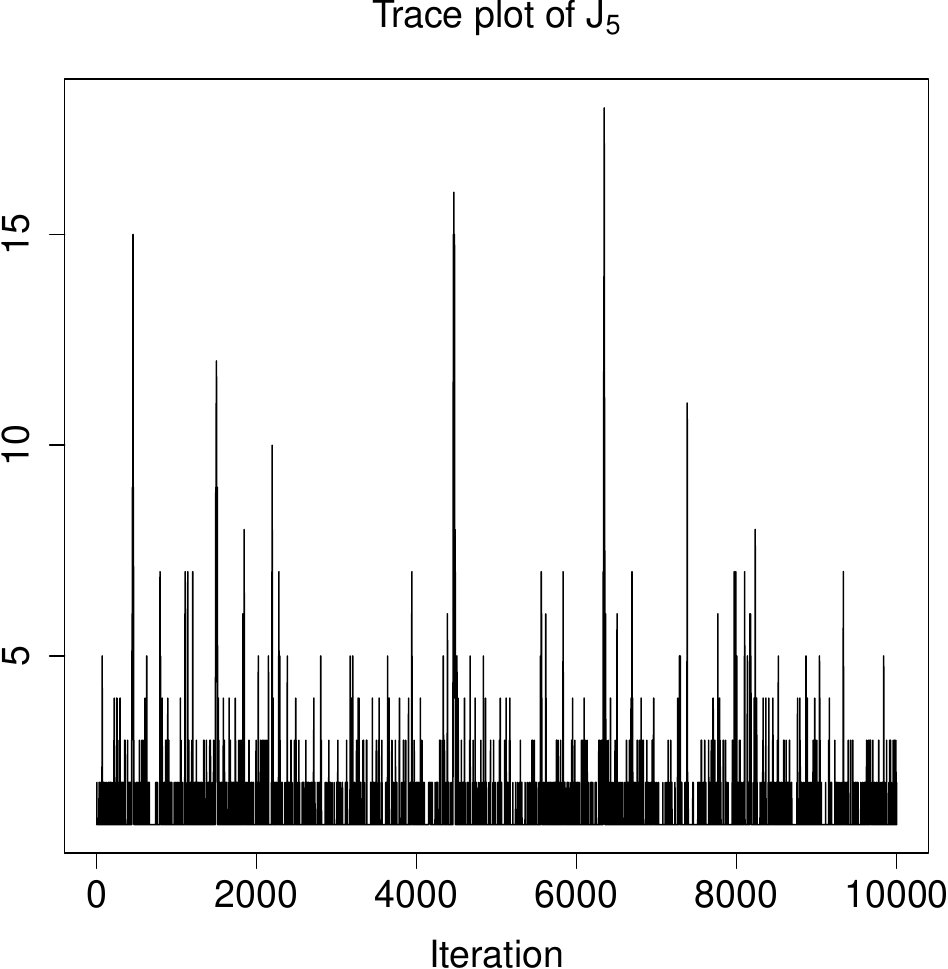}}
	\hspace{2mm}
	\subfigure []{ \label{fig:J15}
	\includegraphics[width=6.5cm,height=5.5cm]{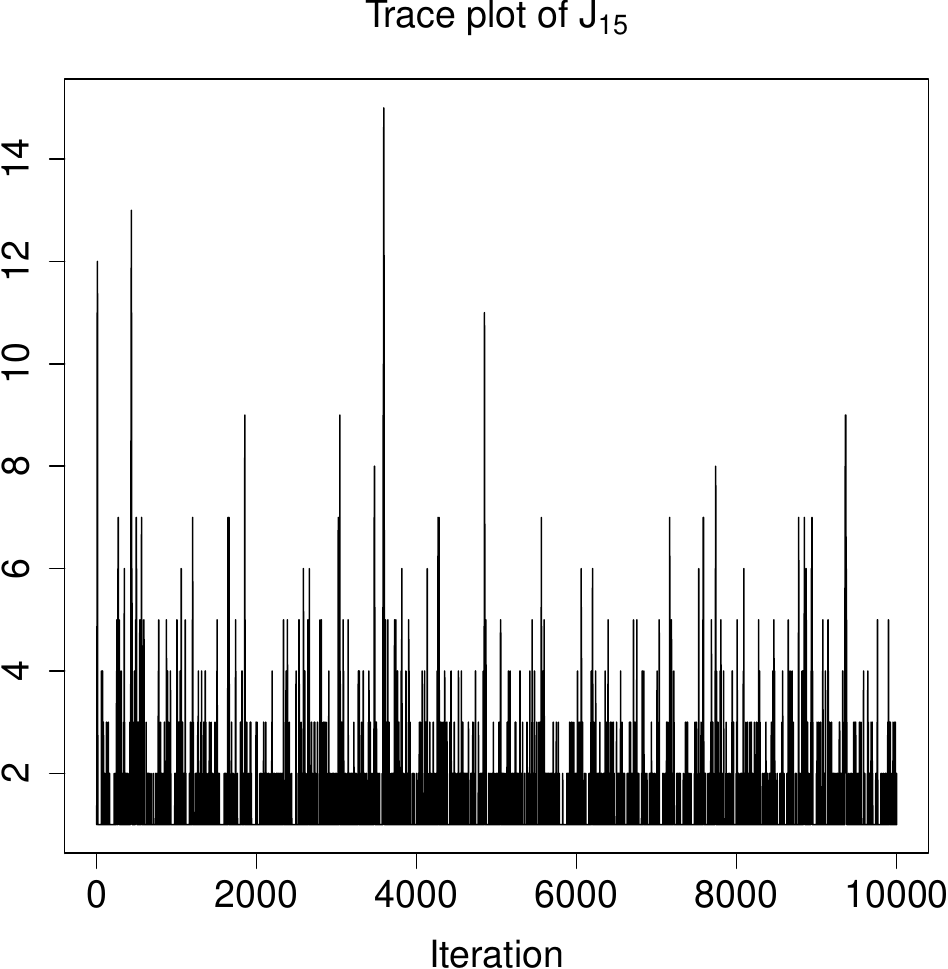}}\\
	\vspace{2mm}
	\subfigure []{ \label{fig:J25}
	\includegraphics[width=6.5cm,height=5.5cm]{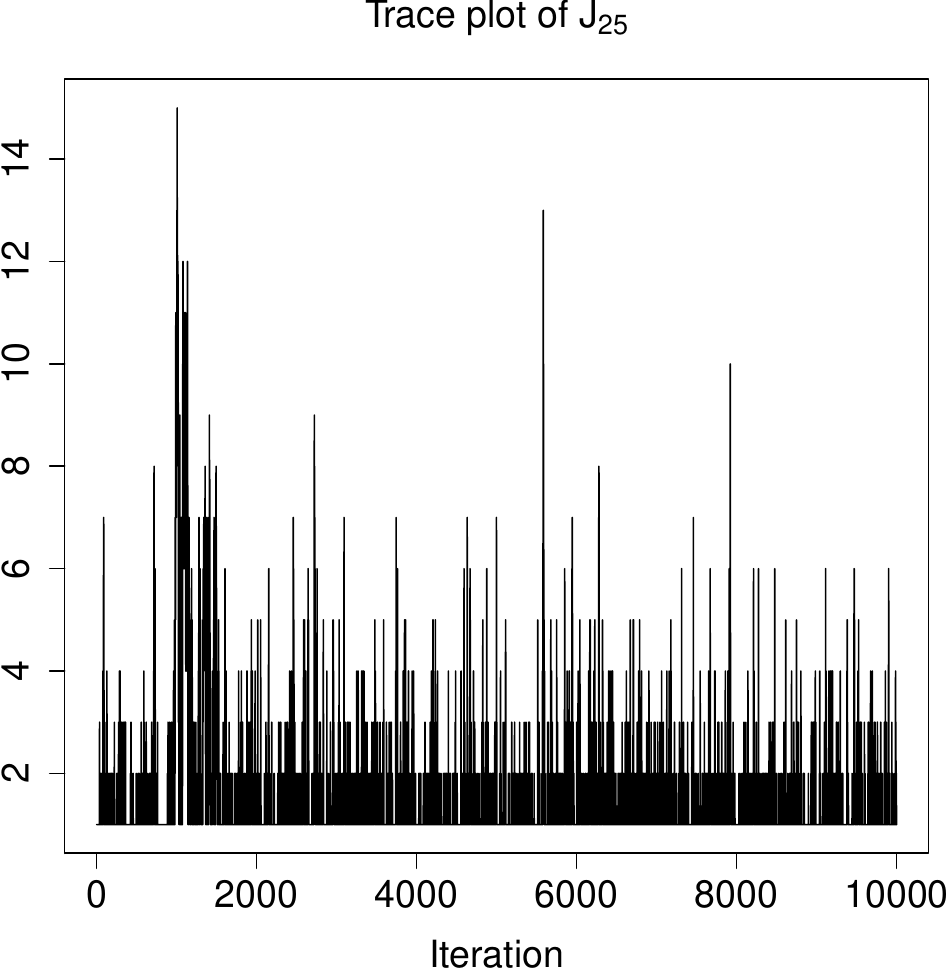}}
	\hspace{2mm}
	\subfigure []{ \label{fig:J35}
	\includegraphics[width=6.5cm,height=5.5cm]{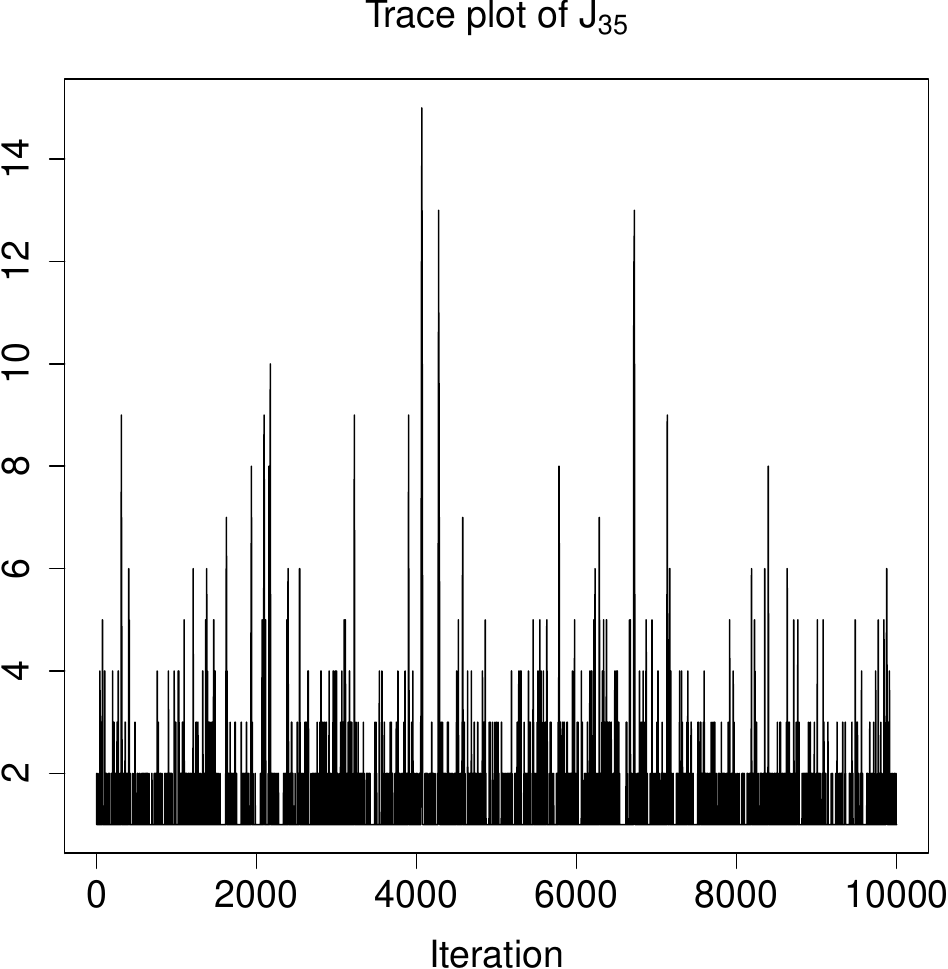}}\\
	\vspace{2mm}
	\subfigure []{ \label{fig:J40}
	\includegraphics[width=6.5cm,height=5.5cm]{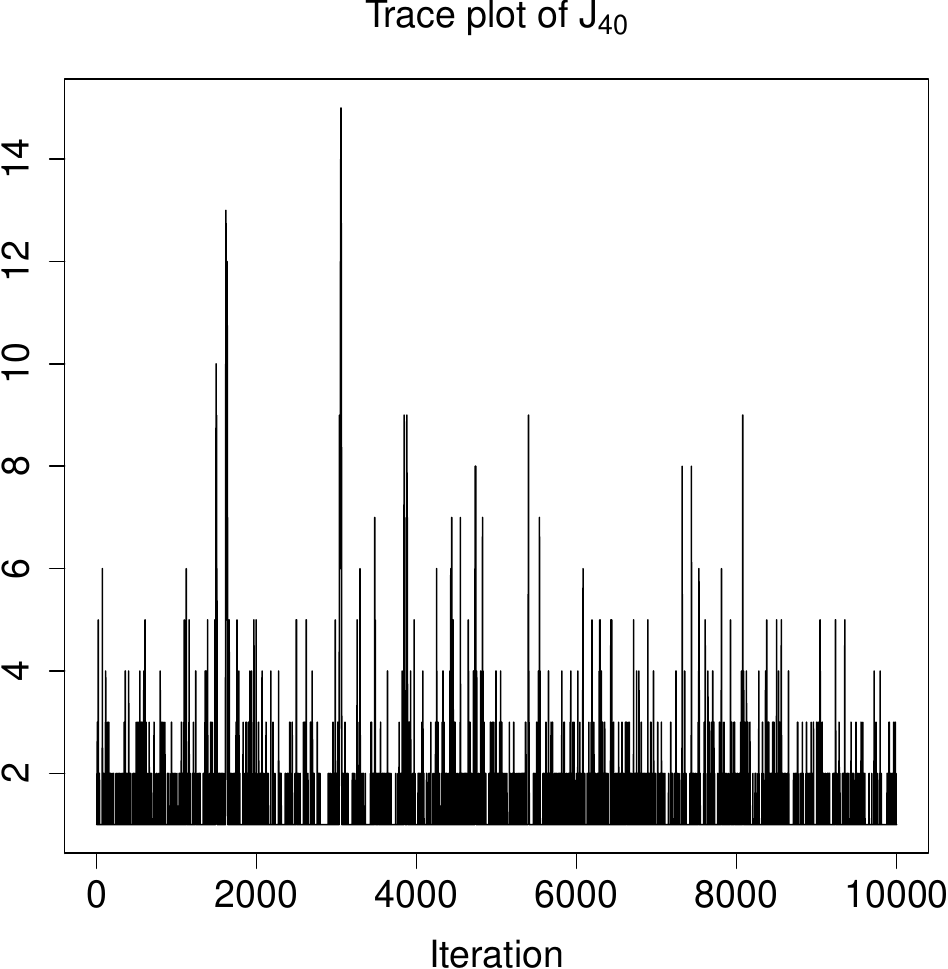}}
	\hspace{2mm}
	\subfigure []{ \label{fig:J50}
	\includegraphics[width=6.5cm,height=5.5cm]{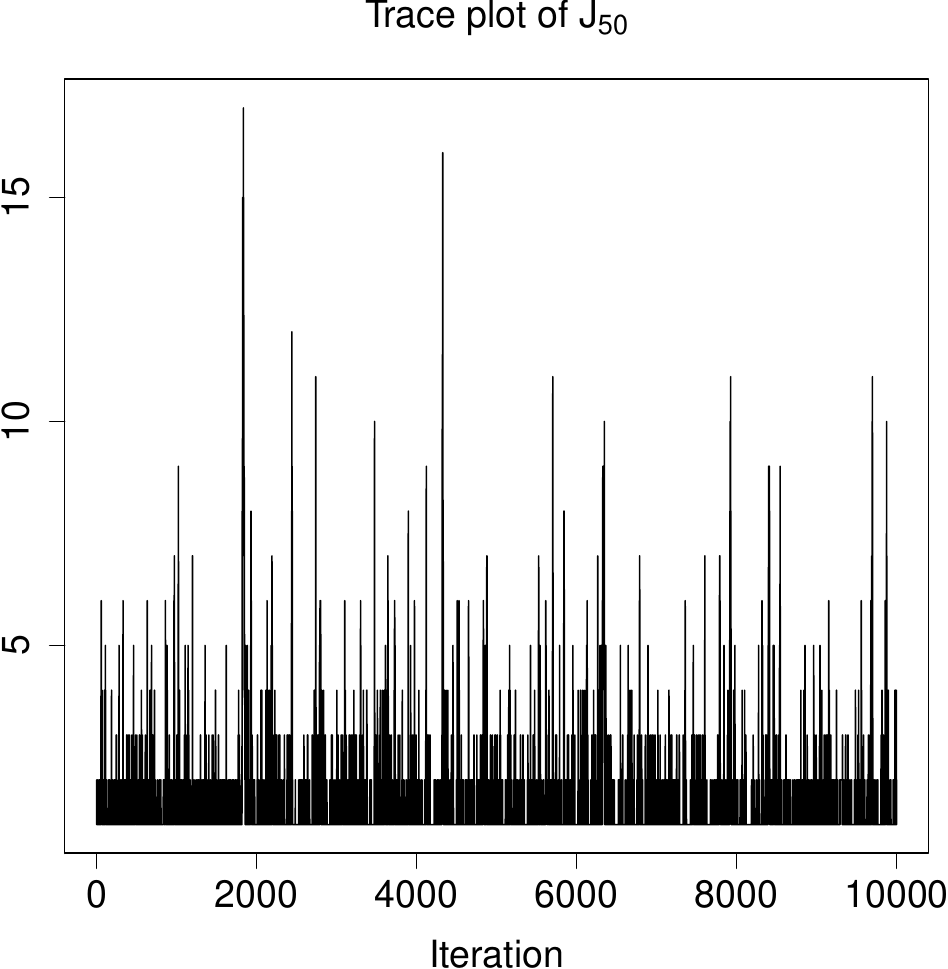}}\\
	\caption{Simulation study: trace plots of $J(k)$ for different $k$}
	\label{fig:J_plots_simstudy1}
\end{figure}

\begin{figure}
	\centering
	\subfigure []{ \label{fig:lambda}
	\includegraphics[width=4.8cm,height=4.8cm]{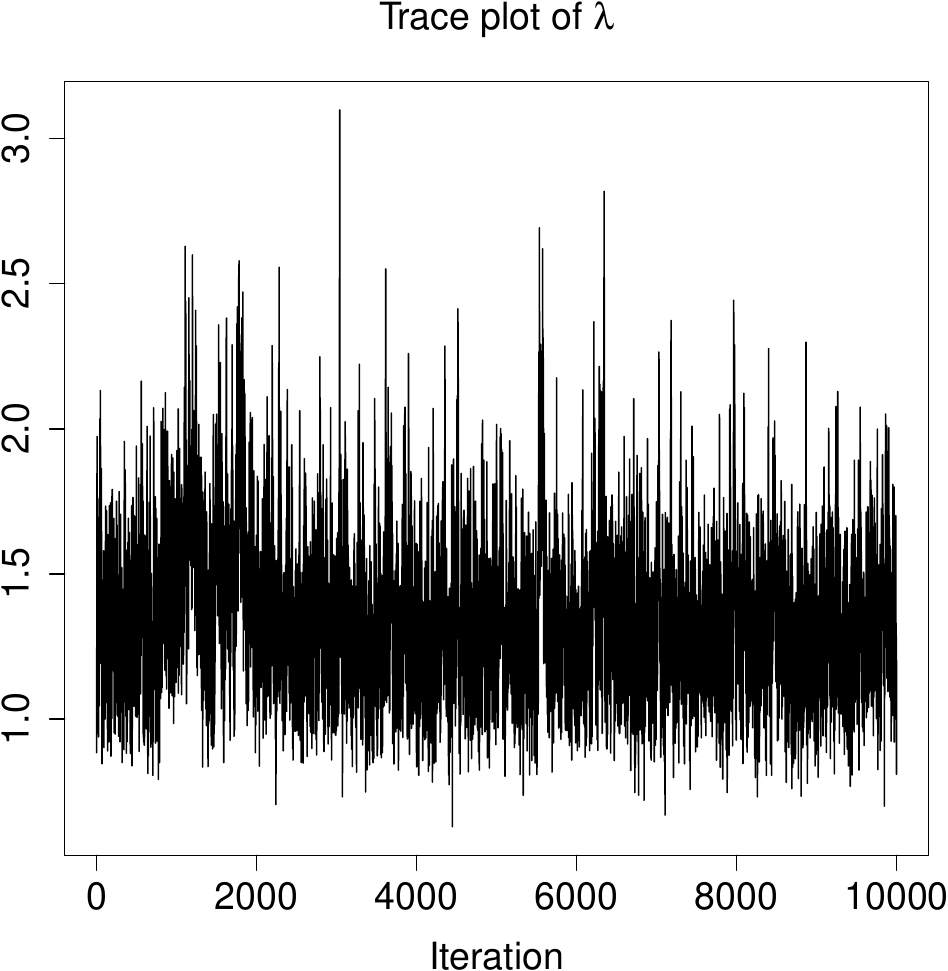}}
	\hspace{2mm}
	\subfigure []{ \label{fig:rho_beta}
	\includegraphics[width=4.8cm,height=4.8cm]{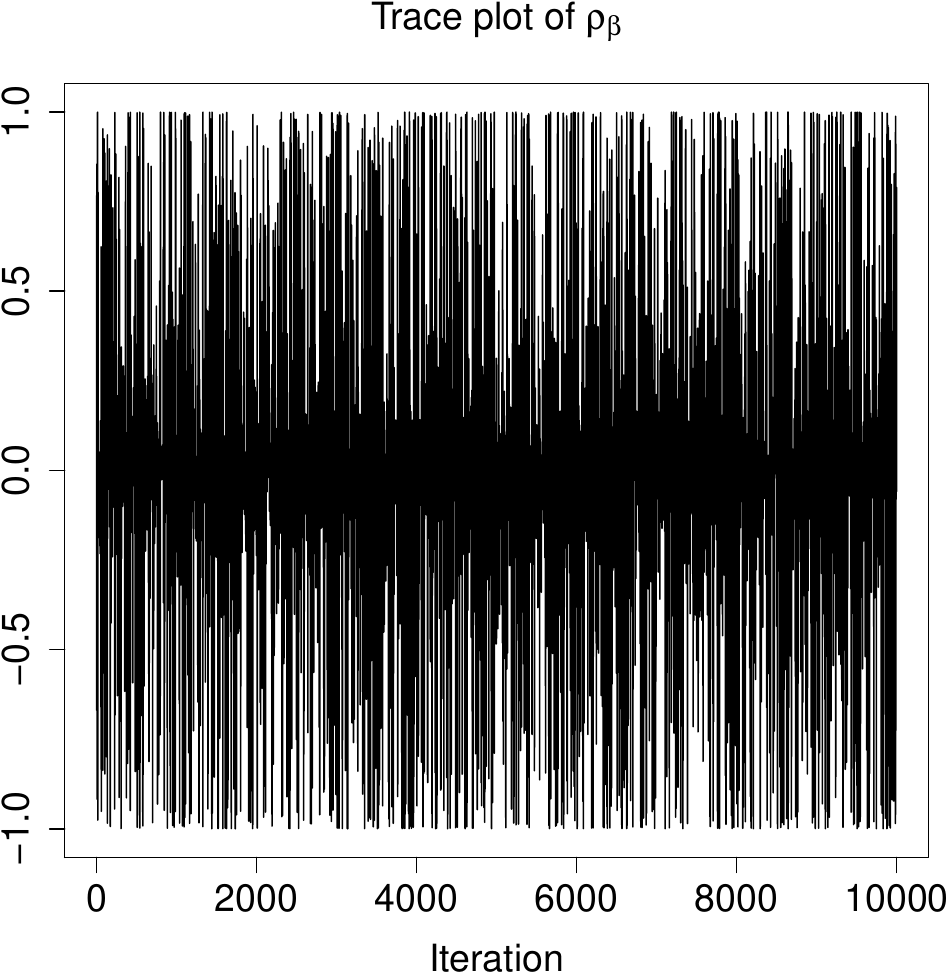}}
	\hspace{2mm}
	\subfigure []{ \label{fig:rho_mu1}
	\includegraphics[width=4.8cm,height=4.8cm]{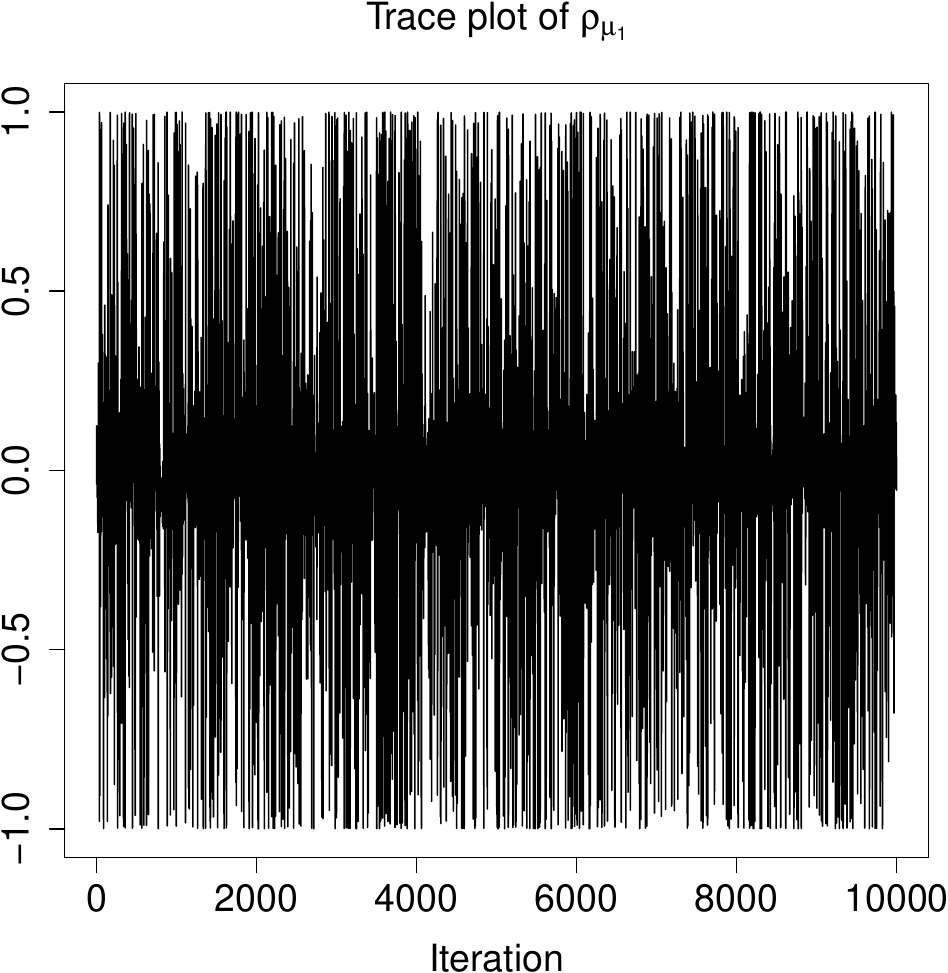}}\\
	\vspace{2mm}
	\subfigure []{ \label{fig:xi}
	\includegraphics[width=4.8cm,height=4.8cm]{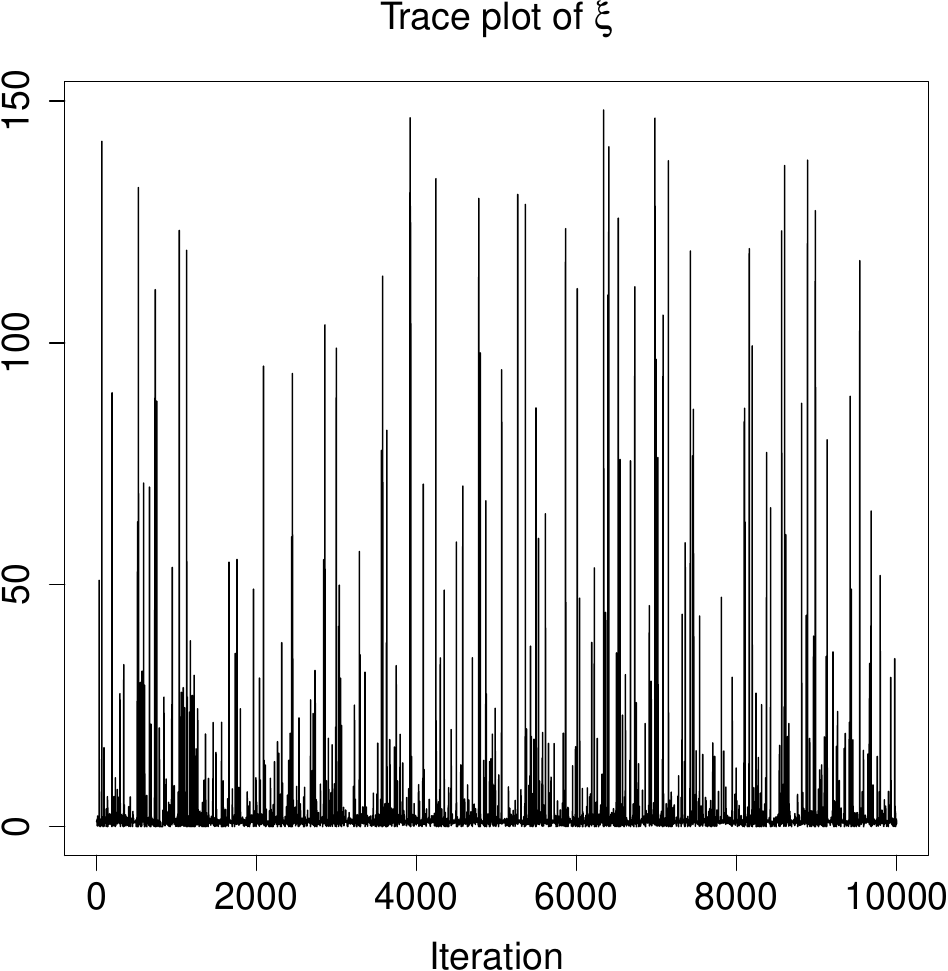}}
	\hspace{2mm}
	\subfigure []{ \label{fig:tau}
	\includegraphics[width=4.8cm,height=4.8cm]{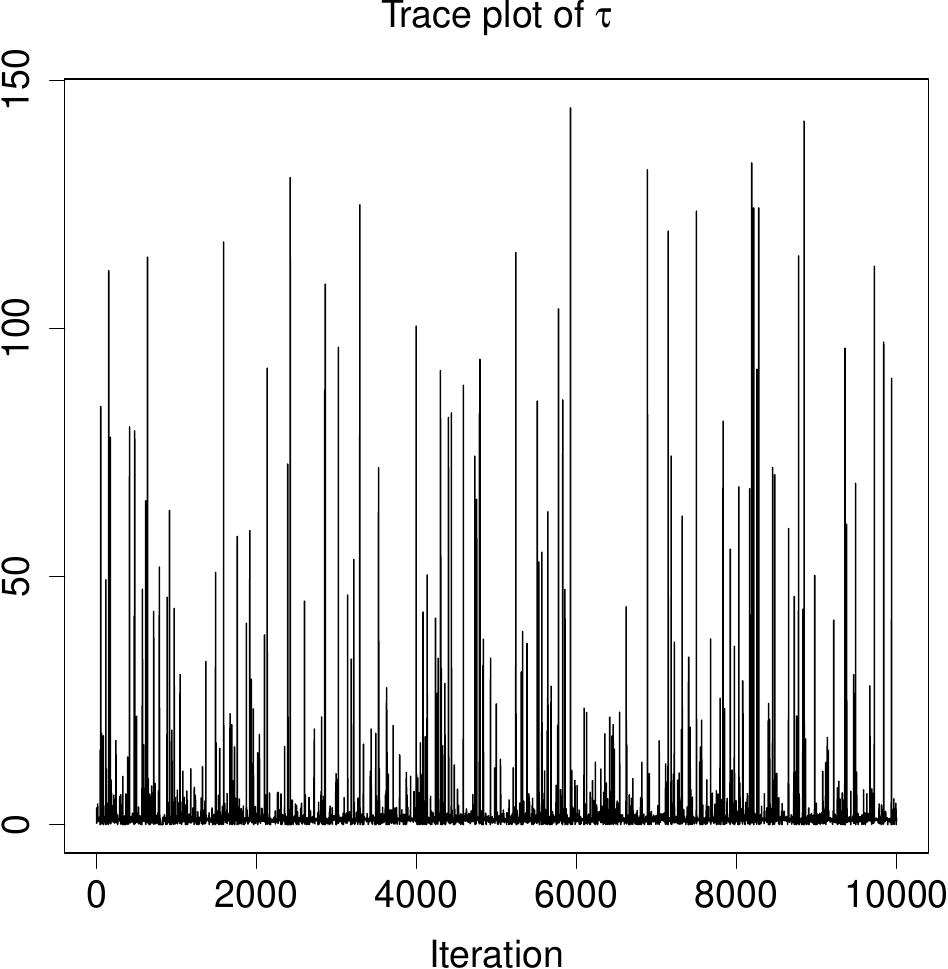}}
	\hspace{2mm}
	\subfigure []{ \label{fig:X1}
	\includegraphics[width=4.8cm,height=4.8cm]{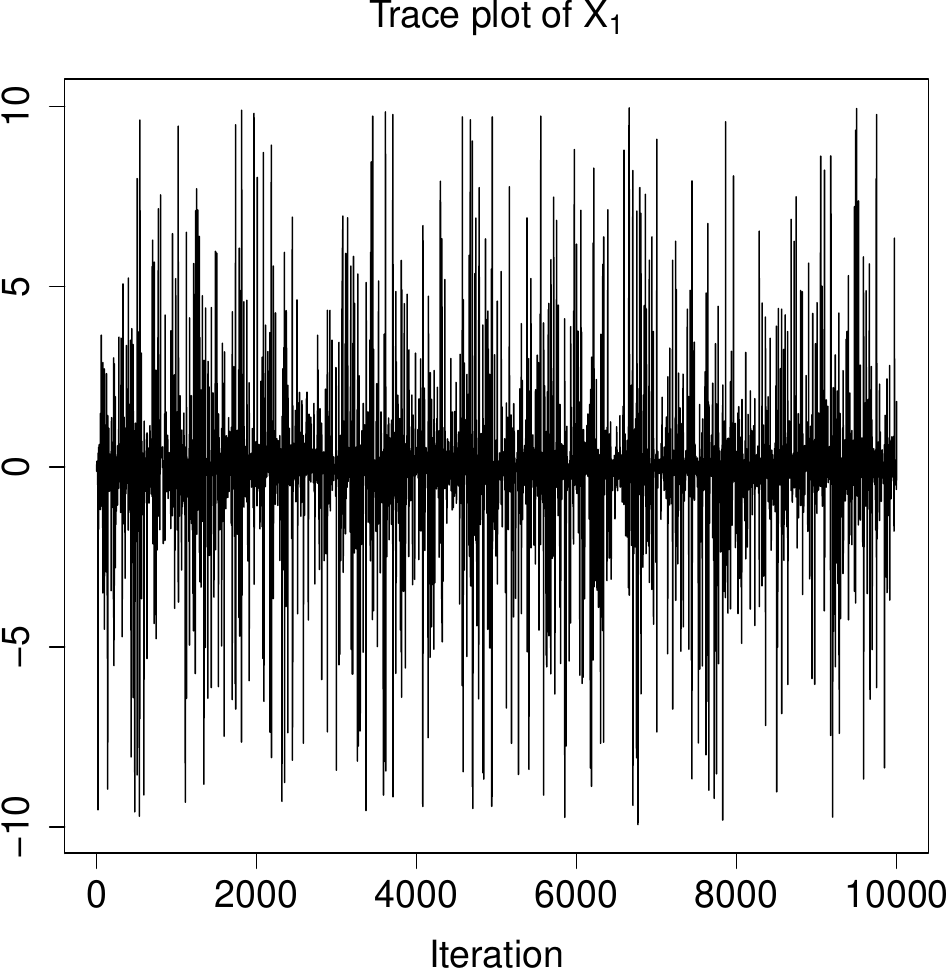}}\\
	\vspace{2mm}
	\subfigure []{ \label{fig:C1}
	\includegraphics[width=4.8cm,height=4.8cm]{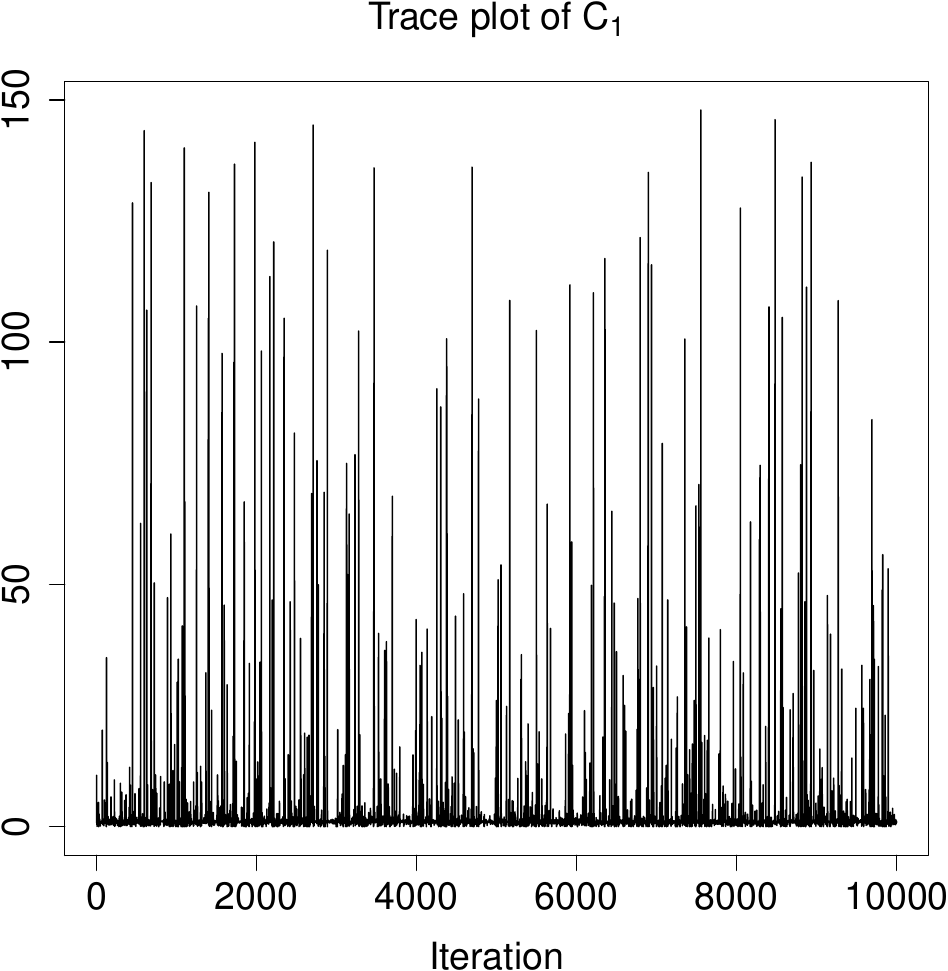}}
	\hspace{2mm}
	\subfigure []{ \label{fig:CSTAR1}
	\includegraphics[width=4.8cm,height=4.8cm]{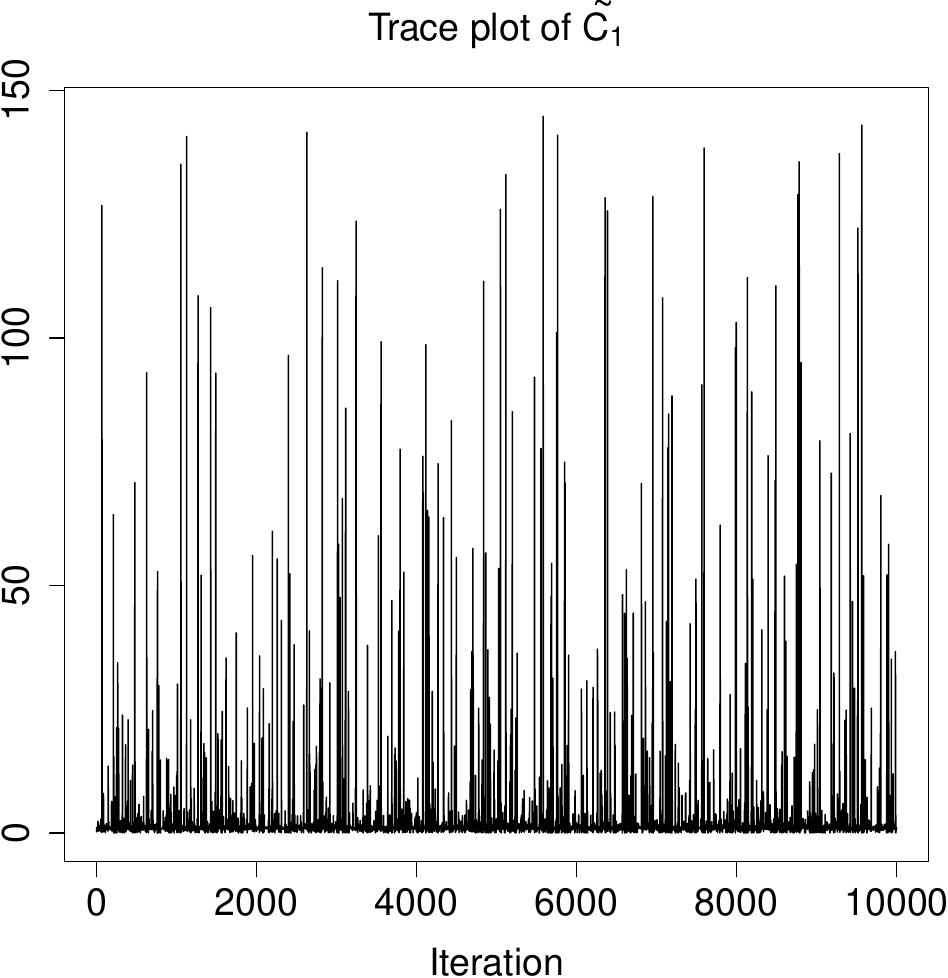}}
	\hspace{2mm}
	\subfigure []{ \label{fig:sigmasq_e}
	\includegraphics[width=4.8cm,height=4.8cm]{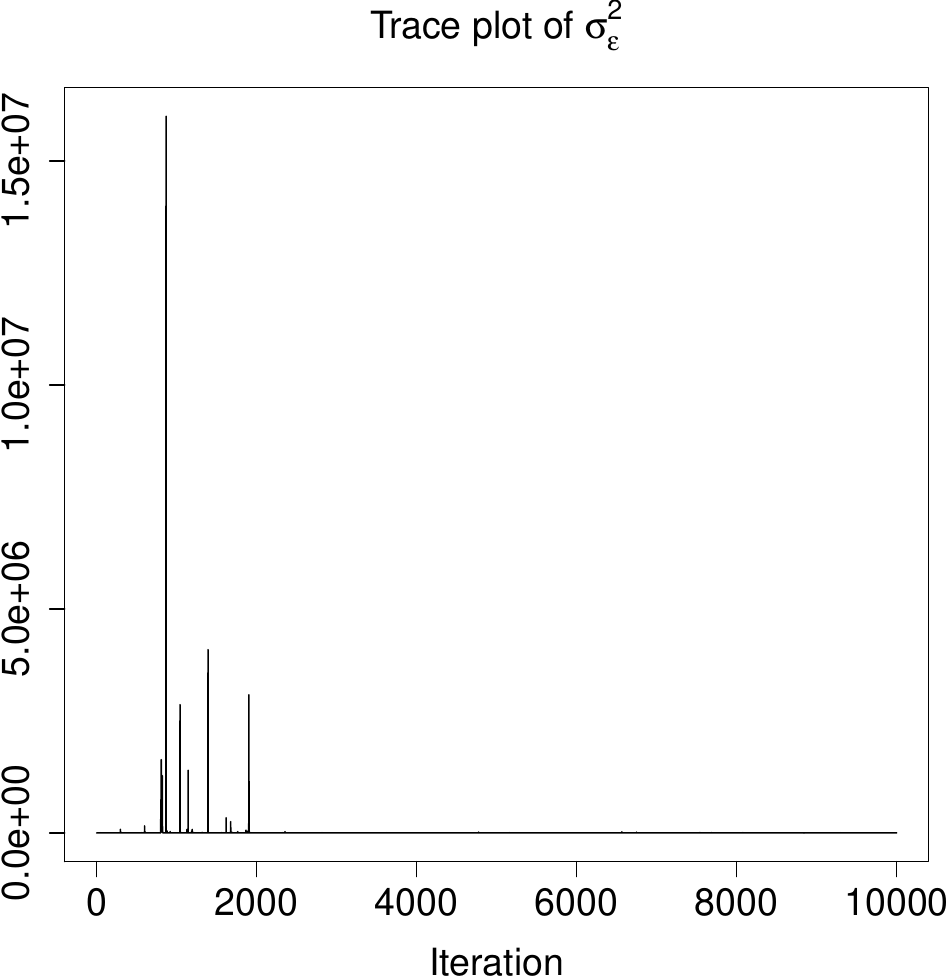}}\\
	\caption{Simulation study: trace plots of some parameters.}
	\label{fig:parameter_plots_simstudy1}
\end{figure}

\begin{figure}
	\centering
	\subfigure [Spatial index $1$.]{ \label{fig:spatial1_0}
	\includegraphics[width=7.5cm,height=5.5cm]{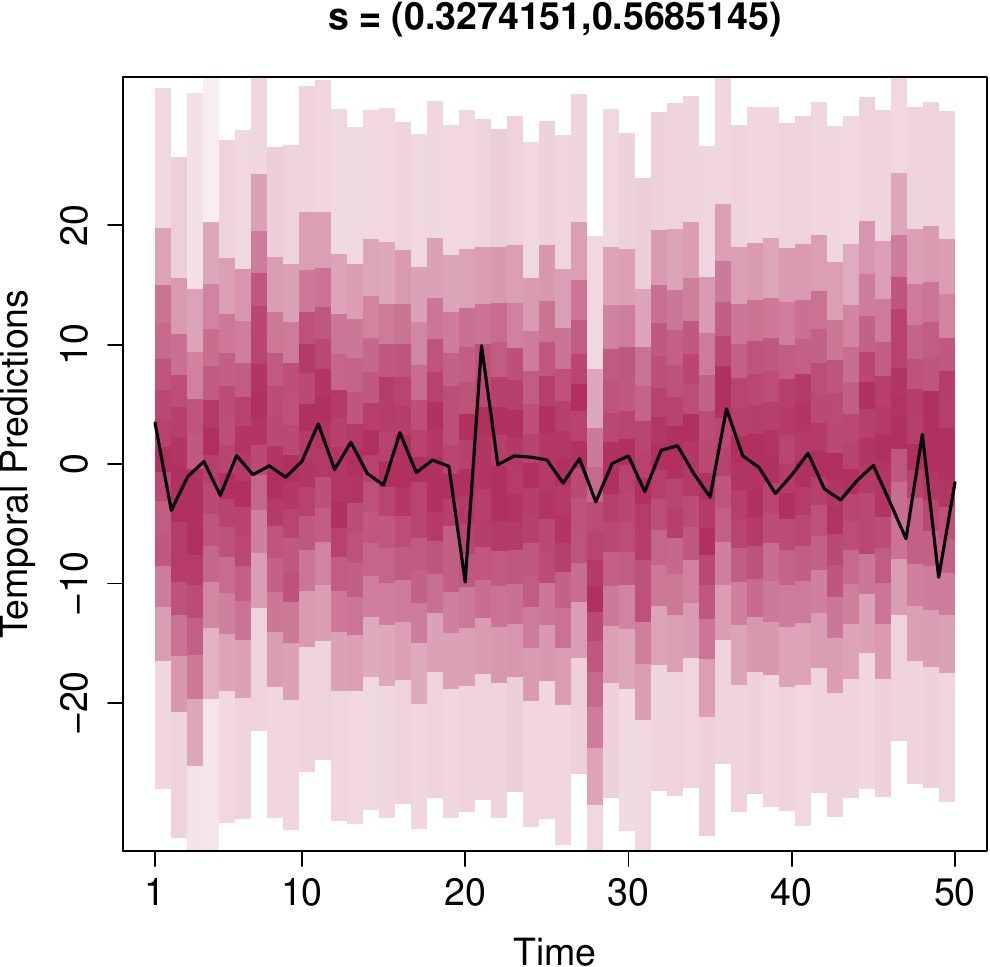}}
	\hspace{2mm}
	\subfigure [Spatial index $5$.]{ \label{fig:spatial5_0}
	\includegraphics[width=7.5cm,height=5.5cm]{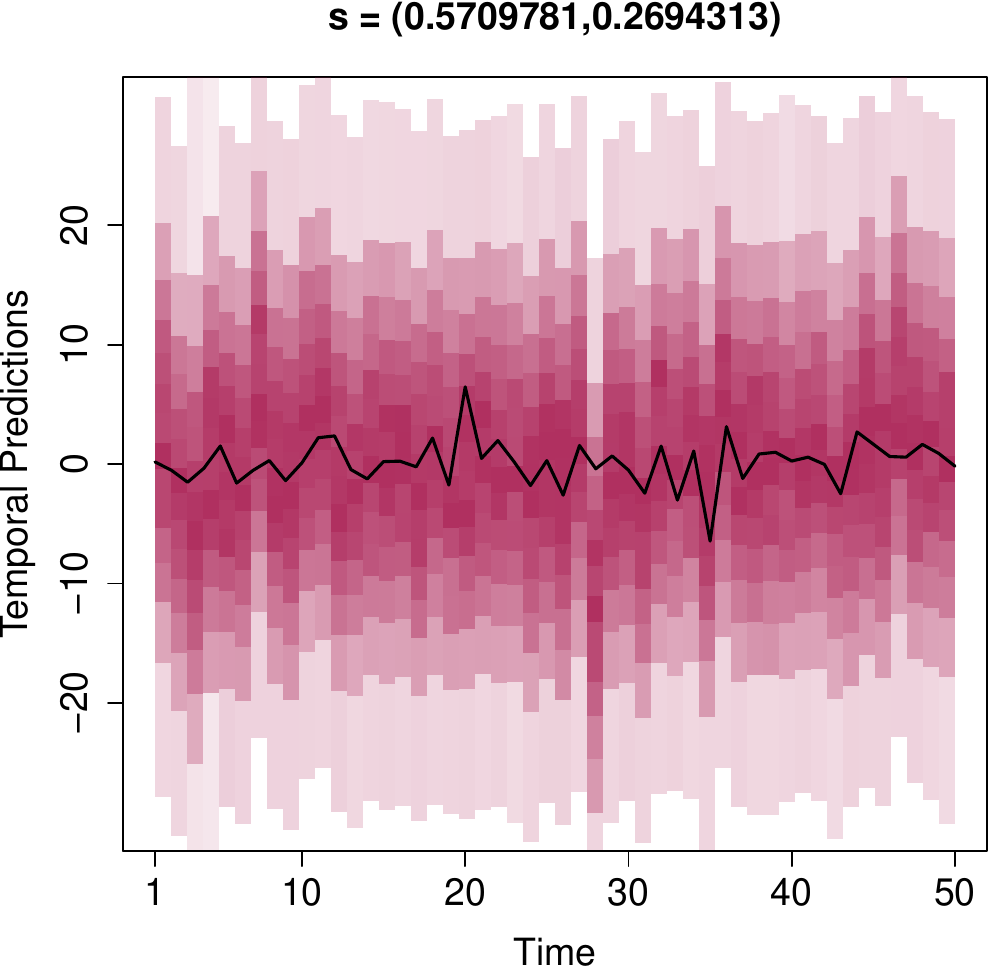}}\\
	\vspace{2mm}
	\subfigure [Spatial index $10$.]{ \label{fig:spatial10_0}
	\includegraphics[width=7.5cm,height=5.5cm]{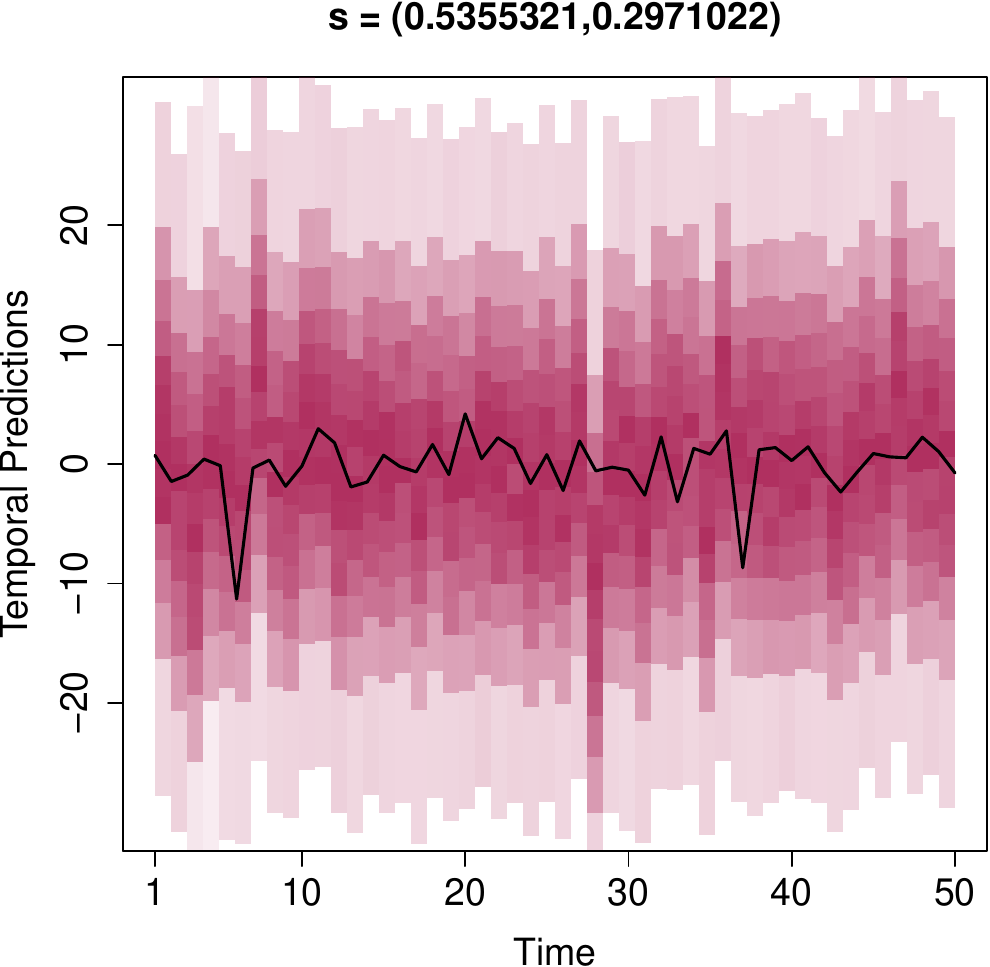}}
	\hspace{2mm}
	\subfigure [Spatial index $15$.]{ \label{fig:spatial15_0}
	\includegraphics[width=7.5cm,height=5.5cm]{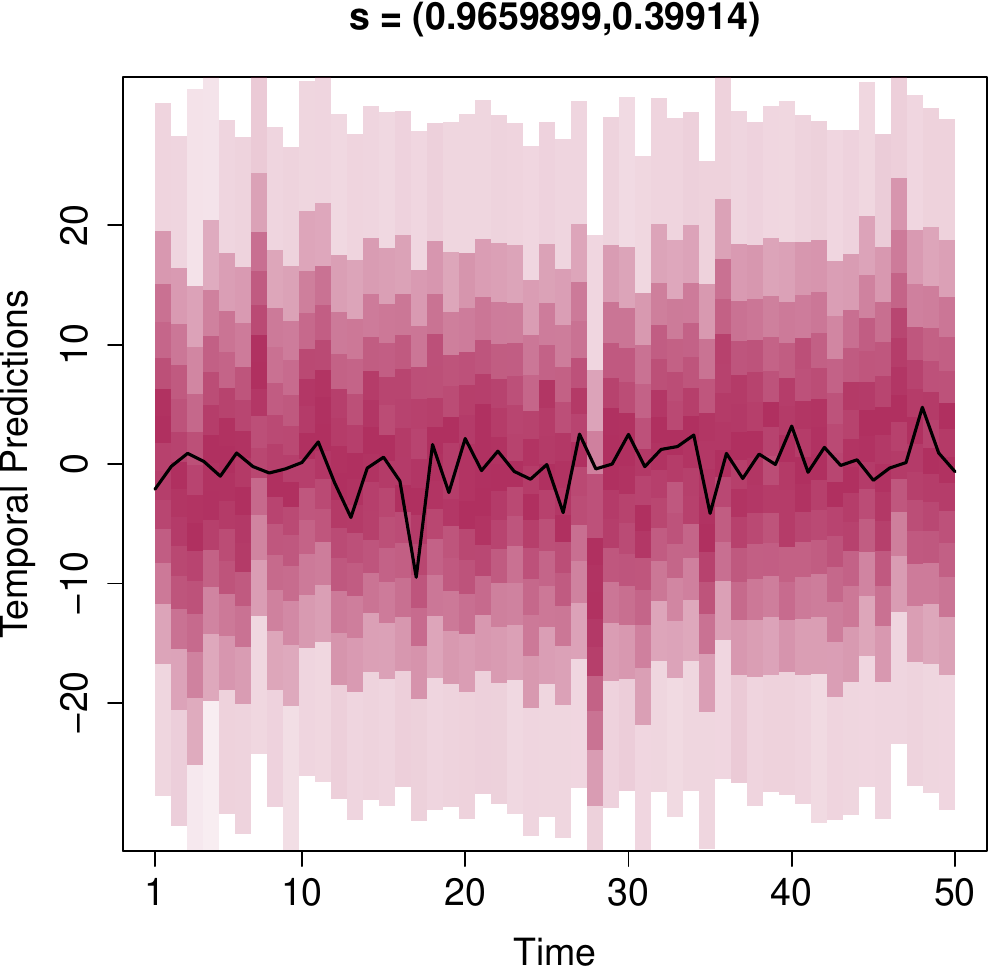}}\\
	\vspace{2mm}
	\subfigure [Spatial index $19$.]{ \label{fig:spatial19_0}
	\includegraphics[width=7.5cm,height=5.5cm]{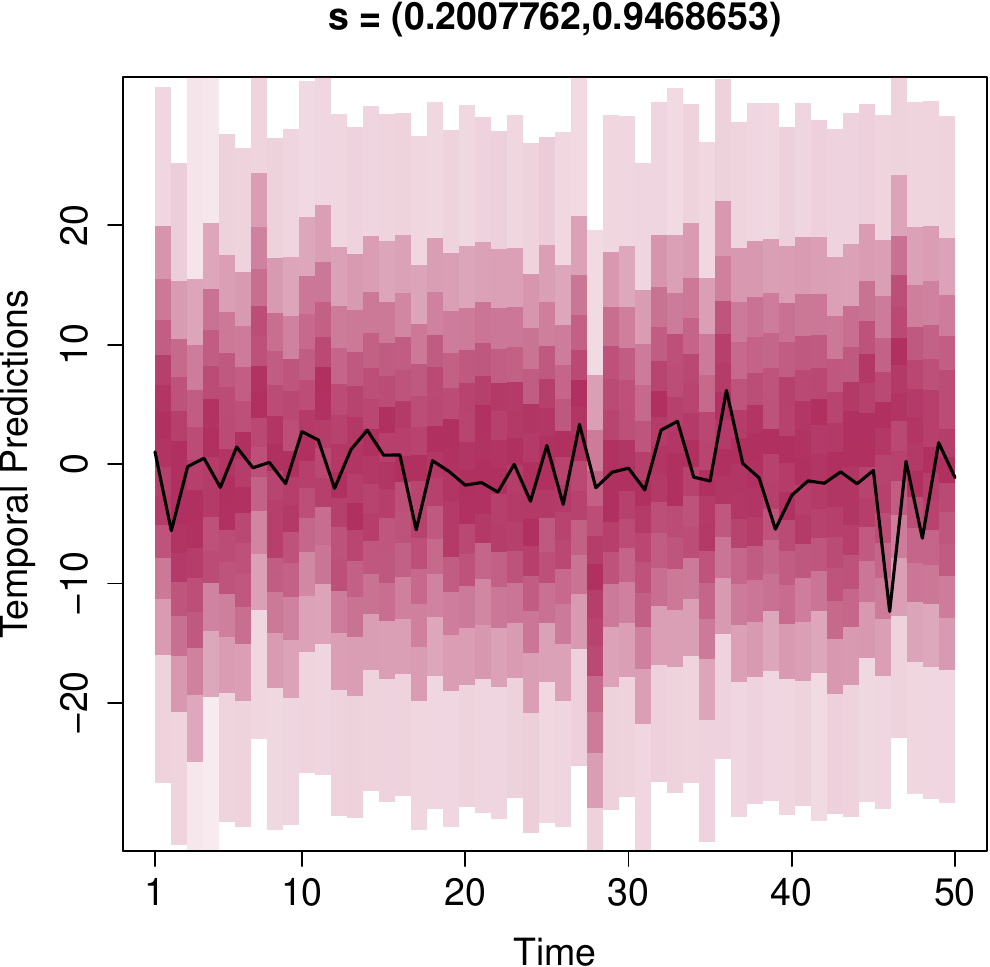}}
	\hspace{2mm}
	\subfigure [Spatial index $20$.]{ \label{fig:spatial20_0}
	\includegraphics[width=7.5cm,height=5.5cm]{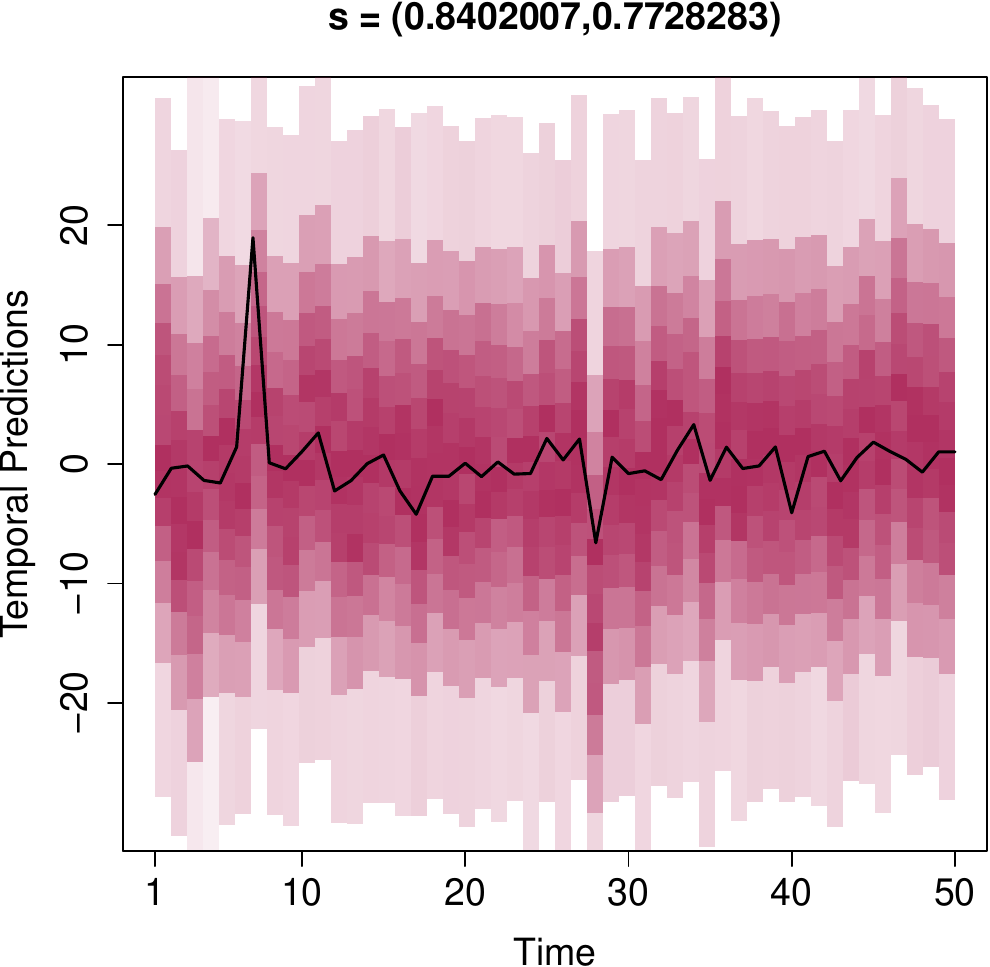}}\\
	\caption{Simulation study: posterior temporal predictions assuming $\phi(\bs_i,t_k)=\phi_0(\bs_i,t_k)$ for $i=1,\ldots,n$ and $k=1,\ldots,m$.}
	\label{fig:temporal_plots_simstudy2}
\end{figure}

\begin{figure}
	\centering
	\subfigure [Temporal index $5$.]{ \label{fig:temporal5_0}
	\includegraphics[width=7.5cm,height=5.5cm]{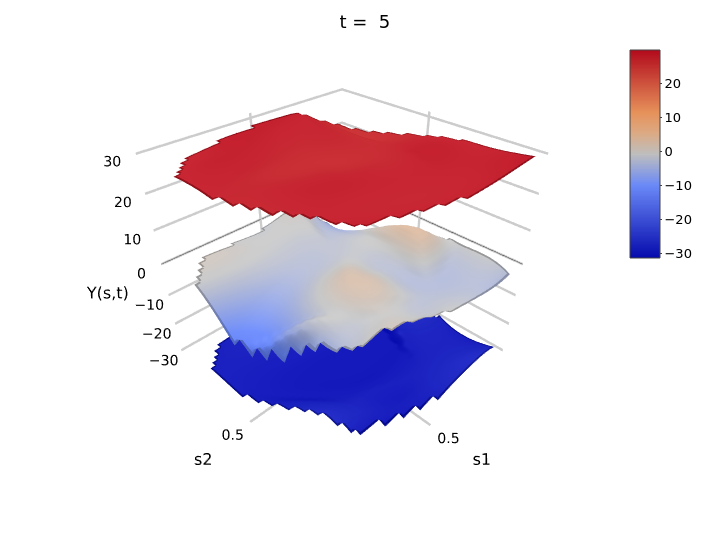}}
	\hspace{2mm}
	\subfigure [Temporal index $10$.]{ \label{fig:temporal10_0}
	\includegraphics[width=7.5cm,height=5.5cm]{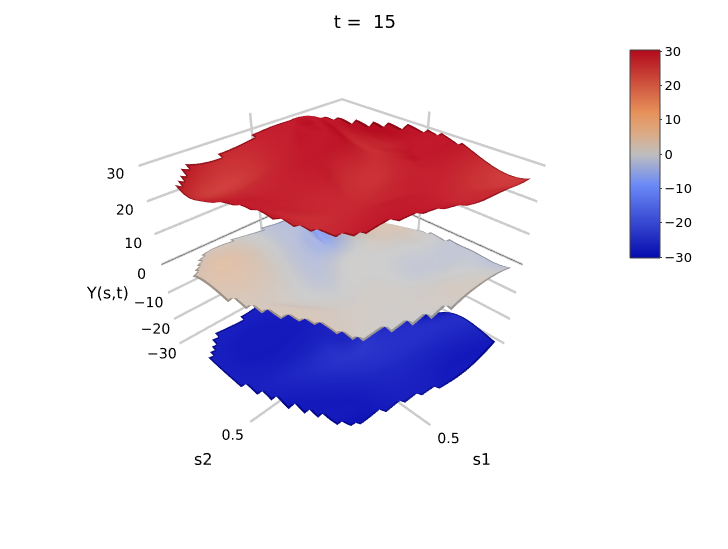}}\\
	\vspace{2mm}
	\subfigure [Temporal index $20$.]{ \label{fig:temporal20_0}
	\includegraphics[width=7.5cm,height=5.5cm]{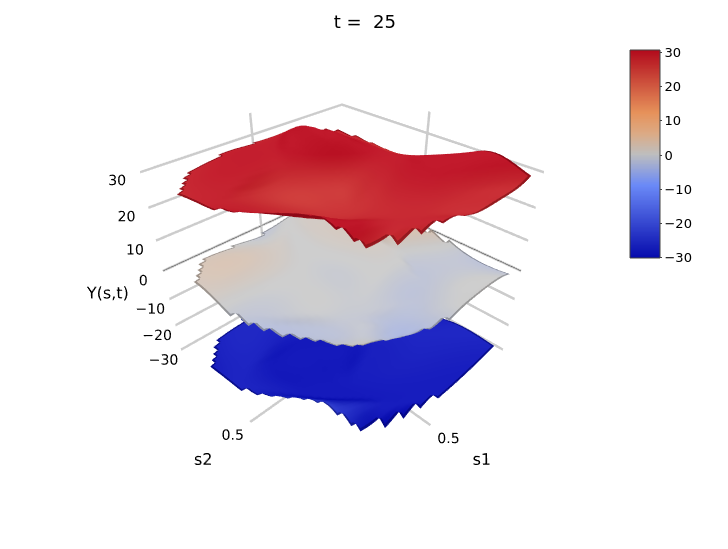}}
	\hspace{2mm}
	\subfigure [Temporal index $30$.]{ \label{fig:temporal30_0}
	\includegraphics[width=7.5cm,height=5.5cm]{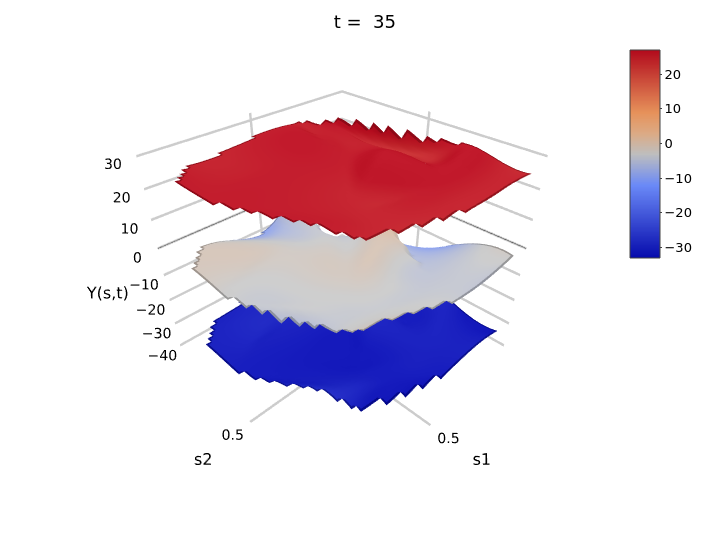}}\\
	\vspace{2mm}
	\subfigure [Temporal index $40$.]{ \label{fig:temporal40_0}
	\includegraphics[width=7.5cm,height=5.5cm]{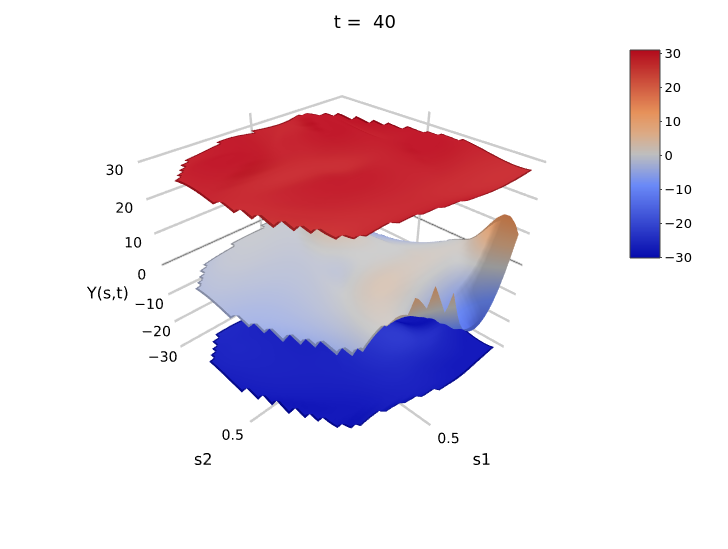}}
	\hspace{2mm}
	\subfigure [Temporal index $50$.]{ \label{fig:temporal50_0}
	\includegraphics[width=7.5cm,height=5.5cm]{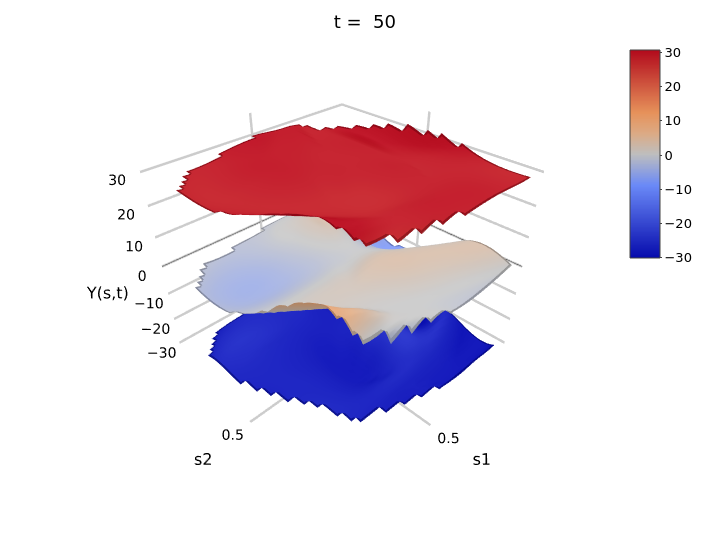}}\\
	\caption{Simulation study: posterior spatial predictions at various time points $t$ assuming $\phi(\bs_i,t_k)=\phi_0(\bs_i,t_k)$ for $i=1,\ldots,n$ and $k=1,\ldots,m$.}
	\label{fig:spatial_plots_simstudy2}
\end{figure}

\begin{figure}
	\centering
	\subfigure [Temporal index $5$.]{ \label{fig:temp5_0}
	\includegraphics[width=7.5cm,height=5.5cm]{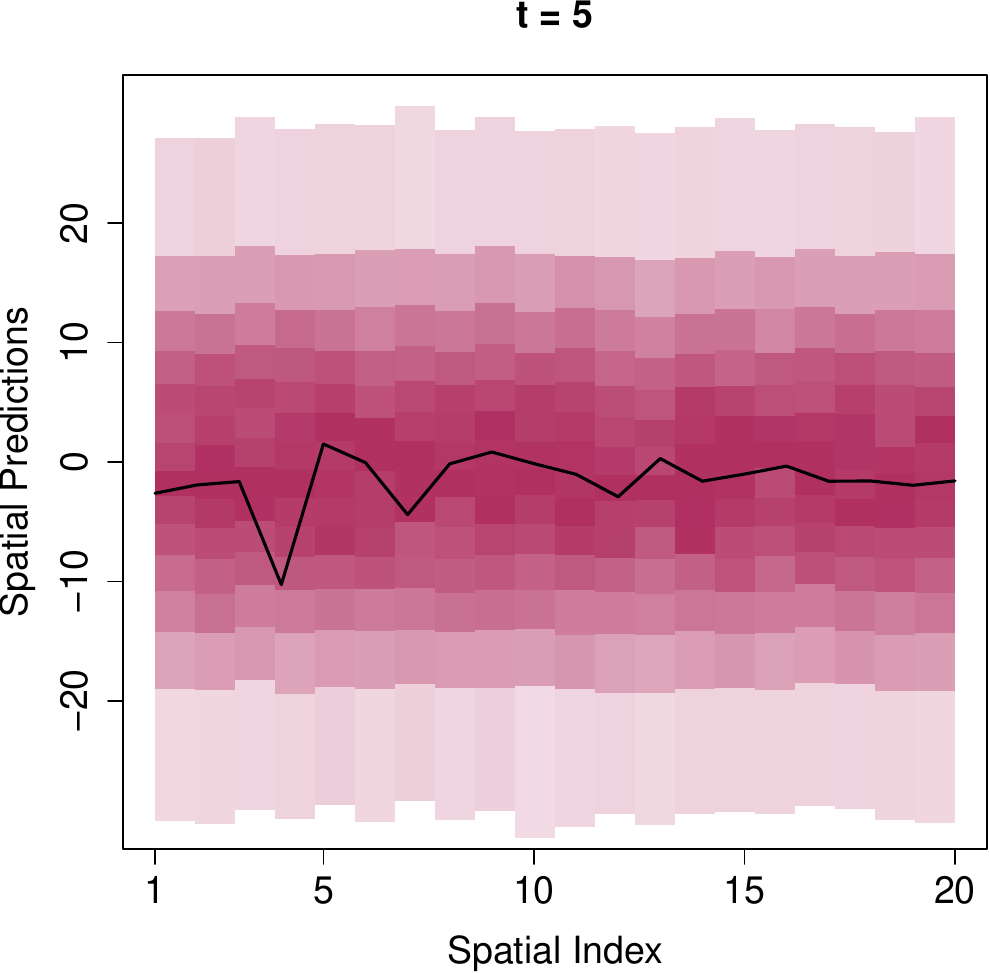}}
	\hspace{2mm}
	\subfigure [Temporal index $15$.]{ \label{fig:temp15_0}
	\includegraphics[width=7.5cm,height=5.5cm]{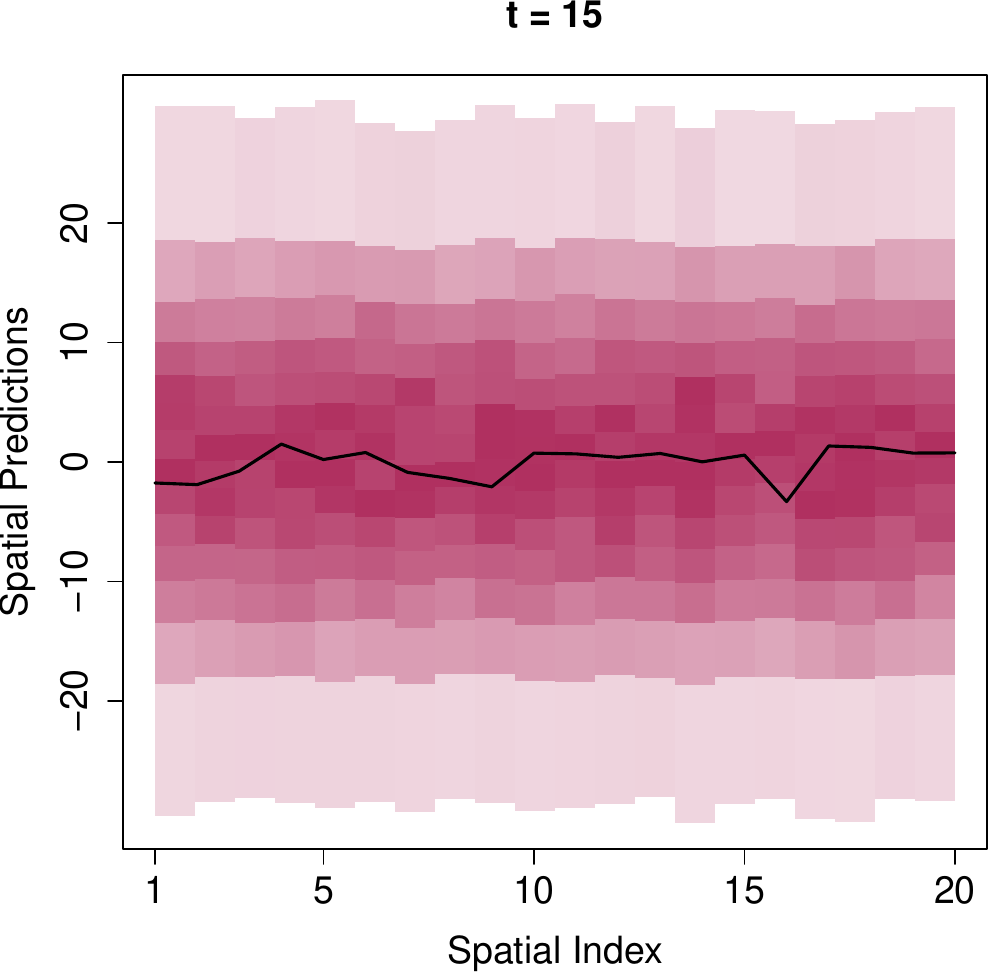}}\\
	\vspace{2mm}
	\subfigure [Temporal index $25$.]{ \label{fig:temp25_0}
	\includegraphics[width=7.5cm,height=5.5cm]{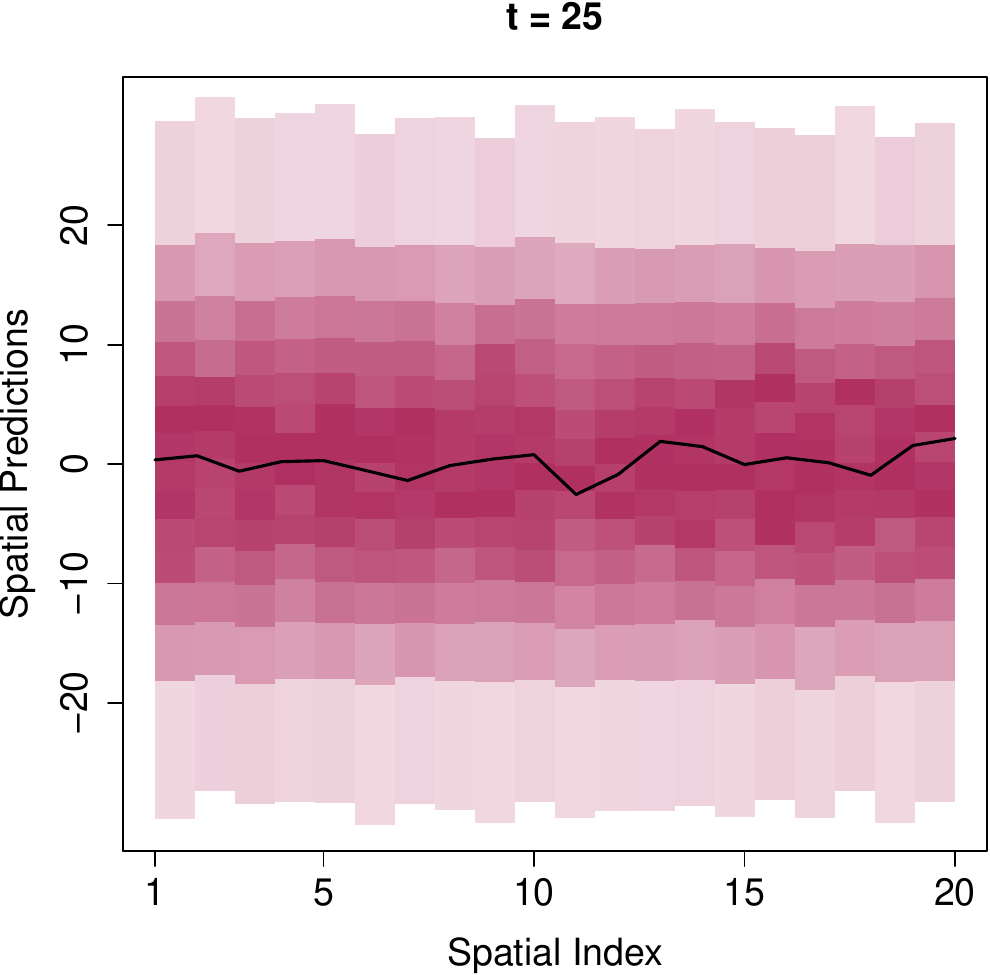}}
	\hspace{2mm}
	\subfigure [Temporal index $35$.]{ \label{fig:temp35_0}
	\includegraphics[width=7.5cm,height=5.5cm]{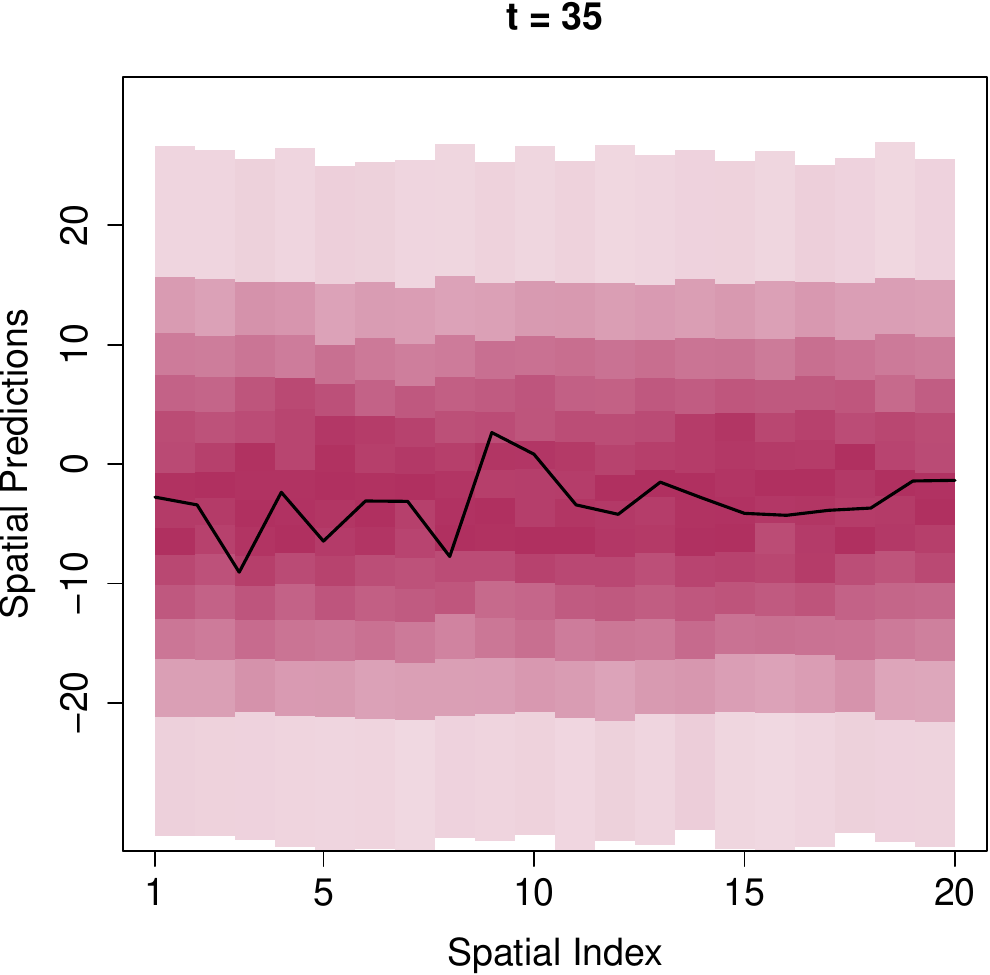}}\\
	\vspace{2mm}
	\subfigure [Temporal index $40$.]{ \label{fig:temp40_0}
	\includegraphics[width=7.5cm,height=5.5cm]{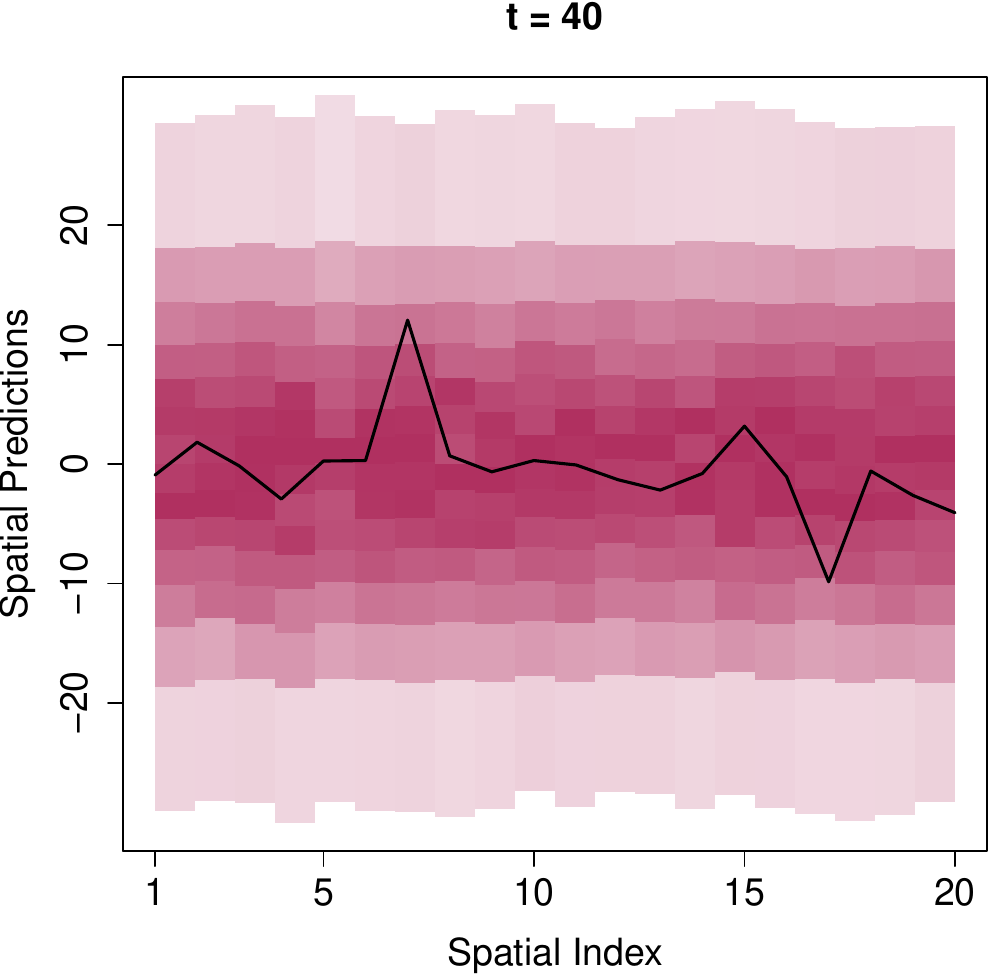}}
	\hspace{2mm}
	\subfigure [Temporal index $50$.]{ \label{fig:temp50_0}
	\includegraphics[width=7.5cm,height=5.5cm]{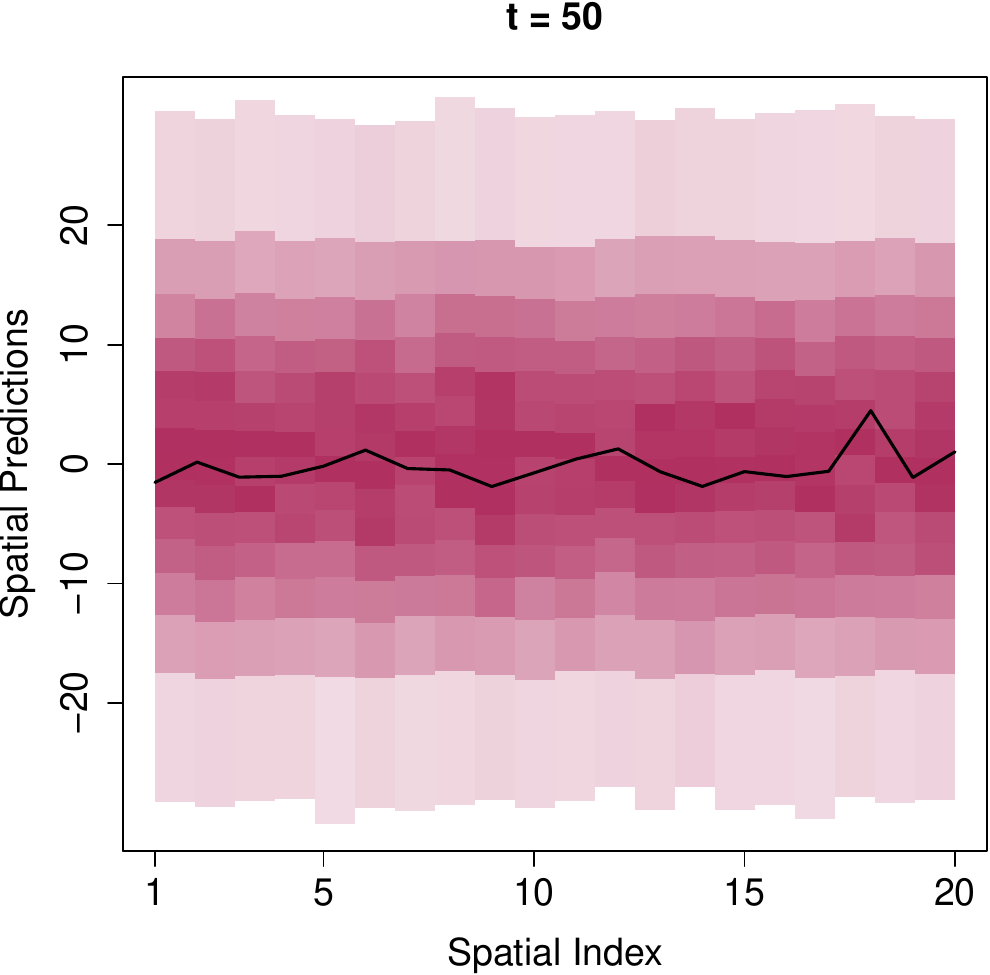}}\\
	\caption{Simulation study: posterior spatial predictions with respect to spatial indices at the time points $t$ as in Figure \ref{fig:spatial_plots_simstudy2} 
	assuming $\phi(\bs_i,t_k)=\phi_0(\bs_i,t_k)$ for $i=1,\ldots,n$ and $k=1,\ldots,m$.}
	\label{fig:spatial_plots2_simstudy2}
\end{figure}

\section{Spatio-temporal nonstationarity of the sea surface temperature data}
\label{sec:nonstationary}
In real life situations, the assumption of stationarity, or even covariance stationarity of the underlying spatio-temporal process, are usually too naive
to be realistic. In our case, Figure \ref{fig:sst_plots} alone vindicates that for different time points, the spatial distributions of temperature are different,
thus empirically ruling out stationarity. Moreover, for different spatial locations, the distributions of the time series are also different, as can be observed
from the thick black lines in Figure \ref{fig:temporal_plots_realdata2}. 

For more formal conclusion regarding nonstationarity of the 
sea surface temperature data, we adopt the recursive Bayesian theory and methods developed by \ctn{Roy20} in this regard. 
In a nutshell, their key idea is to consider the Kolmogorov-Smirnov distance
between distributions of data associated with local and global space-times. 
Associated with the $j$-th local space-time region is an unknown probability $p_j$ of the event 
that the underlying process is stationarity when the observed data corresponds to the $j$-th local region and the Kolmogorov-Smirnov distance falls below $c_j$,
where $c_j$ is any non-negative sequence tending to zero as $j$ tends to infinity. With suitable priors for $p_j$, \ctn{Roy20} constructed recursive posterior
distributions for $p_j$ and proved that the underlying process is stationary if and only if for sufficiently large number of observations
in the $j$-th region, the posterior of $p_j$ converges to one as $j\rightarrow\infty$.
Nonstationarity is the case if and only if the posterior of $p_j$ converges to zero as $j\rightarrow\infty$. 

Covariance stationarity has been treated by \ctn{Roy20} using similar principles, replacing the local and global distributions by local and
global covariances of spatio-temporal lag $\|\bh\|$, where $\bh$ is the difference between the spatial-temporal co-ordinates, and $\|\cdot\|$ is the Euclidean distance. 
The process is covariance stationarity if and only if for sufficiently large number of observations 
in the $j$-th region, the posterior of $p_j$ converges to one as $j\rightarrow\infty$, for all $\|\bh\|>0$. 
On the other hand, the process is covariance nonstationary if and only if
there exists $\|\bh\|>0$ such that the posterior of $p_j$ converges to zero as $j\rightarrow\infty$.

In our implementation of the ideas of \ctn{Roy20}, we set the $j$-th local region to be the entire time series for the spatial location $\bs_j$, for $j=1,\ldots,2520$.
Thus, the size of each local region is $398$, which is sufficiently large for our purpose. The number of regions, $2520$, is also large enough for the Bayesian
recursive theories to be applicable. To check stationarity, we choose $c_j$ to be of the same nonparametric, dynamic and adaptive form as detailed in \ctn{Roy20}. 
The dynamic form requires an initial value for the sequence. 
It is important to remark here that in practice, the choice of the initial value has significant effect on the convergence of the posteriors of $p_j$,
and hence such a choice must be carefully made. However, in our case, for such a large dataset we expect initial values even close enough to zero 
to suffice for inferring stationarity if the underlying phenomenon is indeed stationary. As it turned out, for all initial values less than or equal to $0.26$,
the recursive Bayesian procedure led to the conclusion of nonstationarity of the underlying spatio-temporal process. 

\begin{figure}
	\centering
	\includegraphics[width=7.5cm,height=6.5cm]{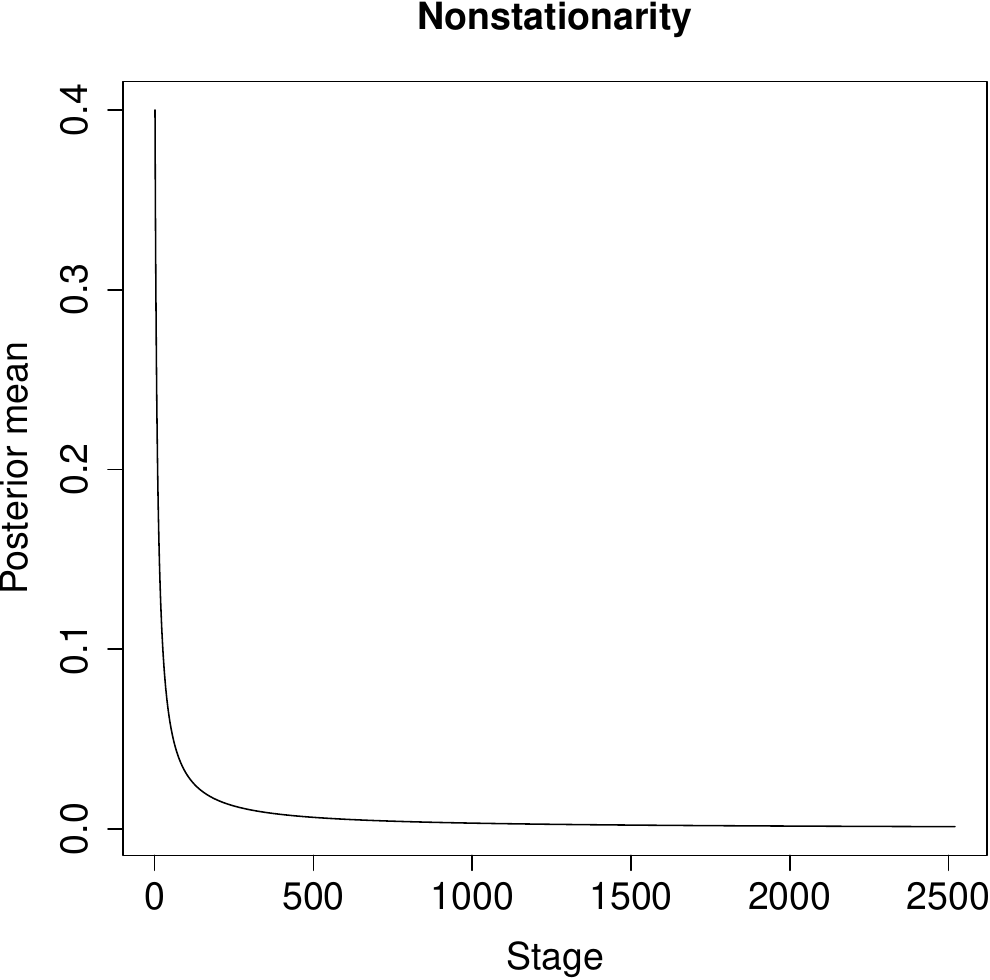}
	\caption{Real data analysis: detection of strict nonstationarity.}
	\label{fig:nonstationarity_realdata2}
\end{figure}

We implemented the idea with our parallelised C code on $2$ parallel processors of our ordinary laptop; the time taken is just $2$ seconds.
For the initial value $0.26$, 
Figure \ref{fig:nonstationarity_realdata2}
displays the means of the posteriors of $p_j$; $j=1,\ldots,2520$, showing clear convergence to zero. The respective posterior variances are negligibly small
and hence not shown.  
Thus, as already anticipated, the spatio-temporal process that generated the sea surface temperature data, can be safely regarded
as nonstationary.

\begin{figure}
	\centering
	\subfigure [$0\leq\|\bh\|<1.5$.]{ \label{fig:covns1}
	\includegraphics[width=6.5cm,height=5.5cm]{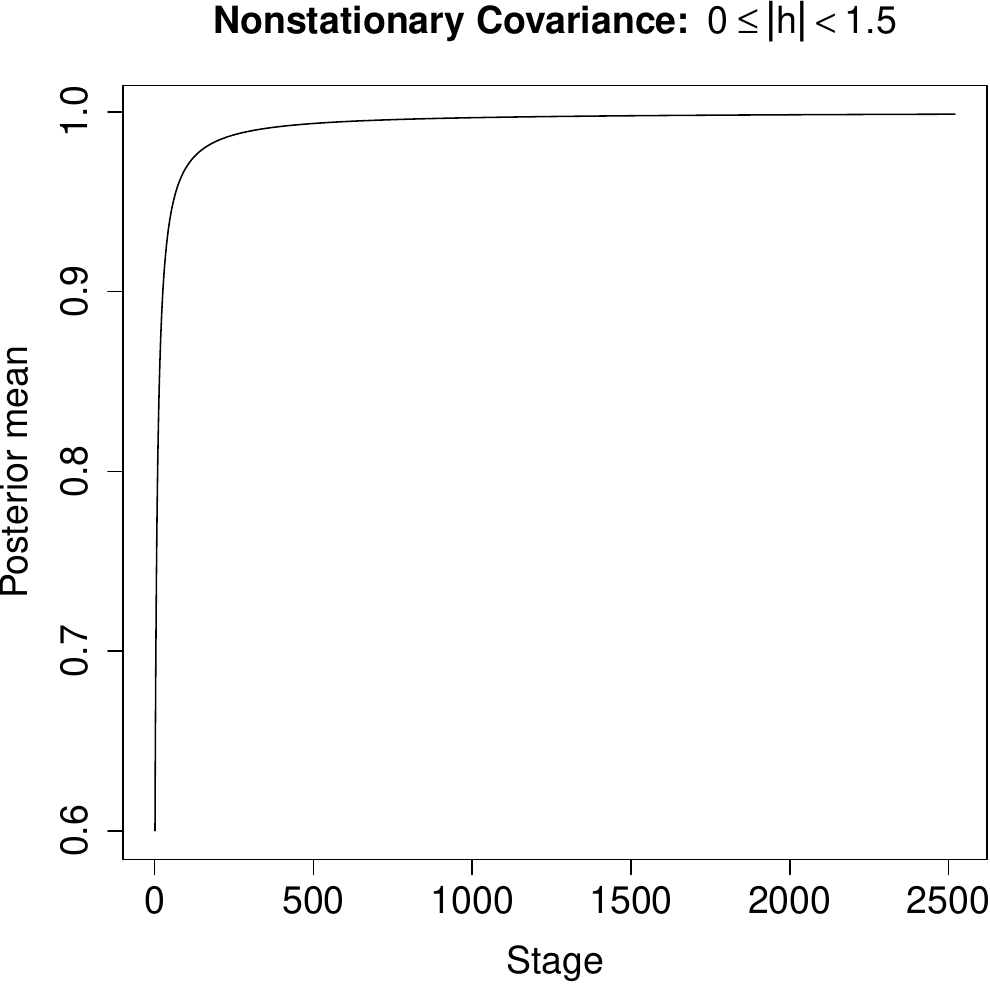}}
	\hspace{2mm}
	\subfigure [$1.5\leq\|\bh\|<2.5$.]{ \label{fig:covns2}
	\includegraphics[width=6.5cm,height=5.5cm]{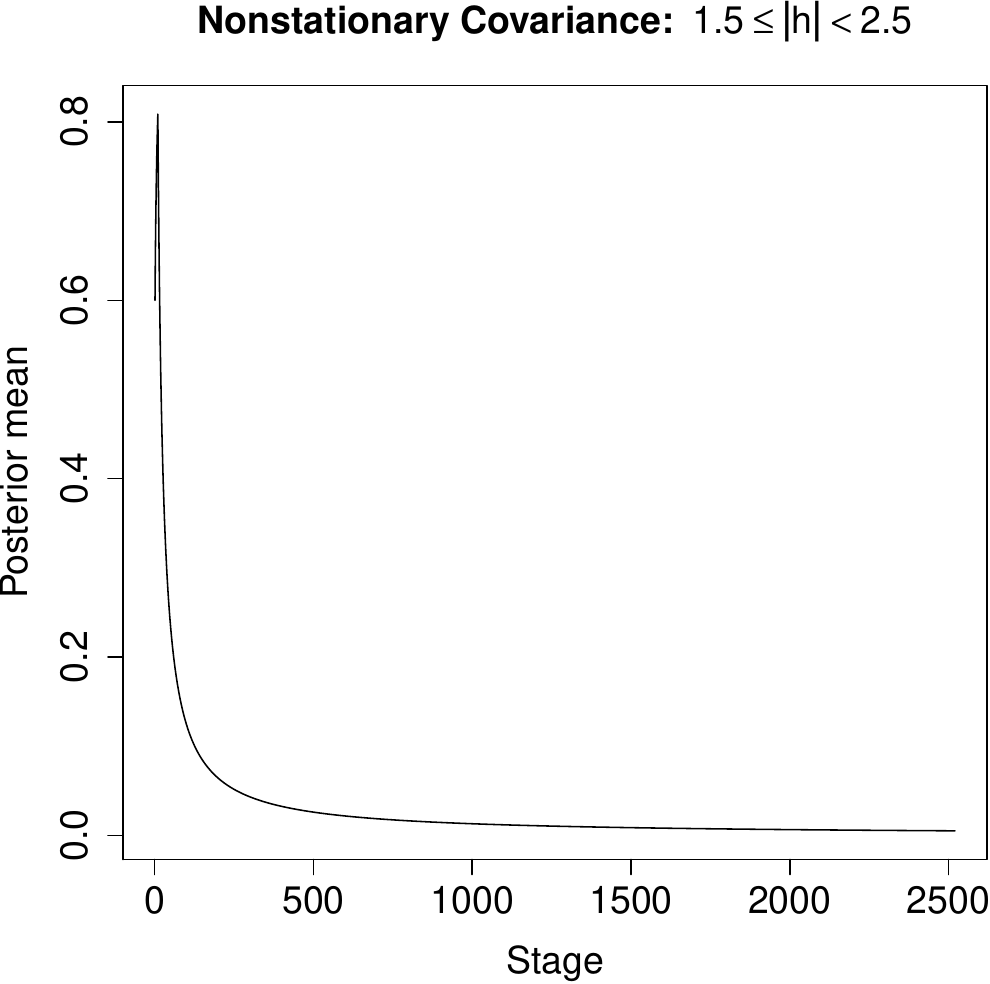}}\\
	\vspace{2mm}
	\subfigure [$2.5\leq\|\bh\|<3.5$.]{ \label{fig:covns3}
	\includegraphics[width=6.5cm,height=5.5cm]{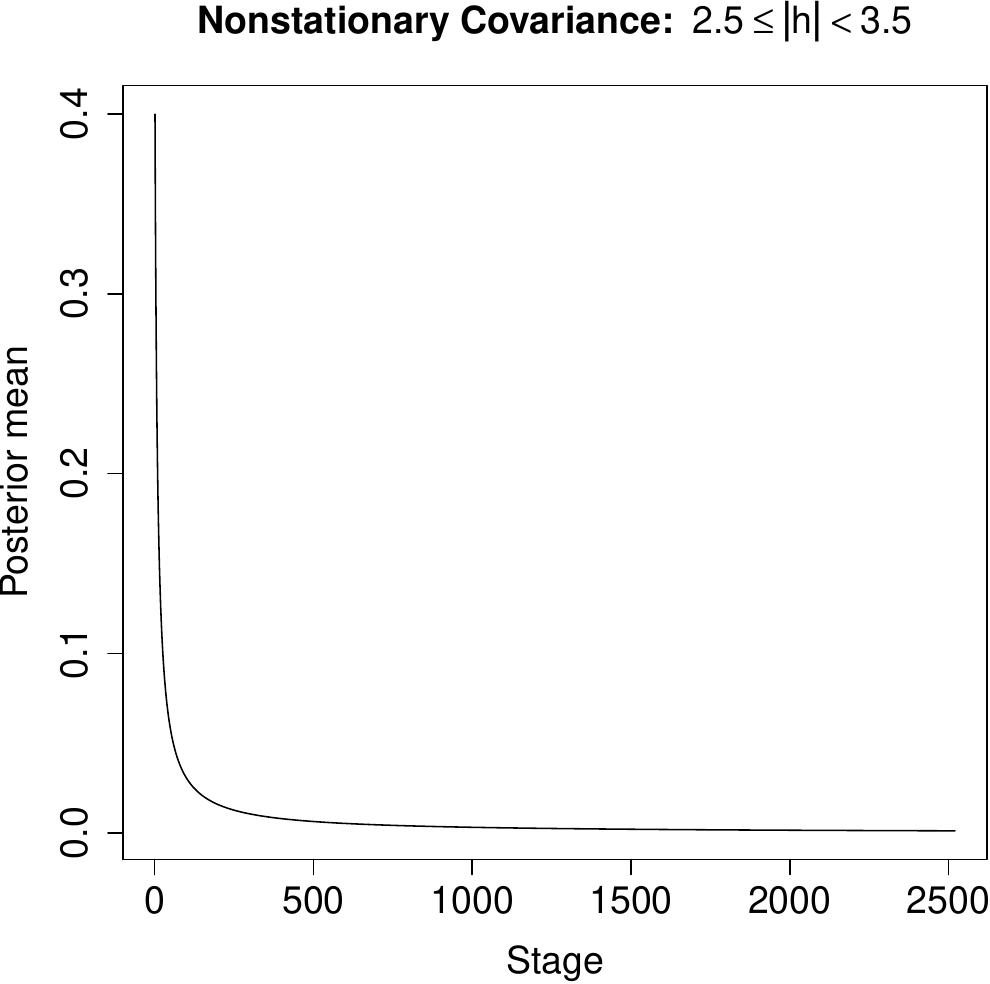}}
	\hspace{2mm}
	\subfigure [$3.5\leq\|\bh\|<4.5$.]{ \label{fig:covns4}
	\includegraphics[width=6.5cm,height=5.5cm]{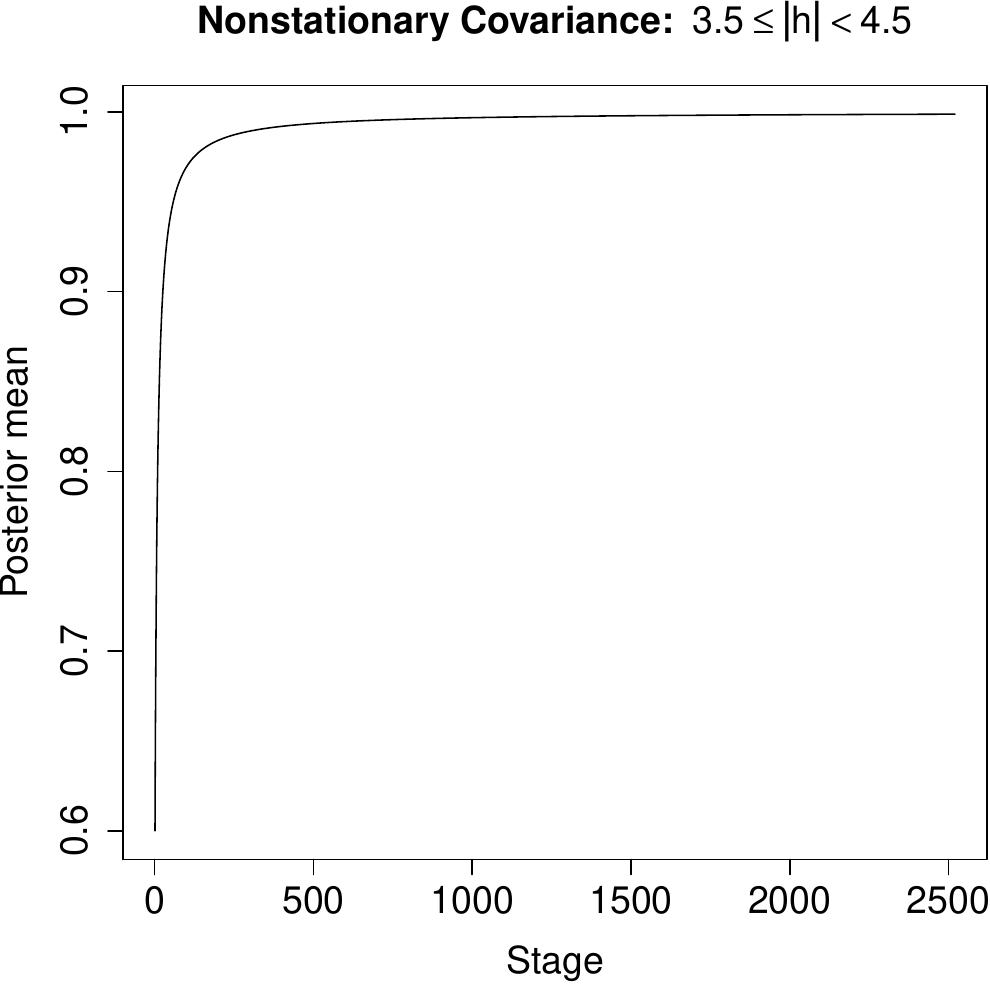}}\\
	\vspace{2mm}
	\subfigure [$4.5\leq\|\bh\|<5.5$]{ \label{fig:covns5}
	\includegraphics[width=6.5cm,height=5.5cm]{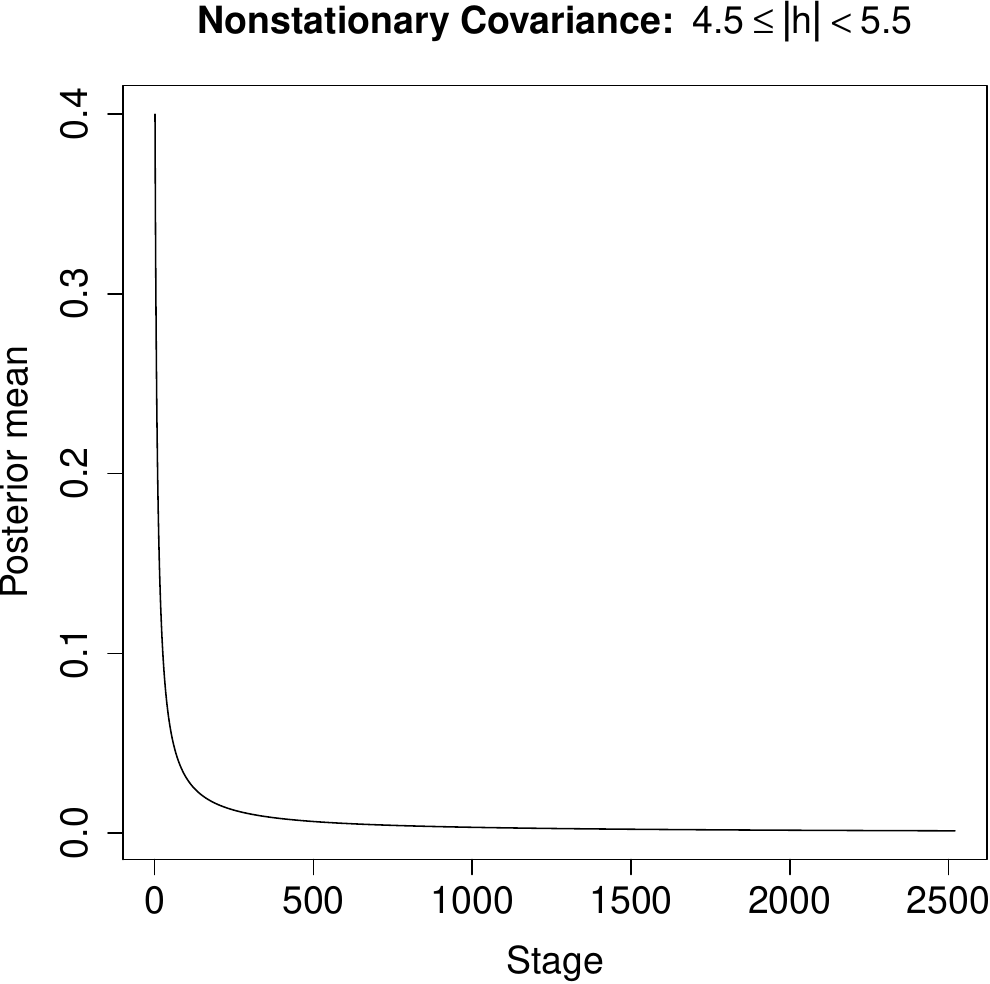}}
	\hspace{2mm}
	\subfigure [$5.5\leq\|\bh\|<6.5$.]{ \label{fig:covns6}
	\includegraphics[width=6.5cm,height=5.5cm]{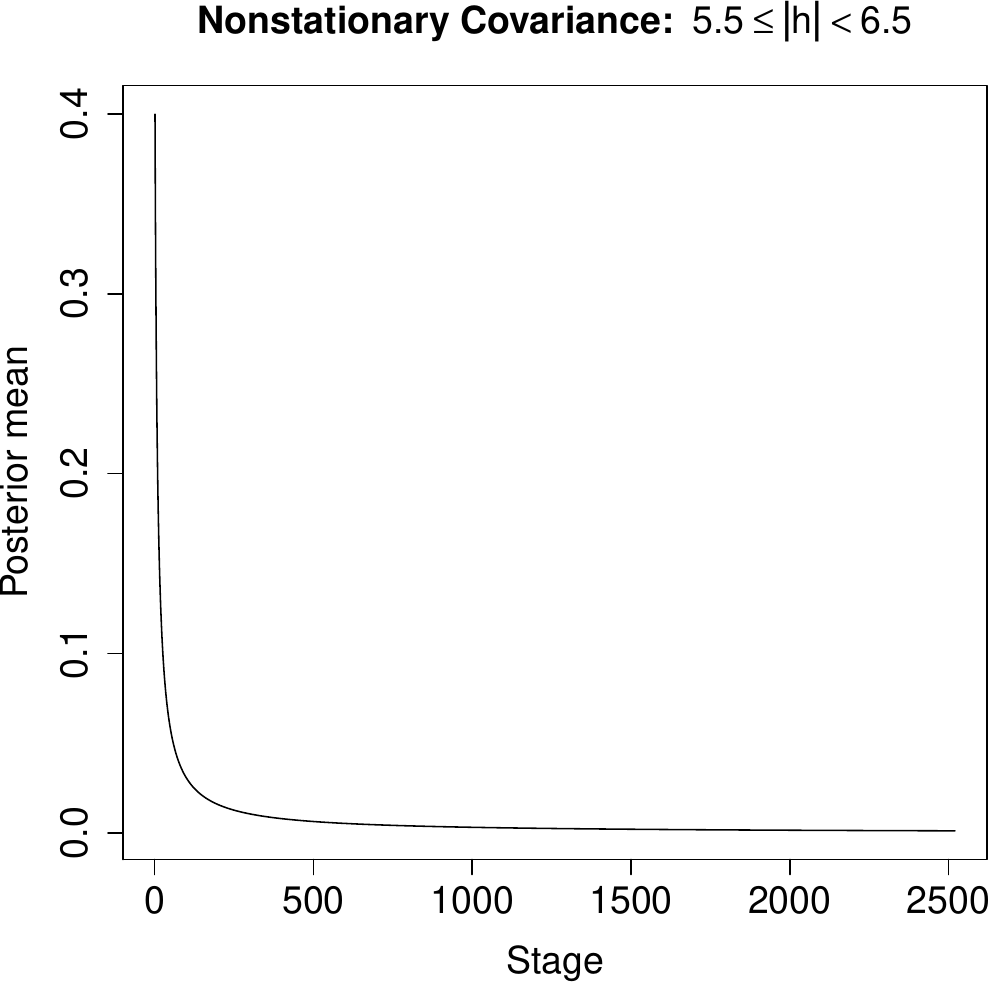}}\\
	\caption{Real data analysis: detection of covariance nonstationarity.}
	\label{fig:cov_nonstatonarity_realdata2}
\end{figure}

Figure \ref{fig:cov_nonstatonarity_realdata2} shows the results of our investigation of covariance stationarity. Panels (b), (c), (e) and (f) show
convergence of the posterior means of $p_j$ in this context to zero for different partitioned intervals of $\|\bh\|$ associated with sufficient data such that 
the covariances are well-defined. The posterior variances are again negligibly small as before.
Thus, covariance nonstationarity of the underlying spatio-temporal process is also clearly indicated. In these cases, we chose the initial values of $c_j$ to be $0.05$.
The time taken for our parallelised C code implementation on our dual-core laptop is only $5$ seconds for each $\|\bh\|$.
Hence, the sea surface temperature phenomenon is not even weakly stationary.

\section{Convergence of lagged spatio-temporal correlations to zero for the sea surface temperature data}
\label{sec:zero_correlation}
Recall that two major purposes of our L\'{e}vy-dynamic spatio-temporal model is to account for nonstationarity of most real-life data and to emulate the property 
of most real datasets that the lagged spatio-temporal correlations tend to zero as the spatio-temporal lag $\|\bh\|$ tends to infinity, in spite of nonstationarity. 
For the sea surface temperature dataset, we have already confirmed strict nonstationarity as well as covariance nonstationarity. It now remains to address the issue
of the lagged correlations. 

For our purpose, we randomly select $25$ spatial locations and consider the entire time series associated with each of them. We further augment the data with the first
$50$ time points of another random location, thus yielding a dataset of size $10,000$. 
We compute the lagged correlations on $80$ parallel processors on our VMWare, each processor computing the correlation for a partitioned interval of lag $\|\bh\|$
such that the interval is associated with sufficient data making the correlation well-defined.
The time taken for this exercise is about $3$ minutes. 

\begin{figure}
	\centering
	\includegraphics[width=7.5cm,height=6.5cm]{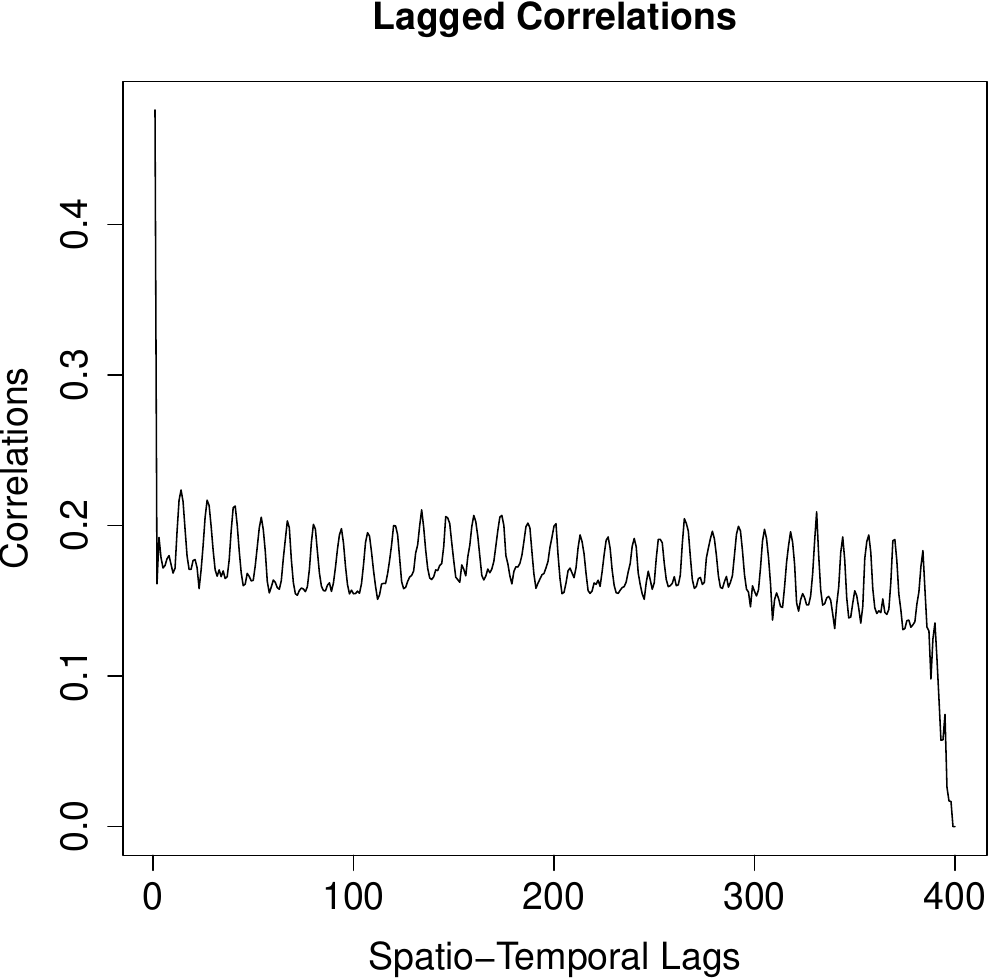}
	\caption{Real data analysis: lagged spatio-temporal correlations converging to zero.}
	\label{fig:zero_correlation_realdata2}
\end{figure}

Figure \ref{fig:zero_correlation_realdata2} shows that with respect to this real dataset, our expectation that the lagged spatio-temporal correlations converge to zero 
in spite of nonstationarity, is not unfounded. Further experiments with larger data sizes, but with much longer implementation times, corroborated this result.


\section{Non-Gaussianity of the sea surface temperature data}
\label{sec:non_gaussian}
Simple quantile-quantile plots (not shown for brevity) revealed that the distributions of the time series data at the spatial locations, distributions of the spatial data
at the time points, and the overall distribution of the entire dataset, are far from normal. Thus, traditional Gaussian process based models of the 
underlying spatio-temporal process are ruled out. Since the temporal distributions at the spatial locations and the spatial distributions at different time points are
also much different, it does not appear feasible to consider parametric stochastic process models for the data. In this regard as well, relevance of our nonparametric 
L\'{e}vy-dynamic process is quite pronounced.

In other words, we have validated that the underlying spatio-temporal process that generated the 
sea surface temperature data is non-Gaussian, strictly and weakly nonstationary, and the lagged correlations converge to zero
as the lags tend to infinity. Moreover, there is no reason to traditionally assume separability of the spatio-temporal covariance structure.
Since our nonparametric L\'{e}vy-dynamic process is endowed with all the aforementioned characteristics, it seems to be a very appropriate candidate for 
analysing the data.






\renewcommand\baselinestretch{1.3}
\normalsize
\bibliography{irmcmc}


\end{document}